\documentclass{ws-ijmpa}

\usepackage{graphicx}
  \usepackage{epsfig}
  \usepackage{amssymb}
  \usepackage{subfigure}
  \usepackage{axodraw}

\def\eqref#1{Eq.(\ref{#1})}

\def\itref#1{(\ref{#1})}

\def\ep{\epsilon}
\def\oo{\infty}

\def\e{\hbox{e}}

\def\ie{{\it i.e.}\phantom{.}}

\begin{document}

\markboth{M. Argeri \& P. Mastrolia}
{Feynman Diagrams \& Differential Equations}

%
\catchline{}{}{}{}{}
%

\title{FEYNMAN DIAGRAMS AND DIFFERENTIAL EQUATIONS}

\author{MARIO ARGERI}

\address{Chemistry Department, Napoli University \\ Via Cinthia, Napoli,
Italy\footnote{{\tt mario.argeri@unina.it}}\\
}

\author{PIERPAOLO MASTROLIA}

\address{Institut f\"ur Theoretische Physik,
Universit\"at Z\"urich \\ CH-8057 Z\"urich, Switzerland\footnote{{\tt pierpaolo.mastrolia@physik.unizh.ch}}\\
}

\maketitle

\begin{history}
\received{Day Month Year}
\revised{Day Month Year}
\end{history}

\begin{abstract}
We review in a pedagogical way the method of differential equations 
for the evaluation of D-dimensionally regulated Feynman integrals. 
After dealing with the general features of the technique,
we discuss its application in the context of 
one- and two-loop corrections to the photon propagator in QED,
by computing the Vacuum Polarization tensor exactly in D.
Finally, we treat two cases of less trivial
differential equations, respectively associated to 
a two-loop three-point, and a four-loop two-point integral.
These two examples are the playgrounds 
for showing more technical aspects 
about: Laurent expansion of the differential equations in D (around D=4); 
the choice of the boundary conditions; and the link among 
differential and difference equations for Feynman integrals.

\end{abstract}

\ccode{PACS numbers: 02.30.Hq, 02.70.Bf, 11.10-z, 12.20.-m}

\newpage
\begin{section}{Preface \label{Preface}}

That ``{\it differentiating} is an operation easier than {\it integrating}'' is
a statement that should not sound too surprising;
while, a more pleasant wonder might result when suitable
{\it differentiations} make us reduce, if not avoid at all, the number of 
direct {\it integrations} - 
of course the two operations, being the inverse of each other,
have not to be thought as performed with respect to the same variable!

\noindent
As paradigmatic example, let us just consider the class of integrals,
$$
I_{n}(\alpha) = \int_0^\infty \e^{- \alpha x^2} \ x^n \ dx \ .
$$
For $n=0$, this is just the Gaussian integral,
$$
I_0(\alpha) = {1 \over 2} \sqrt{ \pi \over \alpha} \ ;
$$
while for $n=1$, the integrand is integrable by quadrature, 
$$
I_1(\alpha) = {1 \over 2 \ \alpha} \ .
$$
To compute $I_n$ with $n>1$, one can use the identity,
$$
- {\partial \over \partial \alpha} I_{n-2} = I_{n} \ .
$$
In fact, 
\begin{itemize}
\item $n=2s$
$$
I_{n} = \bigg(- {\partial \over \partial \alpha} \bigg) I_{n-2} 
      = \bigg(- {\partial \over \partial \alpha} \bigg)^2 I_{n-4}
      = \ldots 
      = \bigg(- {\partial \over \partial \alpha} \bigg)^{n \over 2} I_0 \ ;
$$

\item $n=2s+1$
$$
I_{n} = \bigg(- {\partial \over \partial \alpha} \bigg) I_{n-2} 
      = \bigg(- {\partial \over \partial \alpha} \bigg)^2 I_{n-4}
      = \ldots 
      = \bigg(- {\partial \over \partial \alpha} \bigg)^{(n-1) \over 2} I_1 \ .
$$
\end{itemize}
Therefore the infinite set of integrals $I_n$ can be computed without
any integration, provided
the knowledge of just two basic integrals, namely $I_0$, and $I_1$, 
that in the forthcoming terminology would be defined as the 
{\it master integrals} of the class $I_n$. 

The above example was a too lucky one:
{\it i)} the repeated $\alpha$-derivative did not entangle integrals
having even and odd indices, therefore $I_0$ and $I_1$ 
never appear linked by any differential identity; 
{\it ii)} the value of the master integrals was known,
possibly obtained by direct integrations.

In the more general case masters' are unknown, 
and their evaluation becomes an open problem.
In the following pages, we will see
how the exploitation of integration-by-parts not only yields
algebraic relations among infinite sets of integrals and their masters', 
but as well leads to differential equations satisfied by the master 
integrals themselves. 
Solving these differential equations becomes a tool for computing master 
integrals, when their direct integration is not viable. \\
As it happens to (many) Feynman integrals.

\end{section}

\begin{section}{Introduction \label{Intro}}

A perturbative approach to the quantitative description of 
the scattering of particles in quantum field
theory involves the computation of Feynman diagrams.
For a given number of external particles - the \emph{legs} of diagram - fixed by the process under study, 
and a given order in perturbation theory, the skeletons of diagrams are built up by joining the edges of 
legs and propagators into vertexes, forming \emph{tree} patterns and closed loops.\\
Beyond the tree level, each Feynman diagram represents an integral which has, in general, 
a tensorial structure, induced by the tensorial nature of the 
interacting fields.
Therefore, the result of its evaluation must be a linear combination
of the tensors provided by the theory and by the kinematics of the 
process under study. The coefficients of this linear combination, usually called \emph{form factors}, 
can be always extracted from each Feynman diagram, before performing any evaluation, by means of 
suitably chosen projectors.\\
These form factors are \emph{scalar integrals}  closely connected to 
the original Feynman diagram: the numerator of their integrand may contain all the possible scalar 
products formed by external momenta and loop variables; whereas its denominator is formed by 
the denominators of propagators present in the diagram itself.\\
Due to the bad convergence of loop 
integrals in four dimensions, regularization prescriptions are mandatory.
Hereafter the integrals are regularised within the framework of  
't Hooft-Veltman  continuous-dimensional regularisation scheme\cite{DimReg}. 
Accordingly, the dimension $D$ of an 
extended integration space is used as a regulator for  
both infrared (IR) and ultraviolet (UV) divergences, 
which finally do appear as poles in $(D-4)$ when $D$ goes to 4 
\cite{DimReg}. \\ 
The aim of a precise calculation is to 
compute Feynman diagrams for any value of the available kinematic invariants.
Except in case of simple configurations
(e.g. very few legs and/or few scales),
quite generally, approximations have to be taken
by limiting the result to a specific kinematics domain,
and, thus, looking for a hierarchy among the scales, to get rid of the ones
which anyhow would give a negligible contribution in that domain. 

The puzzling complexity of the Feynman diagrams calculation arises 
because of two different sources: either multi-leg or 
multi-loop processes. In recent years the progress in the evaluation 
of higher loop radiative
corrections in quantum field theory has received a strong boost, due to
the optimisation and automatising of various techniques
(see refs. in \cite{Buttar:2006zd,Smirnov:2004ym,Weinzierl:2006qs}). 
In this work we review one of the most effective computational tools
which have been developed in the framework of the dimensional regularization:
the method of {\it differential equations for Feynman integrals}.\\

The method was first proposed by Kotikov\cite{Kotikov:1990kg} 
in the early nineties, 
while dealing with the evaluation of 2- and 3-point functions.
The basic idea was to consider a given unknown integral as a 
function of one of the 
propagator masses, and to write for it a differential equation in 
that variable. 
Thus, instead of addressing its direct integration,
the value of the integral could be found by solving the differential equation.

The advantages of the novel ideas were soon realized 
\cite{Boos:1990rg,Kotikov:1991hy,Kotikov:1991hm,Lunev:1994sz,Scharf:1994},
and generalised at a later stage by Remiddi \cite{Remiddi:1997ny}, who proposed the 
differentiation with respect to any other available kinematics invariants formed by the
external momenta. That enabled the application of the differential equation method also to integral
with massless propagators (provided the existence of any other non-trivial scale).

Finally, Remiddi and Gehrmann \cite{Gehrmann:1999as,Gehrmann:2000zt,Gehrmann:2001ck}
fully developed the method by showing its effectiveness through the systematic application to 
a non-trivial class of two-loop four-point functions, whose 
result is still considered as state-of-the-art. 

From that moment on, the method of differential equations 
became to be widely used in different contexts \cite{Ford:1991hw,Caffo:1998yd,Caffo:1998du,Bonciani:1999mj,Kotikov:2000ye,Czachor:2001mv,Caffo:2002we,Caffo:2002wm,Mastrolia:2002gt,Aglietti:2004vs,Bonciani:2004dz,Smirnov:2006ry}.
The lists of unprecedented results obtained through its application spans among 
multi-loop functions from zero to four external legs,
\cite{Gracey:1992ew,Tarasov:1997kx,Fleischer:1997bw,Berends:1997vk,Chung:1998mz,Fleischer:1998nb,Davydychev:1999ic,Fleischer:1999aa,Fleischer:1999hp,Fleischer:1999tu,Fleischer:1999mp,Fleischer:2000vb,Smirnov:2002kq,Smirnov:2002je,Onishchenko:2002ri,Martin:2003qz,Remiddi:2003ci,Remiddi:2004kv,Actis:2004bp,Birthwright:2004kk,Kniehl:2005yc,Bogdan:2005rg,Kniehl:2005bc,Tarasov:2006nk,Dunne:2003tr,Martin:2003it,Martin:2005qm,Dunne:2006sx,Gracey:2006jc} and within the most interesting sectors of particle phenomenology,
like jets physics \cite{Gehrmann:1999as,Gehrmann:2000zt,Gehrmann:2001ck,Anastasiou:2000mf,Gehrmann:2000xj,Smirnov:2000ie,Gehrmann:2001ih,Gehrmann:2001ru,Giele:2002hx,Frixione:2002kn,allonshell,oneoffshell}; 
QED corrections to lepton form factors \cite{Bonciani:2003te,Bonciani:2003ai,Bonciani:2004xp}; 
Bhabha Scattering \cite{Smirnov:2001cm,Bonciani:2003cj,Bonciani:2004gi,Heinrich:2004iq,Bonciani:2004qt,Czakon:2004wm,Bonciani:2005im,Bonciani:2006qu,Czakon:2006hb};
QCD corrections to lepton form factors and top-physics \cite{Bonciani:2003hc,Bernreuther:2004ih,Bernreuther:2004th,Bernreuther:2005rw,Bernreuther:2005gq}, 
forward-backward asymmetry of heavy-quark \cite{Neubert:1993mb,Juste:2006sv,Bernreuther:2006yt,Bernreuther:2006vp,Bonciani:2006eu};  
Higgs Physics \cite{Maher:1993vj,Aglietti:2004nj,Aglietti:2004ki,Bernreuther:2005gw,Anastasiou:2006hc,Aglietti:2006tp}; 
Electroweak sector \cite{DeFazio:2000up,Jegerlehner:2002em,Awramik:2004ge,Awramik:2004qv,Freitas:2005vb,Awramik:2006ar}; 
Sudakov form factors \cite{Aglietti:2003yc,Aglietti:2004tq,Aglietti:2007as};
semileptonic decay \cite{Seidel:2004jh,Melnikov:2005bx,Asatrian:2006sm}; 
static parameters and gauge boson properties \cite{Kotikov:2001ct,Kotikov:2002ab,Jegerlehner:2003py,Anastasiou:2003ds,Chetyrkin:2004fq,Schroder:2005va,Faisst:2006sr}.

The efforts to achieve analytic solutions of differential equations for Feynman integrals has
stimulated new developments on the more mathematical side \cite{Broadhurst:1998ke,Davydychev:1998fk},
especially concerning the properties of transcendental functions 
\cite{Moch:2001zr,Weinzierl:2002hv,Blumlein:2003gb,Kalmykov:2004kg,Kalmykov:2006hu}.
In particular, a novel set of functions that generalize 
Nielsen polylogarithms, the so called Harmonic Polylogarithms (HPL) 
\cite{HPL,Maitre:2005uu,Maitre:2007kp}, 
have been found suitable for casting the result in {\it analytic} form - 
that means without ambiguities due to zeroes
hidden in functional relations, and supplied with series expansions.
While HPL's can be considered as iterative integral of rational kernels,
recently it has been pointed out that the solution of differential equations
for generic integrals with massive loops demands as well for irrational 
integration kernels, yielding elliptic functions \cite{Laporta:2004rb,Aglietti:2007as}.

The range of applicability of the differential equations technique is 
broadened by the possibility of a natural
switch toward a semi-numerical approach, since,
whenever the analytic integration were not required or not viable, 
the differential equation(s), analytically obtained, could be 
solved with numerical techniques \cite{Caffo:2003gh,Caffo:2003ma}.

Nowadays we are not at the point to have {\it the} method for evaluating any Feynman integral,
but certainly we dispose of several tools \cite{Davydychev:1996fn,Anastasiou:1999ui,Binoth:2000ps,Kotikov:2000yd,Laporta:2001dd,Passarino:2001wv,Kotikov:2001sd,Laporta:2002pg,Ferroglia:2002mz,Binoth:2003ak,Duplancic:2003tv,Suzuki:2003jn,Smirnov:2004ip,Czakon:2005rk,Anastasiou:2005cb,Smirnov:2006vt,Weinzierl:2006qs} to attack successfully many problems in perturbation theory, and usually 
a combination of them is necessary for the achievement of the final answer.
Therefore let us discuss in detail how to build ans solve differential equations for integral 
associated to Feynman graphs.

The computational strategy is twofold.

\begin{itemize}
\item In a preliminary stage, 
by exploiting some remarkable properties of the dimensionally regularised 
integrals, 
namely {\it{integration-by-parts identities}} (IBP), {\it{Lorentz invariance identities}} (LI),
and further sets of identities due to kinematic symmetry specific of each diagram, 
one establishes several relations among the whole set of scalar integrals 
associated to the original Feynman diagram.

By doing so, one reduces the result, initially demanding for a large number of  scalar integrals 
(from hundreds to billions, according to the case), to  
a combination of a minimal set (usually of the order of tens) of independent functions, 
the so called {\it{master integrals}} (MI's).
\\
\item The second phase consists of the actual evaluation of the MI's. 
By using the set of identities previously 
obtained, it is also possible to write Differential Equations in the 
kinematic invariants which are satisfied by the MI's themselves. 
When possible, these equations can be solved exactly in $D$ dimensions. 
Alternatively, they can be Laurent-expanded 
around suitable values of the dimensional parameter up to the required order, 
obtaining a system of chained differential equations 
for the coefficients of the expansions, which, in the most general case, are finally integrated
by Euler's {\it variation of constants} method.
\end{itemize}

One of the key advantages of the method is that it yields a clear separation
between 
the merely algebraic part of the work - which is, not surprisingly, 
always very heavy in multiloop calculations, and can be 
most conveniently processed by a computer algebra program\cite{ibpsolver}-, 
from the actual analytic issues of the problem, 
which can then be better investigated without the disturbance of the 
algebraic complexity.\\
The paper is organised as follows.
In section 2, it is described how to reduce a generic 
(combination of) Feynman integrals to a limited set of MI's
and how to write the system of {\it differential equations} they fulfil. 
As illustrative applications, respectively in section 3-4, the one-loop and two-loop 
vacuum polarization functions in QED are explicitly computed. 
In the further two sections, we discuss some less obvious example of differential equations.
In section 5, we describe the solution of a system of three coupled first-order differential equations, 
to compute three MI's associated to a class of two-loop 3-point functions,
addressing as well the problem of finding their boundary conditions.
While in section 5, we describe the solution of a fourth-order 
differential equation involving the MI's
of a 4-loop 2-point diagram, and it will be considered the link between 
differential and {\it difference equation} for Feynman integrals
\cite{Laporta:2001dd}.
\end{section}

\begin{section}{Reduction to Master Integrals}

\begin{subsection}{Topologies and Integrals}

Let us consider a Feynman diagram with $\ell$ loops, $g$ external legs, 
$d$ internal lines and a given tensorial structure.
Such a diagram, when not representing a scalar quantity, 
can be decomposed as a combination of products 
of a scalar form factor times a tensor. 
Thus, computing the contribution of any diagram to a given process is equivalent to the 
computation of scalar factors, which,
after some preliminary algebra 
(evaluation of Dirac traces and contraction of Lorentz indices) 
reads as a combination of scalar integrals like,
\begin{equation}\label{INTRO:eq1}
\int{d^D k_{1} \over (2 \pi)^{D-2}}
{d^D k_{2} \over (2 \pi)^{D-2}}
\cdots
{d^D k_{\ell} \over (2 \pi)^{D-2}}
\frac{\Pi_{i=1}^{N_{sp}}S_{i}^{n_{i}}}
{\Pi_{j=1}^{d}{\cal{D}}_{j}},
\end{equation} 
 where
\begin{itemize}
\item $S_{i}\,\,\,(i=1,\cdots,N_{sp})$ is any scalar product formed by 
either one of independent external momentum and an 
internal loop momentum, or by two internal loop momenta ($n_{i}$ is an integer exponent such that $n_{i}\geq 0$) and $N_{sp}$ is the total number of such scalar products, given by
\begin{equation}\label{INTRO:eq2}
N_{sp}=\underbrace{\ell(g-1)}_{s.p.\,\,external-internal}+\underbrace{\frac{\ell(\ell+1)}{2}}_{s.p.\,\,internal-internal}=\ell\left(g+\frac{\ell}{2}-\frac{1}{2}\right)
\end{equation}
\item ${\cal{D}}_{j}=q_{j}^2+m_{j}^2\,\,\,(j=1,\cdots,d)$ is the denominator of the $j$-th propagator being  $q_{j}$ and $m_{j}$, respectively, the corresponding momentum and mass. From now, we call ${\cal{D}}_{j}$ \emph{propagator}. 
\end{itemize} 

The first task relies on a suitable classification of the integrals, in order to
minimize the number of the ones which have to be actually integrated.

As a preliminary set up, one is invited to consider
not anymore diagrams but \emph{topologies}, that, drawn exactly as 
scalar Feynman diagrams, contain only and all different 
propagators (and scalar vertices).
In view of settling down a correspondence between a given topology 
and the class of integrals it represents, one proceeds 
by simplifying as well the number of scalar products in the numerator.
The number $N_{sp}$ is a redundant quantity and
can be easily reduced, with a procedure 
commonly called \emph{trivial tensor reduction}, according to which
some (when not all) of the scalar products can be expressed 
in terms of the denominators of the topology. 

In general, if $t$ is the 
number of propagators of a given topology, we can express $t$ of the $N_{sp}$ scalar 
products containing the loop momenta in terms of the denominators - which will 
be later on simplified 
against the corresponding term in the denominator. 
Thus the most general integral associated to every topology does contain only  
the remaining $q \ (=N_{sp}-t)$ \emph{irreducible} scalar products and reads as
\begin{equation}\label{INTRO:eq9}
\int{d^D k_{1} \over (2 \pi)^{D-2}}{d^D k_{2} \over (2 \pi)^{D-2}}\cdots
{d^D k_{\ell} \over (2 \pi)^{D-2}}
\frac{\Pi_{i=1}^{q}S_{i}^{n_{i}}}
{\Pi_{j=1}^{t}{\cal{D}}_{j}^{m_{j}}},
\end{equation} 
where $n_{i}\geq 0$ and $m_{j}\geq 1$.\\

One can denote with $I_{t,r,s}$ the class of the integrals with:
a given set of $t$ denominators; 
$q =N_{sp}-t$ irreducible scalar products;
a total of $r=\sum_{i}(m_{i}-1)$ powers of the $t$ denominators; 
and $s =\sum_{j}n_{j}$ powers of the $q$ scalar products. It can be shown 
that the number of the integrals belonging to the class $I_{t,r,s}$ is
\begin{equation}\label{INTRO:eq10}
N(I_{t,r,s})=\left(\begin{array}{c}r+t-1\\ t-1\end{array}\right)\left(\begin{array}{c}s+q-1\\ q-1\end{array}\right).
\end{equation}

In the next sections, we will see that  
integrals of the type (\ref{INTRO:eq9}), belonging to a given topology, 
therefore differing for the values
of the indices $m_{i},n_{i}$, are not independent.
Algebraic relations among them, can be written in the form of a sum of a 
finite number of terms set equal to zero, where each term is given by 
the product of a polynomial 
(of finite order and with integer coefficients in the variable $D$, 
masses and Mandelstam invariants) and one of the integrals belonging 
to $I_{t,r,s}$. They can be used recursively to express as many as possible integrals 
of a given class in terms of as few as possible (suitably chosen) ones.
The way to generate those kind of relations goes through integration-by-parts, 
Lorentz invariance and symmetry considerations.

Before going on with the discussion let us see explicitly how the general definition 
we have introduced do apply in practice.

{\footnotesize{
\subsubsection{Example: from the diagram to the topology}

Let us consider the Feynman diagram showed in Fig.\ref{INTRO:fig1}

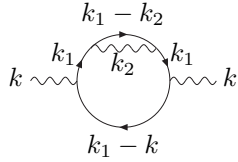
\begin{figure}[ht]
\begin{center}
\begin{picture}(150,50)(0,0)
\SetWidth{0.5}
\SetScale{0.7}
\Photon(25,25)(50,25){2}{3}    
\Photon(100,25)(125,25){2}{3}      
\ArrowArcn(75,25)(25,0,180)
\ArrowArcn(75,25)(25,180,135.5)
\ArrowArcn(75,25)(25,135.5,44.5)
\ArrowArcn(75,25)(25,44.5,0)
\Photon(57.5,42.5)(92.5,42.5){2}{4}
\Text(12,19)[]{$k$}
\Text(93,19)[]{$k$}
\Text(52.5,-7)[]{$k_{1}-k$}
\Text(52.5,43)[]{$k_{1}-k_{2}$}
\Text(30,28)[]{$k_{1}$}
\Text(75,28)[]{$k_{1}$}
\Text(52.5,24)[]{$k_{2}$}
\end{picture}
\end{center}
\caption{Feynman diagram with $g$=2 legs and $\ell=2$ loops}
\label{INTRO:fig1}
\end{figure}
\noindent
with $g=2$ external legs, $\ell=2$ loops and $d=5$ internal 
lines, which gives a number of scalar products amounting to,
\begin{displaymath}
N_{sp}=2\left(2+1-\frac{1}{2}\right)=5 \ .
\end{displaymath}
The most general set of scalar integrals possibly arising from its
computation has the following representation,
\begin{equation}\label{INTRO:eq3}
\int{d^D k_{1} \over (2 \pi)^{D-2}}{d^D k_{2} \over (2 \pi)^{D-2}}\frac{\Pi_{i=1}^{5}S_{i}^{n_{i}}}{\Pi_{j=1}^{5}{\cal{D}}_{j}},
\end{equation} 
where
\begin{displaymath}
\left\{\begin{array}{l}{\cal{D}}_{1}=k_{1}^2+m^2\\
{\cal{D}}_{2}=k_{2}^2\\
{\cal{D}}_{3}=(k_{1}-k_{2})^2+m^2\\
{\cal{D}}_{4}=(k_{1}-k)^2+m^2\\
{\cal{D}}_{5}={\cal{D}}_{1}=k_{1}^2+m^2\\
\end{array}\right.\hspace{1.5cm}
\left\{\begin{array}{l}S_{1}=k_{1}^2\\
S_{2}=k_{2}^2\\
S_{3}=k_{1}\cdot k_{2}\\
S_{4}=k\cdot k_{1}\\
S_{5}=k\cdot k_{2}
\end{array}\right.
\end{displaymath}
The original diagram contains five internal lines, but two propagators 
are indeed equal, so there are 
only four different propagators. 
Therefore, the integrals (\ref{INTRO:eq3}) actually belong to the {\it simpler} 
set,
\begin{equation}\label{INTRO:eq4}
\int{d^D k_{1} \over (2 \pi)^{D-2}}{d^D k_{2} \over (2 \pi)^{D-2}}\frac{\Pi_{i=1}^{5}S_{i}^{n_{i}}}{{\cal{D}}_{1}^2{\cal{D}}_{2}{\cal{D}}_{3}{\cal{D}}_{4}},
\end{equation}
which can be represented by the topology in Fig. \ref{INTRO:fig2}.
\begin{figure}[t]
\begin{center}
\begin{picture}(150,50)(0,0)
\SetWidth{0.5}
\SetScale{0.7}
\Photon(25,25)(50,25){2}{3}    
\Photon(100,25)(125,25){2}{3}      
\Photon(75,50)(100,25){2}{4}
\ArrowArcn(75,25)(25,0,180)
\ArrowArcn(75,25)(25,180,90)
\ArrowArcn(75,25)(25,90,0)
\Text(12,19)[]{$k$}
\Text(93,19)[]{$k$}
\Text(52.5,-7)[]{$k_{1}-k$}
\Text(32,32)[]{$k_{1}$}
\Text(84,32)[]{$k_{1}-k_{2}$}
\Text(54,24)[]{$k_{2}$}
\end{picture}
\end{center}
\caption{4-denominators topology
}
\label{INTRO:fig2}
\end{figure}
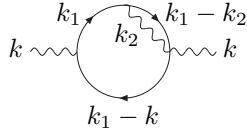

\noindent


The trivial tensor reduction of the scalar product can be realized
according to the following table,
where one chooses to express four (out of five)
scalar products in terms of the denominators.
\begin{table}[h]
\begin{center}
\begin{tabular}{||c|c|c||}
\hline
Scalar product $S_{i}$ & Corresponding propagator & Relationship  \\
\hline
$S_{1}=k_{1}^2$ & ${\cal{D}}_{1}=k_{1}^2+m^2$ & $k_{1}^2={\cal{D}}_{1}-m^2$\\
\hline
$S_{2}=k_{2}^2$ & ${\cal{D}}_{2}=k_{2}^2$ & $k_{2}^2={\cal{D}}_{2}$\\
\hline
$S_{3}=k_{1}\cdot k_{2}$ & ${\cal{D}}_{3}=(k_{1}-k_{2})^2+m^2$ & $k_{1}\cdot k_{2}=\frac{1}{2}({\cal{D}}_{1}+{\cal{D}}_{2}-{\cal{D}}_{3})$\\
\hline 
$S_{4}=k\cdot k_{1}$ & ${\cal{D}}_{4}=(k_{1}-k)^2+m^2$ & $k\cdot k_{1}=\frac{1}{2}({\cal{D}}_{1}-{\cal{D}}_{4}+k^2)$\\
\hline 
\end{tabular}
\end{center}
\end{table}
%
In the end, only one of the five scalar products involving the loop 
momenta is left over as irreducible, arbitrary chosen to be $k_2 \cdot k$.
Therefore, the integrals in (\ref{INTRO:eq3}), represented by the
topology in Fig.\ref{INTRO:fig2}, indeed 
belongs to the class,
%
%
%
%
%
\begin{equation}\label{INTRO:eq8}
{\cal{I}}(n_{1};m_{1},m_{2},m_{3},m_{4})\int{d^D k_{1} \over (2 \pi)^{D-2}}{d^D k_{2} \over (2 \pi)^{D-2}}\frac{(k_{2}\cdot k)^{n_{1}}}{{\cal{D}}_{1}^{m_{1}}{\cal{D}}_{2}^{m_{2}}{\cal{D}}_{3}^{m_{3}}{\cal{D}}_{4}^{m_{4}}}.
\end{equation}
The trivial tensor reduction might as well lead to the complete cancellation of some
denominator. Should it be the case, the resulting integral can be classified as belonging
to the subtopology obtained by pinching the corresponding internal line.\\

}} 

\begin{subsection}{Integration-by-parts Identities}
Integration-by-parts identities (IBP-Id's) are among the most remarkable properties of dimensionally regularized integrals and they were first proposed 
in the eighties by Chetyrkin and Tkachov \cite{ibp}.
The basic idea underlying IBP-Id's is an extension to $D$-dimensional spaces of Gauss' theorem. For each of the integrals defined in equation (\ref{INTRO:eq9}) one can write 
the vanishing of the integral of a divergence given by, 
\begin{equation}\label{INTRO:eq11}
\int{d^D k_{1} \over (2 \pi)^{D-2}}\cdots
{d^D k_{\ell} \over (2 \pi)^{D-2}}
\frac{\partial}{\partial k_{i,\mu}}\left\{v_{\mu}\frac{S_{1}^{n_{1}}\cdots S_{q}^{n_{q}}}{{\cal{D}}_{1}^{m_{1}}\cdots {\cal{D}}_{t}^{m_{t}}}\right\}=0.
\end{equation}
In the above identities the index $i$ runs over the number of loops $(i=1,2,...,\ell)$, and the vector $v_{\mu}$ can be any of the $(\ell+g-1)$ independent vectors of the problem:$k_{1},\cdots,k_{\ell},p_{1},\cdots,p_{g-1}$; in such way, for each integrand, $\ell(\ell+g-1)$ IBP-Id's can be established. When evaluating explicitly the derivatives, one obtains a combination of integrands with a total power of the irreducible scalar products equal to $(s-1)$, $s$ and $(s+1)$ and total powers of the propagators in the denominator equal to $(t+r)$ and $(t+r+1)$, therefore involving, besides the integrals of the class $I_{r,s,t}$, also the classes $I_{t,r,s-1}$, $I_{t,r+1,s}$ and $I_{t,r+1,s+1}$.
Simplifications between reducible scalar products and propagators in the denominator may occur, 
lowering the powers of the propagators. During that simplification, some propagator might disappear, 
generating an integral belonging to a \emph{subtopology}, with $t-1$ propagators.\\
\end{subsection}

\begin{subsection}{Lorentz Invariance Identities}
Another class of identities can be derived by exploiting a general properties of the integrals in (\ref{INTRO:eq10}),
namely their nature as Lorentz scalars. 
If we consider an infinitesimal Lorentz transformation on the external momenta, $p_{i}\rightarrow p_{i}+\delta p_{i}$, where $\delta p_{i}=\omega_{\mu\nu}p_{i,\nu}$ with $\omega_{\mu\nu}$ a totally antisymmetric tensor, we have

\begin{equation}\label{INTRO:eq22}
I(p_{i}+\delta p_{i})=I(p_{i}).
\end{equation}  
Because of the antisymmetry of $\omega_{\mu\nu}$ and because 

\begin{equation}\label{INTRO:eq23}
I(p_{i}+\delta p_{i})=I(p_{i})+\sum_{n}\frac{\partial I(p_{i})}{\partial p_{n,\mu}}=I(p_{i})+\omega_{\mu\nu}\sum_{n}p_{n,\nu}\frac{\partial I(p_{i})}{\partial p_{n,\mu}},
\end{equation}
we can write the following relation

\begin{equation}\label{INTRO:eq24}
\sum_{n}\left(p_{n,\nu}\frac{\partial}{\partial p_{n,\mu}}-p_{n,\mu}\frac{\partial}{\partial p_{n,\nu}}\right)I(p_{i})=0.
\end{equation}
Eq. (\ref{INTRO:eq24}) can be contracted with all possible antisymmetric combinations of the external momenta $p_{i,\mu}p_{j,\nu}$, to obtain other identities for the considered integrals.\\
In case of integral associated to any \emph{vertex} topologies 
with two independent external momenta, $p_{1}$ and $p_{2}$, we can build up the identity 
\begin{equation}\label{INTRO:eq25}
\left[(p_{1}\cdot p_{2})\left(p_{1,\mu}\frac{\partial}{\partial p_{1,\mu}}-p_{2,\mu}\frac{\partial}{\partial p_{2,\mu}}+p_{2}^{2}p_{1,\mu}\frac{\partial}{\partial p_{2,\mu}}-p_{1}^{2}p_{2,\mu}\frac{\partial}{\partial p_{1,\mu}}\right)\right]\parbox{10mm}{\begin{picture}(300,20)(0,0)
\SetScale{.4}
\Line(25,50)(45,40)
\Line(25,0)(45,10)
\Line(80,25)(100,25)
\GOval(60,25)(20,20)(0){.5}
\Text(6,23)[]{\footnotesize{$p_{1}$}}
\Text(6,-3)[]{\footnotesize{$p_{2}$}}
\Text(45,10)[]{\footnotesize{$p_{3}$}}
\end{picture}}\hspace{.7cm}=0
\end{equation} 
In the case of a richer kinematics, like in the case of integrals
associated to \emph{box} topologies with three independent external momenta, 
$p_{1}$, $p_{2}$ and $p_{3}$, we ca write down three LI-id's

\begin{equation}\label{INTRO:eq26}
(p_{1,\mu}p_{2,\mu}-p_{1,\nu}p_{2,\mu})\sum_{n}\left(p_{n,\nu}\frac{\partial}{\partial p_{n,\mu}}-p_{n,\mu}\frac{\partial}{\partial p_{n,\nu}}\right)\parbox{10mm}{\begin{picture}(300,20)(0,0)
\SetScale{.4}
\Line(25,50)(45,40)
\Line(25,0)(45,10)
\Line(75,40)(95,50)
\Line(75,10)(95,0)
\GOval(60,25)(20,20)(0){.5}
\Text(6,23)[]{\footnotesize{$p_{1}$}}
\Text(6,-3)[]{\footnotesize{$p_{2}$}}
\Text(44,23)[]{\footnotesize{$p_{3}$}}
\Text(44,-3)[]{\footnotesize{$p_{4}$}}
\end{picture}}\hspace{.75cm}=0,
\end{equation}

\vspace{.3cm}

\begin{equation}\label{INTRO:eq27}
(p_{1,\mu}p_{3,\mu}-p_{1,\nu}p_{3,\mu})\sum_{n}\left(p_{n,\nu}\frac{\partial}{\partial p_{n,\mu}}-p_{n,\mu}\frac{\partial}{\partial p_{n,\nu}}\right)\parbox{10mm}{\begin{picture}(300,20)(0,0)
\SetScale{.4}
\Line(25,50)(45,40)
\Line(25,0)(45,10)
\Line(75,40)(95,50)
\Line(75,10)(95,0)
\GOval(60,25)(20,20)(0){.5}
\Text(6,23)[]{\footnotesize{$p_{1}$}}
\Text(6,-3)[]{\footnotesize{$p_{2}$}}
\Text(44,23)[]{\footnotesize{$p_{3}$}}
\Text(44,-3)[]{\footnotesize{$p_{4}$}}
\end{picture}}\hspace{.75cm}=0,
\end{equation}

\vspace{.3cm}

\begin{equation}\label{INTRO:eq28}
(p_{2,\mu}p_{3,\mu}-p_{2,\nu}p_{3,\mu})\sum_{n}\left(p_{n,\nu}\frac{\partial}{\partial p_{n,\mu}}-p_{n,\mu}\frac{\partial}{\partial p_{n,\nu}}\right)\parbox{10mm}{\begin{picture}(300,20)(0,0)
\SetScale{.4}
\Line(25,50)(45,40)
\Line(25,0)(45,10)
\Line(75,40)(95,50)
\Line(75,10)(95,0)
\GOval(60,25)(20,20)(0){.5}
\Text(6,23)[]{\footnotesize{$p_{1}$}}
\Text(6,-3)[]{\footnotesize{$p_{2}$}}
\Text(44,23)[]{\footnotesize{$p_{3}$}}
\Text(44,-3)[]{\footnotesize{$p_{4}$}}
\end{picture}}\hspace{.75cm}=0 \ .
\end{equation}

\end{subsection}

\subsection{Symmetry relations}

In general, further identities among Feynman integrals can arise when
it is possible to redefine the loop momenta so that the value of the integral itself does not change, but the integrand transforms into a combination of different integrands. By imposing the identity of the original integral to the combination of integrals resulting from the change of loop momenta, one obtains a set of identities relating integrals belonging to the same topology.\\

More identities, originally found by Larin, 
may arise as well when there is a sub-loop diagram
depending - after its integration - on a specific combination of momenta. 
By equating the original integral and the
one obtained by projecting onto such a combination of momenta,
one gets additional relations
(see \cite{Grozin:2003ak} for details).

{\footnotesize{
\subsubsection{Example 1: IBP-id's}

Let us take again the two-loop topology of Fig. 2, and its
class of integrals (\ref{INTRO:eq8}). The corresponding
IBP-Id's can be established by the vanishing of the following integral of a divergence,

\begin{equation}\label{INTRO:eq12}
\int{d^D k_{1} \over (2 \pi)^{D-2}}{d^D k_{2} \over (2 \pi)^{D-2}}\frac{\partial}{\partial k_{i,\mu}}\left\{v_{\mu}\frac{(k\cdot k_{2})^{n_{1}}}{{\cal{D}}_{1}^{m_{1}}{\cal{D}}_{2}^{m_{2}}{\cal{D}}_{3}^{m_{3}}{\cal{D}}_{4}^{m_{4}}}\right\}=0\,\,\,\,\,
(i=1,2),
\end{equation}
where $v_{\mu}=k_{1},k_{2},k$, 
which amount to $3\times 2=6$ identities.\\
For simplicity let us choose $n_{1}=0$ and $m_{1}=...=m_{4}=1$. We have 
\begin{equation}\label{INTRO:eq13}
\int{d^D k_{1} \over (2 \pi)^{D-2}}{d^D k_{2} \over (2 \pi)^{D-2}}\frac{\partial}{\partial k_{i,\mu}}\left\{\frac{v_{\mu}}{{\cal{D}}_{1}{\cal{D}}_{2}{\cal{D}}_{3}{\cal{D}}_{4}}\right\}=0\,\,\,\,\,(i=1,2).
\end{equation}
The six IBP-Id's are resumed in the following table
\begin{figure}[h]
\begin{center}
\begin{tabular}{||c|c|c||}
\hline
\emph{derivative} $\frac{\partial}{\partial k_{i,\mu}}$ & \emph{vector} $v_{\mu}$ & \emph{corresponding} IBP-Id \\
\hline
$\frac{\partial}{\partial k_{1,\mu}}$ & $k_{1,\mu}$ & $\,\,\int{d^D k_{1} \over (2 \pi)^{D-2}}{d^D k_{2} \over (2 \pi)^{D-2}}\frac{\partial}{\partial k_{1,\mu}}\left(\frac{k_{1,\mu}}{{\cal{D}}_{1}{\cal{D}}_{2}{\cal{D}}_{3}{\cal{D}}_{4}}\right)=0$\\
\hline
$\frac{\partial}{\partial k_{1,\mu}}$ & $k_{2,\mu}$ & $\,\,\int{d^D k_{1} \over (2 \pi)^{D-2}}{d^D k_{2} \over (2 \pi)^{D-2}}\frac{\partial}{\partial k_{1,\mu}}\left(\frac{k_{2,\mu}}{{\cal{D}}_{1}{\cal{D}}_{2}{\cal{D}}_{3}{\cal{D}}_{4}}\right)=0$\\
\hline
$\frac{\partial}{\partial k_{1,\mu}}$ & $k_{\mu}$ & $\,\,\int{d^D k_{1} \over (2 \pi)^{D-2}}{d^D k_{2} \over (2 \pi)^{D-2}}\frac{\partial}{\partial k_{1,\mu}}\left(\frac{k_{\mu}}{{\cal{D}}_{1}{\cal{D}}_{2}{\cal{D}}_{3}{\cal{D}}_{4}}\right)=0$\\
\hline 
$\frac{\partial}{\partial k_{2,\mu}}$ & $k_{1,\mu}$ & $\,\,\int{d^D k_{1} \over (2 \pi)^{D-2}}{d^D k_{2} \over (2 \pi)^{D-2}}\frac{\partial}{\partial k_{2,\mu}}\left(\frac{k_{1,\mu}}{{\cal{D}}_{1}{\cal{D}}_{2}{\cal{D}}_{3}{\cal{D}}_{4}}\right)=0$\\
\hline 
$\frac{\partial}{\partial k_{2,\mu}}$ & $k_{2,\mu}$ & $\,\,\int{d^D k_{1} \over (2 \pi)^{D-2}}{d^D k_{2} \over (2 \pi)^{D-2}}\frac{\partial}{\partial k_{2,\mu}}\left(\frac{k_{2,\mu}}{{\cal{D}}_{1}{\cal{D}}_{2}{\cal{D}}_{3}{\cal{D}}_{4}}\right)=0$\\
\hline
$\frac{\partial}{\partial k_{2,\mu}}$ & $k_{\mu}$ & $\,\,\int{d^D k_{1} \over (2 \pi)^{D-2}}{d^D k_{2} \over (2 \pi)^{D-2}}\frac{\partial}{\partial k_{2,\mu}}\left(\frac{k_{\mu}}{{\cal{D}}_{1}{\cal{D}}_{2}{\cal{D}}_{3}{\cal{D}}_{4}}\right)=0$\\
\hline
\end{tabular}
\end{center}
\caption{IBP for the topology of Fig.\ref{INTRO:fig2}}
\label{INTRO:fig3}
\end{figure}
\\
Let us perform explicitly the calculation of the last of above identities.\\
We have
\begin{equation}\label{INTRO:eq14}
\int{d^D k_{1} \over (2 \pi)^{D-2}}{d^D k_{2} \over (2 \pi)^{D-2}}\frac{\partial}{\partial k_{2,\mu}}\left(\frac{k_{\mu}}{{\cal{D}}_{1}{\cal{D}}_{2}{\cal{D}}_{3}{\cal{D}}_{4}}\right)=0 \ . 
\end{equation}
By performing the derivative with respect  to $k_{2,\mu}$, we can rewrite the integrand as follows
\begin{eqnarray}\label{INTRO:eq15}
\lefteqn{\frac{\partial}{\partial k_{2,\mu}}\left(\frac{k_{\mu}}{{\cal{D}}_{1}{\cal{D}}_{2}{\cal{D}}_{3}{\cal{D}}_{4}}\right)=\frac{\partial k_{\mu}}{\partial k_{2,\mu}}\frac{1}{{\cal{D}}_{1}{\cal{D}}_{2}{\cal{D}}_{3}{\cal{D}}_{4}}+}\nonumber\\ & & +k_{\mu}\frac{1}{{\cal{D}}_{1}{\cal{D}}_{4}}\frac{\partial}{\partial k_{2,\mu}}\left(\frac{1}{{\cal{D}}_{2}{\cal{D}}_{3}}\right)
=k_{\mu}\frac{1}{{\cal{D}}_{1}{\cal{D}}_{4}}\left[-\frac{2 k_{2,\mu}}{{\cal{D}}_{2}^2{\cal{D}}_{3}}+\frac{2(k_{1,\mu}-k_{2,\mu})}{{\cal{D}}_{2}{\cal{D}}_{3}^2}\right]=\nonumber\\ & &
=-\frac{2k\cdot k_{2}}{{\cal{D}}_{1}{\cal{D}}_{2}^2{\cal{D}}_{3}{\cal{D}}_{4}}+\frac{2k\cdot k_{1}}{{\cal{D}}_{1}{\cal{D}}_{2}{\cal{D}}_{3}^2{\cal{D}}_{4}}-\frac{2k\cdot k_{2}}{{\cal{D}}_{1}{\cal{D}}_{2}{\cal{D}}_{3}^2{\cal{D}}_{4}}.
\end{eqnarray} 
After expressing the scalar product $k\cdot k_{1}$ in terms of the propagators 
${\cal{D}}_{1}=k_{1}^2+m^2$ and ${\cal{D}}_{4}=(k_{1}-k)^2+m^2$

\begin{equation}\label{INTRO:eq16}
2k\cdot k_{1}={\cal{D}}_{1}-{\cal{D}}_{4}+k^2,
\end{equation}
and substituting Eq. (\ref{INTRO:eq16}) in Eq. (\ref{INTRO:eq15}), we obtain,

\begin{eqnarray}\label{INTRO:eq17}
\lefteqn{-\int{d^D k_{1} \over (2 \pi)^{D-2}}{d^D k_{2} \over (2 \pi)^{D-2}}\frac{1}{{\cal{D}}_{1}{\cal{D}}_{2}{\cal{D}}_{3}^2}+\int{d^D k_{1} \over (2 \pi)^{D-2}}{d^D k_{2} \over (2 \pi)^{D-2}}\frac{1}{{\cal{D}}_{2}{\cal{D}}_{3}^2{\cal{D}}_{4}}+}\nonumber\\ & &
+k^2\int{d^D k_{1} \over (2 \pi)^{D-2}}{d^D k_{2} \over (2 \pi)^{D-2}}\frac{1}{{\cal{D}}_{1}{\cal{D}}_{2}{\cal{D}}_{3}^2{\cal{D}}_{4}}-2\int{d^D k_{1} \over (2 \pi)^{D-2}}{d^D k_{2} \over (2 \pi)^{D-2}}\frac{k\cdot k_{2}}{{\cal{D}}_{1}{\cal{D}}_{2}{\cal{D}}_{3}^2{\cal{D}}_{4}}+\nonumber\\ & &
-2\int{d^D k_{1} \over (2 \pi)^{D-2}}{d^D k_{2} \over (2 \pi)^{D-2}}\frac{k\cdot k_{2}}{{\cal{D}}_{1}{\cal{D}}_{2}^2{\cal{D}}_{3}{\cal{D}}_{4}}=0
\end{eqnarray}
Eq. (\ref{INTRO:eq17}) can be pictorially represented by

\begin{eqnarray}\label{INTRO:eq18}
\lefteqn{-\hspace{-.5cm}\parbox{10mm}{\begin{picture}(300,20)(0,0)
\SetScale{0.1}
\SetWidth{2.5}
%
%
%
\GOval(300,100)(100,100)(0){1.0}
\Photon(200,100)(400,100){6}{4}
\Vertex(300,200){10}
\end{picture}}\hspace{.5cm}
+\parbox{10mm}{\begin{picture}(300,20)(0,0)
\SetScale{0.1}
\SetWidth{2.5}
\Photon(100,100)(200,100){6}{2}    
\Photon(400,100)(500,100){6}{2}      
%
\GOval(300,100)(100,100)(0){1.0}
\Photon(200,100)(400,100){6}{4}
\Vertex(300,200){10}
\end{picture}}\hspace{.9cm}+k^2\hspace{-.2cm}\parbox{10mm}{\begin{picture}(300,20)(0,0)
\SetScale{0.1}
\SetWidth{2.5}
\Photon(100,100)(200,100){6}{2}    
\Photon(400,100)(500,100){6}{2}      
%
\GOval(300,100)(100,100)(0){1.0}
\Photon(300,200)(400,100){6}{3}
\Vertex(370,165){10}
%
\end{picture}}\hspace{1cm}+}\nonumber\\ & &
-2\parbox{10mm}{\begin{picture}(300,20)(0,0)
\SetScale{0.1}
\SetWidth{2.5}
\Photon(100,100)(200,100){6}{2}    
\Photon(400,100)(500,100){6}{2}      
%
\GOval(300,100)(100,100)(0){1.0}
\Photon(300,200)(400,100){6}{3}
\Vertex(370,165){10}
\Text(53,15)[]{{\tiny $(k\cdot k_{2})$}}
%
\end{picture}}\hspace{1.2cm}
-2\parbox{10mm}{\begin{picture}(300,20)(0,0)
\SetScale{0.1}
\SetWidth{2.5}
\Photon(100,100)(200,100){6}{2}    
\Photon(400,100)(500,100){6}{2}      
%
\GOval(300,100)(100,100)(0){1.0}
\Photon(300,200)(400,100){6}{3}
\Vertex(345,150){10}
\Text(53,15)[]{{\tiny $(k\cdot k_{2})$}}
%
\end{picture}}\hspace{1.2cm}=0
\end{eqnarray}
where a dot on a propagator line means that the propagator is squared; irreducible scalar 
products in the numerator are explicitly indicated.\\

\subsubsection{Example 2: IBP-id's}

Let us consider the case of the two-loop three-leg topology of 
Fig.\ref{INTRO:fig4}.

\vspace{-1cm}

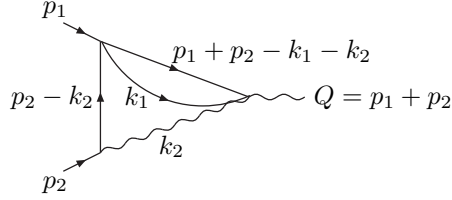
\begin{figure}[ht]
\begin{center}
\begin{picture}(150,100)(0,0)
\SetWidth{0.7}
\SetScale{0.7}
\ArrowLine(20,10)(40,20)    
\ArrowLine(20,90)(40,80)      
\ArrowLine(40,20)(40,80)
\ArrowLine(40,80)(120,50)
\Photon(40,20)(120,50){1.5}{6}
\Photon(120,50)(150,50){1.5}{2}
\ArrowArc(95.1,105.2)(60,205,295)
\Text(11,68)[]{$p_{1}$}
\Text(11,2)[]{$p_{2}$}
\Text(10,35)[]{$p_{2}-k_{2}$}
\Text(42,35)[]{$k_{1}$}
\Text(55,16.5)[]{$k_{2}$}
\Text(93,52.5)[]{$p_{1}+p_{2}-k_{1}-k_{2}$}
\Text(135,35)[]{$Q=p_{1}+p_{2}$}
\end{picture}
\end{center}
\caption{Example of a two-loop three-leg four-denominators topology: $p_{1}^2=p_{2}^2=-m^2$, and $(p_{1}+p_{2})^2=Q^2=-s$}
\label{INTRO:fig4}
\end{figure}

The integrals associated to this topology can have 
three irreducible scalar products, arbitrary chosen to be
$(p_{1}\cdot k_{1})$, $(p_{2}\cdot k_{1})$ and $(k_{1}\cdot k_{2})$.
In doing so, Eq. (\ref{INTRO:eq11}), for generic values of the indices $m_{i}$ and $n_{i}$, reads

\begin{equation}\label{INTRO:eq19}
\int{d^D k_{1} \over (2 \pi)^{D-2}}{d^D k_{2} \over (2 \pi)^{D-2}}\frac{\partial}{\partial k_{i,\mu}}\left\{v_{\mu}\frac{(p_{1}\cdot k_{1})^{n_{1}}(p_{2}\cdot k_{1})^{n_{2}}(k_{1}\cdot k_{2})^{n_{3}}}{{\cal{D}}_{1}^{m_{1}}{\cal{D}}_{2}^{m_{2}}{\cal{D}}_{3}^{m_{3}}{\cal{D}}_{4}^{m_{4}}}\right\}=0
\end{equation}
where ${\cal{D}}_{1}=k_{1}^2+m^2,{\cal{D}}_{2}=k_{2}^2,{\cal{D}}_{3}=(p_{2}-k_{2})^2+m^2,{\cal{D}}_{4}=(p_{1}+p_{2}-k_{1}-k_{2})^2+m^2$ and $i=1,2$.
By setting, for simplicity, 
$m_{1}=\cdots=m_{4}=1$, 
$n_{1}=\cdots=n_{3}=0$, 
$v_{\mu}=p_{1}$ and  $i=1$, Eq. (\ref{INTRO:eq19}) becomes


\begin{equation}\label{INTRO:eq20}
\int{d^D k_{1} \over (2 \pi)^{D-2}}{d^D k_{2} \over (2 \pi)^{D-2}}
\frac{\partial}{\partial k_{1,\mu}}\left\{\frac{p_{1,\mu}}{{\cal{D}}_{1}{\cal{D}}_{2}{\cal{D}}_{3}{\cal{D}}_{4}}\right\}=0.
\end{equation}
After taking the derivative with respect to $k_{1}$ and simplifying the reducible scalar products against the corresponding propagators, one can write Eq.(\ref{INTRO:eq20}) as follows

\begin{eqnarray}\label{INTRO:eq21}
\lefteqn{-2\hspace{-.3cm}\parbox{10mm}{\begin{picture}(300,20)(0,0)
\SetWidth{1}
\SetScale{0.4}
\Line(20,-20)(40,-10)    
\Line(20,60)(40,50)      
\Line(40,-10)(40,50)
\Line(40,50)(120,20)
\Photon(40,-10)(120,20){1.5}{6}
\Photon(120,20)(150,20){1.5}{2}
\CArc(95.1,75.2)(60,205,295)
\Vertex(71,21){3}
\Text(63,13)[]{{\tiny $(p_{1}\cdot k_{1})$}}
\end{picture}}\hspace{2cm}-2\parbox{10mm}{\begin{picture}(300,20)(0,0)
\SetWidth{1}
\SetScale{0.4}
\Line(20,-20)(40,-10)    
\Line(20,60)(40,50)      
\Line(40,-10)(40,50)
\Line(40,50)(120,20)
\Photon(40,-10)(120,20){1.5}{6}
\Photon(120,20)(150,20){1.5}{2}
\CArc(95.1,75.2)(60,205,295)
\Vertex(79,33){3}
\Text(63,13)[]{{\tiny $(k_{1}\cdot k_{2})$}}
\end{picture}}\hspace{1.7cm}+}\nonumber\\ & &
\nonumber\\ & &
\nonumber\\ & &
\vspace{.2cm} 
+2\parbox{10mm}{\begin{picture}(300,20)(0,0)
\SetWidth{1}
\SetScale{0.4}
\Line(20,-20)(40,-10)    
\Line(20,60)(40,50)      
\Line(40,-10)(40,50)
\Line(40,50)(120,20)
\Photon(40,-10)(120,20){1.5}{6}
\Photon(120,20)(150,20){1.5}{2}
\CArc(95.1,75.2)(60,205,295)
\Vertex(79,33){3}
\Text(63,13)[]{{\tiny $(p_{2}\cdot k_{1})$}}
\end{picture}}\hspace{1.7cm}
+\hspace{-.5cm}\parbox{10mm}{\begin{picture}(300,20)(0,0)
\SetWidth{1}
\SetScale{0.4}
\Line(20,-20)(40,-10)    
\Line(20,60)(40,50)      
\Line(40,-10)(40,50)
\Line(40,50)(120,20)
\Photon(40,-10)(120,20){1.5}{6}
\Photon(120,20)(150,20){1.5}{2}
\CArc(95.1,75.2)(60,205,295)
\end{picture}}\hspace{1.3cm}+\nonumber\\ & &
\nonumber\\ & & 
\nonumber\\ & &
-\parbox{10mm}{\begin{picture}(300,20)(0,0)
\SetScale{0.1}
\SetWidth{2.5}
\Photon(100,100)(200,100){6}{2}    
\Photon(400,100)(500,100){6}{2}      
%
\GOval(300,100)(100,100)(0){1.0}
\Photon(200,100)(400,100){6}{4}
\Vertex(300,0){10}
\end{picture}}\hspace{1cm}
+\frac{(D-2)}{2m^2}\parbox{10mm}{\begin{picture}(300,20)(0,0)
\SetWidth{1}
\SetScale{0.4}
\Photon(25,28)(50,28){1.5}{2}         
\CArc(75,28)(25,0,180)
\PhotonArc(75,28)(25,180,360){2}{6}
\GOval(125,28)(25,25)(0){1}
\Photon(100,-5)(100,28){3}{2}
\end{picture}}\hspace{1.2cm}=0
\end{eqnarray}

\subsubsection{Example: LI-id's}

The Lorentz Invariance identity (\ref{INTRO:eq25})
for the integral of the topology in Fig. \ref{INTRO:fig4}
with $n_{1}=n_{2}=n_{3}=0$ and $m_{1}=m_{2}=m_{3}=m_{4}=1$, 
reads,
%
%
\begin{eqnarray}\label{INTRO:eq30}
\lefteqn{0 = 4m^2\left\{\parbox{10mm}{\begin{picture}(300,20)(0,0)
\SetWidth{1}
\SetScale{0.4}
\Line(20,-20)(40,-10)    
\Line(20,60)(40,50)      
\Line(40,-10)(40,50)
\Line(40,50)(120,20)
\Photon(40,-10)(120,20){1.5}{6}
\Photon(120,20)(150,20){1.5}{2}
\CArc(95.1,75.2)(60,205,295)
\Vertex(40,20){3}
\Text(63,13)[]{{\tiny $(k_{1}\cdot k_{2})$}}
\end{picture}}\hspace{2cm}-\parbox{10mm}{\begin{picture}(300,20)(0,0)
\SetWidth{1}
\SetScale{0.4}
\Line(20,-20)(40,-10)    
\Line(20,60)(40,50)      
\Line(40,-10)(40,50)
\Line(40,50)(120,20)
\Photon(40,-10)(120,20){1.5}{6}
\Photon(120,20)(150,20){1.5}{2}
\CArc(95.1,75.2)(60,205,295)
\Vertex(40,20){3}
\Text(63,13)[]{{\tiny $(p_{1}\cdot k_{1})$}}
\end{picture}}\hspace{2cm}-\parbox{10mm}{\begin{picture}(300,20)(0,0)
\SetWidth{1}
\SetScale{0.4}
\Line(20,-20)(40,-10)    
\Line(20,60)(40,50)      
\Line(40,-10)(40,50)
\Line(40,50)(120,20)
\Photon(40,-10)(120,20){1.5}{6}
\Photon(120,20)(150,20){1.5}{2}
\CArc(95.1,75.2)(60,205,295)
\Vertex(40,20){3}
\Text(63,13)[]{{\tiny $(p_{2}\cdot k_{1})$}}
\end{picture}}\hspace{2cm}\right\}}\nonumber\\ & &
\nonumber\\ & &
\nonumber\\ & &
-2m^2 s \parbox{10mm}{\begin{picture}(300,20)(0,0)
\SetWidth{1}
\SetScale{0.4}
\Line(20,-20)(40,-10)    
\Line(20,60)(40,50)      
\Line(40,-10)(40,50)
\Line(40,50)(120,20)
\Photon(40,-10)(120,20){1.5}{6}
\Photon(120,20)(150,20){1.5}{2}
\CArc(95.1,75.2)(60,205,295)
\Vertex(40,20){3}
\end{picture}}\hspace{1.2cm}+2s\left\{\parbox{10mm}{\begin{picture}(300,20)(0,0)
\SetWidth{1}
\SetScale{0.4}
\Line(20,-20)(40,-10)    
\Line(20,60)(40,50)      
\Line(40,-10)(40,50)
\Line(40,50)(120,20)
\Photon(40,-10)(120,20){1.5}{6}
\Photon(120,20)(150,20){1.5}{2}
\CArc(95.1,75.2)(60,205,295)
\Vertex(79,33){3}
\Text(63,13)[]{{\tiny $(k_{1}\cdot k_{2})$}}
\end{picture}}\hspace{1.8cm}-2\parbox{10mm}{\begin{picture}(300,20)(0,0)
\SetWidth{1}
\SetScale{0.4}
\Line(20,-20)(40,-10)    
\Line(20,60)(40,50)      
\Line(40,-10)(40,50)
\Line(40,50)(120,20)
\Photon(40,-10)(120,20){1.5}{6}
\Photon(120,20)(150,20){1.5}{2}
\CArc(95.1,75.2)(60,205,295)
\Vertex(79,33){3}
\Text(63,13)[]{{\tiny $(p_{2}\cdot k_{1})$}}
\end{picture}}\hspace{2cm}\right\}+\nonumber\\ & &
\nonumber\\ & &
\nonumber\\ & &
-s^2\parbox{10mm}{\begin{picture}(300,20)(0,0)
\SetWidth{1}
\SetScale{0.4}
\Line(20,-20)(40,-10)    
\Line(20,60)(40,50)      
\Line(40,-10)(40,50)
\Line(40,50)(120,20)
\Photon(40,-10)(120,20){1.5}{6}
\Photon(120,20)(150,20){1.5}{2}
\CArc(95.1,75.2)(60,205,295)
\Vertex(79,33){3}
\end{picture}}\hspace{1.2cm}+
-(s-2m^2) \parbox{10mm}{\begin{picture}(300,20)(0,0)
\SetScale{0.1}
\SetWidth{2.5}
\Line(100,100)(200,100)
\Line(400,100)(500,100)      
%
\GOval(300,100)(100,100)(0){1.0}
\Line(200,100)(400,100)
\Vertex(300,100){10}
\end{picture}}\hspace{.8cm}-s\parbox{10mm}{\begin{picture}(300,20)(0,0)
\SetScale{0.1}
\SetWidth{2.5}
\Line(100,100)(200,100)
\Line(400,100)(500,100)      
%
\GOval(300,100)(100,100)(0){1.0}
\Line(200,100)(400,100)
\Vertex(300,0){10}
\end{picture}}\hspace{.8cm}+2s\parbox{10mm}{\begin{picture}(300,20)(0,0)
\SetScale{0.1}
\SetWidth{2.5}
\Photon(100,100)(200,100){6}{2}
\Photon(400,100)(500,100){6}{2}      
%
\GOval(300,100)(100,100)(0){1.0}
\Photon(200,100)(400,100){6}{4}
\Vertex(300,0){10}
\end{picture}}\hspace{1.2cm}+\nonumber\\ & &
\nonumber\\ & &
\nonumber\\ & &
-\frac{s(D-2)}{2m^2}\parbox{10mm}{\begin{picture}(300,20)(0,0)
\SetWidth{1}
\SetScale{0.4}
\Photon(25,28)(50,28){1.5}{2}         
\CArc(75,28)(25,0,180)
\PhotonArc(75,28)(25,180,360){2}{6}
\GOval(125,28)(25,25)(0){1}
\Photon(100,-5)(100,28){3}{2}
\end{picture}}\hspace{1.2cm}
\end{eqnarray}

\subsubsection{Example: Symmetry relations}

One can exploit the invariance of the integrals belonging to 
the topology of Fig. \ref{INTRO:fig4}, 

\begin{equation}\label{INTRO:eq31}
\int{d^D k_{1} \over (2 \pi)^{D-2}}{d^D k_{2} \over (2 \pi)^{D-2}}\frac{(p_{1}\cdot k_{1})^{n_{1}}(p_{2}\cdot k_{1})^{n_{2}}(k_{1}\cdot k_{2})^{n_{3}}}{{\cal{D}}_{1}^{m_{1}}{\cal{D}}_{2}^{m_{2}}{\cal{D}}_{3}^{m_{3}}{\cal{D}}_{4}^{m_{4}}}.
\end{equation}
under the redefinition of the momenta running in the nested electron-loop.
In fact the two denominators with momentum $k_{1}$ and $(p_{1}+p_{2}-k_{1}-k_{2})$, i.e. ${\cal{D}}_{1}=k_{1}^2+m^2$ and ${\cal{D}}_{4}=(p_{1}+p_{2}-k_{1}-k_{2})+m^2$ have the same mass. 
 Therefore the following redefinition of the integration momentum 

\begin{equation}\label{INTRO:eq32}
k_{1}=p_{1}+p_{2}-k'_{1}-k_{2},
\end{equation}
that consists in the interchange of the two denominators in the closed electron loop, does not affect the 
topology of the integral; nevertheless the explicit form of the integrand can change generating non trivial identities. Taking for instance $n_{1}=n_{2}=n_{3}=0$ and $m_{1}=m_{2}=m_{4}=1$ and $m_{3}=2$ in Eq. (\ref{INTRO:eq31}), the substitution (\ref{INTRO:eq32}) gives the following very simple relation

\vspace{-1cm}

\begin{equation}\label{INTRO:eq33}
 \parbox{10mm}{\begin{picture}(150,100)(0,0)
\SetWidth{1}
\SetScale{0.4}
\Line(20,90)(40,100)    
\Line(20,170)(40,160)      
\Line(40,100)(40,160)
\Line(40,160)(120,130)
\Photon(40,100)(120,130){1.5}{6}
\Photon(120,130)(150,130){1.5}{2}
\CArc(95.1,185.2)(60,205,295)
\Vertex(71,129){3}
\end{picture}}\hspace{1.5cm}-\parbox{10mm}{\begin{picture}(300,20)(0,0)
\SetWidth{1}
\SetScale{0.38}
\Line(20,-20)(40,-10)    
\Line(20,60)(40,50)      
\Line(40,-10)(40,50)
\Line(40,50)(120,20)
\Photon(40,-10)(120,20){1.5}{6}
\Photon(120,20)(150,20){1.5}{2}
\CArc(95.1,75.2)(60,205,295)
\Vertex(79,33){3}
\end{picture}}\hspace{1.2cm}=0.
\end{equation}
\vspace{-1cm}

\noindent
By choosing $n_{1}=n_{2}=0$, $n_{3}=1$ and $m_{1}=m_{2}=m_{4}=1$ and $m_{3}=2$, 
we get a less trivial identity

\begin{eqnarray}\label{INTRO:eq34}
\lefteqn{0=\parbox{10mm}{\begin{picture}(300,20)(0,0)
\SetWidth{1}
\SetScale{0.4}
\Line(20,-20)(40,-10)    
\Line(20,60)(40,50)      
\Line(40,-10)(40,50)
\Line(40,50)(120,20)
\Photon(40,-10)(120,20){1.5}{6}
\Photon(120,20)(150,20){1.5}{2}
\CArc(95.1,75.2)(60,205,295)
\Vertex(40,20){3}
\Text(63,13)[]{{\tiny $(k_{1}\cdot k_{2})$}}
\end{picture}}\hspace{2cm}+\parbox{10mm}{\begin{picture}(300,20)(0,0)
\SetWidth{1}
\SetScale{0.4}
\Line(20,-20)(40,-10)    
\Line(20,60)(40,50)      
\Line(40,-10)(40,50)
\Line(40,50)(120,20)
\Photon(40,-10)(120,20){1.5}{6}
\Photon(120,20)(150,20){1.5}{2}
\CArc(95.1,75.2)(60,205,295)
\Vertex(40,20){3}
\Text(63,13)[]{{\tiny $(p_{1}\cdot k_{1})$}}
\end{picture}}\hspace{2cm}+}\nonumber\\ & &
\nonumber\\ & &
\nonumber\\ & &
+\parbox{10mm}{\begin{picture}(300,20)(0,0)
\SetWidth{1}
\SetScale{0.4}
\Line(20,-20)(40,-10)    
\Line(20,60)(40,50)      
\Line(40,-10)(40,50)
\Line(40,50)(120,20)
\Photon(40,-10)(120,20){1.5}{6}
\Photon(120,20)(150,20){1.5}{2}
\CArc(95.1,75.2)(60,205,295)
\Vertex(40,20){3}
\Text(63,13)[]{{\tiny $(p_{2}\cdot k_{1})$}}
\end{picture}}\hspace{2cm}
+\frac{s}{2}\parbox{10mm}{\begin{picture}(300,20)(0,0)
\SetWidth{1}
\SetScale{0.4}
\Line(20,-20)(40,-10)    
\Line(20,60)(40,50)      
\Line(40,-10)(40,50)
\Line(40,50)(120,20)
\Photon(40,-10)(120,20){1.5}{6}
\Photon(120,20)(150,20){1.5}{2}
\CArc(95.1,75.2)(60,205,295)
\Vertex(40,20){3}
\end{picture}}\hspace{1.2cm}+\frac{1}{2}\parbox{10mm}{\begin{picture}(300,20)(0,0)
\SetScale{0.1}
\SetWidth{2.5}
\Line(100,100)(200,100)    
\Line(400,100)(500,100)      
%
\GOval(300,100)(100,100)(0){1.0}
\Line(200,100)(400,100)
\Vertex(300,100){10}
\end{picture}}
\end{eqnarray}

\vspace{1.0cm}

}} 

\begin{subsubsection}{Master Integrals}

For each of the $N(I_{t,r,s})$ integrals of the class $I_{t,r,s}$ (see Eq.(\ref{INTRO:eq10}))
we dispose of IBP-id's, LI-id's and symmetry relations, involving integrals of the 
families up to $I_{t,r+1,s+1}$. For $r=s=0$ the number of all the integrals involved in the identities exceeds the number of equations; but, in writing down systematically all the equations for increasing values of $r$ and $s$ 
the number of equations grows faster than the number of the integrals,
generating an apparently overconstrained system of equations 
--- as realized first by Stefano Laporta \cite{Laporta:1996mq}.
After its generation one is left with the problem of solving such a linear system of identities which is trivial in principle, but algebraically very lengthy, so that some organization is needed for achieving the 
solution\cite{Mastrolia:2000va}.\\
To this end, one can order the integrals in an appropriate way. In particular a "weight" is assigned to each integral: the weight can be almost any increasing function of the indices $n_{i}$ and $m_{j}$, such that integrals with higher indices have bigger weights. The system can be solved by Gauss' substitution rule,
by considering the equations of the system one by one, and using each equation for expressing the integral with the highest weight in terms of the integrals of lower weight.
Then the result is substituted in the leftover equations.
The algorithm, by now known as {\it Laporta algorithm}, is straightforward, but its execution 
requires a great amount of algebra, which has been implemented in several computer codes\cite{ibpsolver}. 

One finds that several equations are identically satisfied, and the original unknown integrals are expressed in terms of very few independent integrals, the {\it master integrals} (MI's). In doing so, the resulting MI's correspond to the integrals of lowest weight; but as the choice of the weight is to a large extent arbitrary, there is also some freedom in the choice of the integrals to pick up as actual MI's (not in their number of course!).\\
There are several cases in which more than one MI is found for a given topology, while sometimes only one MI is present. It may also happen that no MI for the considered topology is left over in the reduction --- 
{\it i.e.} all the integrals corresponding to a given topology, with $t$ propagators, can be expressed in terms of the MI's of its subtopologies with $(t-1)$, and/or less, propagators. In such a case one speaks about \emph{reducible topologies}.

Finally, three points are worth mentioning. \\
First, any given topology with $t$ propagators has $(t-1)$ subtopologies with $(t-1)$ propagators, $(t-1)(t-2)$ subtopologies with $(t-2)$ propagators etc., down to as many propagators as the number of loops. It turns out, however, that many subtopologies are in fact equivalent, up to a translation of the loop variable, and the subtopologies coming from the contribution of different graphs do overlap to a great extent. 
For these reasons the compilation of the system for the actual number 
of all \emph{independent} topologies, 
namely those that cannot be transformed one into the others by a redefinition of internal momenta, is relatively small.

The second remark is about the {\it independence} of the identities for a given 
topology. 
Under an idealistic point of view, it might happen that the infinite set if IBP-id's,
obtained by considering the infinite choices of indices of denominators and scalar
products for the considered topology, plus the infinite set of IBP-id's of all 
the {\it parent} topologies containing it as a subtopology, could include
LI-id's and symmetry relations as subset. 
Certainly, at the practical level, since one is working with finite 
sets of indices,
LI-id's and symmetry relations can be considered as {\it additional}  
to the IBP-id's, which surely speed up the solution of the system
and work as further checks of the reduction procedure. \\
Third, we are not able to prove analytically that the MI's one finds are really the minimal set of MI's, {\it i.e.} that they are independent from each other; in any case the final number of the MI's is quite small, so that reducing the several hundred integrals occurring in a typical calculation to a few of them is after all a great progress.
Studies on the {\it apriori} determination of whether a topology could 
have or not MI's, have been performed by 
Baikov \cite{Baikov:2005nv} and Smirnov \cite{Smirnov:2005ky}. Although a general algorithm is still lacking,
we think that the analysis of leading and subleading Landau singularities 
(see \cite{SMatrix} for an extensive treatment)
could be related to this issue:
since a reducible topology without MI's is expressible in terms of the MI's of 
its subtopologies,
it should mean that the leading singularities of the reducible topology are not independent
of the subleading ones, which are leading singularities for the 
subtopologies.

As a last remark, let us recall that, at the end of the reduction, there is 
some freedom for choosing the basis of MI's and
usually the choice is in general motivated by convenience. For example  
the behaviour of the functions in the $D$-to-4 expansion might determine
to select:
{\it i)} simpler integrands, in view of a successive analytic computation;
{\it ii)} more complicated integrands, but with a better convergence, 
should their numerical evaluation be of interest.
\end{subsubsection}

\end{subsection}
\end{section}

\section{Differential Equations for Master Integrals}

The outcome of the {\it reduction} procedure, previously discussed, 
is a collection of identities thanks to which
any expression, demanding originally for the evaluation of a very large number of integrals, 
is simplified and written as linear combination of few MI's with rational coefficients.
The completion of the analytic achievement of the result proceeds with the evaluation
of the yet unknown MI's. 
As we will see in a moment, the same collection of identities is as well necessary to write
Differential Equations satisfied by the MI's.

\subsection{System of differential equations}

Once all the MI's of a given topology are identified, 
the problem of their calculation arises. 
Exactly at this stage of the computation, 
{\it differential equations} enter the game.
The use of differential equations in one of the internal masses was first proposed out by Kotikov\cite{Kotikov:1990kg}, 
then extended to more general differential equations in any of Mandelstam variables by Remiddi\cite{Remiddi:1997ny}. \\
Let us point out the basic idea of the method. 
To begin with, consider any scalar integral defined as

\begin{equation}\label{INTRO:eq35}
M(s_{1},s_{2},\cdots,s_{{\cal{N}}})
=\int{d^D k_{1} \over (2 \pi)^{D-2}}...{d^D k_{\ell} \over (2 \pi)^{D-2}}\frac{S_{1}^{n_{1}}...S_{q}^{n_{q}}}{{\cal{D}}_{1}...{\cal{D}}_{t}},
\end{equation}
where $\{s_{1},s_{2},\cdots,s_{{\cal{N}}}\}$ is any set of kinematic invariants of the topology and ${\cal{N}}$ is the number of such invariants.  

Let us denote the set $\{s_{1},s_{2},\cdots,s_{{\cal{N}}}\}={\bf{s}}$ and consider the following quantities 

\begin{equation}\label{INTRO:eq36}
O_{jk}({\bf{s}})=p_{j,\mu}\frac{\partial M({\bf{s}})}{\partial p_{k,\mu}}\,\,\,\,(j,k=1,2,\cdots,g-1),
\end{equation}
where $g-1$ is the number of independent external momenta. 
By the chain differentiation rule we have

\begin{equation}\label{INTRO:eq37}
O_{jk}({\bf s})=p_{j,\mu}\cdot\sum_{\alpha=1}^{{\cal{N}}} \frac{\partial {s_{\alpha}}}{\partial p_{k,\mu}}\frac{\partial M({\bf s})}{\partial s_{\alpha}}
=\sum_{\alpha=1}^{{\cal{N}}}\left(p_{j,\mu}\cdot \frac{\partial {s_{\alpha}}}{\partial p_{k,\mu}}\right)\frac{\partial M({\bf s})}{\partial s_{\alpha}}.
\end{equation}
According to the available number of the kinematic invariants, 
the \emph{r.h.s.} of Eq. (\ref{INTRO:eq36}) and the \emph{r.h.s.} of Eq. (\ref{INTRO:eq37}) may be equated to form the following system

\begin{equation}
 \sum_{\alpha=1}^{{\cal{N}}}\left(p_{j,\mu}\cdot \frac{\partial {s_{\alpha}}}{\partial p_{k,\mu}}\right)\frac{\partial M({\bf s})}{\partial s_{\alpha}}= 
p_{j,\mu}\frac{\partial M({\bf s})}{\partial p_{k,\mu}},
\label{INTRO:eq38}
\end{equation}
which can be solved in order to express $\frac{\partial M({\bf s})}{\partial s_{\alpha}}$ in terms of 
$p_{j,\mu}\frac{\partial M({\bf s})}{\partial p_{k,\mu}}$, so the corresponding identity,
can be finally 
read as a {\it differential equation} for $M$.\\
Examples of such equations are the following.

\begin{itemize}
\item \emph{2-point case.}

$\bullet$ Differentiation with respect to a mass \\
\begin{eqnarray}\label{INTRO:eq39}
\frac{\partial}{\partial m^2}\left\{\parbox{10mm}{\begin{picture}(300,20)(0,0)
\SetScale{.4}
\Line(15,25)(45,25)
\Line(85,25)(115,25)
\GOval(65,25)(20,20)(0){.5}
\GBoxc(65,15)(40,20){1}
\SetColor{White}
\SetWidth{3}
\Line(45,25)(45,5)
\SetColor{White}
\SetWidth{3}
\Line(45,5)(85,5)
\SetColor{White}
\SetWidth{3}
\Line(85,5)(85,25)
\SetColor{Black}
\SetWidth{1}
\CArc(65,25)(20,180,360)
\Text(2,9)[]{\footnotesize{$p$}}
\Text(50,9)[]{\footnotesize{$p$}}
\end{picture}}\hspace{.8cm}\right\}=-\left\{\parbox{10mm}{\begin{picture}(300,20)(0,0)
\SetScale{.4}
\Line(15,25)(45,25)
\Line(85,25)(115,25)
\GOval(65,25)(20,20)(0){.5}
\GBoxc(65,15)(40,20){1}
\SetColor{White}
\SetWidth{3}
\Line(45,25)(45,5)
\SetColor{White}
\SetWidth{3}
\Line(45,5)(85,5)
\SetColor{White}
\SetWidth{3}
\Line(85,5)(85,25)
\SetColor{Black}
\SetWidth{1}
\CArc(65,25)(20,180,360)
\Vertex(65,5){3}
\Text(2,9)[]{\footnotesize{$p$}}
\Text(50,9)[]{\footnotesize{$p$}}
\end{picture}}\hspace{.8cm}\right\}\hspace{1cm} \\
\hspace{-1cm} \nonumber
\end{eqnarray}
where, for simplicity, we assumed there is just one propagator of mass $m$.

$\bullet$ Differentiation with respect to the squared momentum \\
\begin{eqnarray}\label{INTRO:eq40}
p^2\frac{\partial}{\partial p^2}\left\{\parbox{10mm}{\begin{picture}(300,20)(0,0)
\SetScale{.4}
\Line(15,25)(45,25)
\Line(85,25)(115,25)
\GOval(65,25)(20,20)(0){.5}
\Text(2,9)[]{\footnotesize{$p$}}
\Text(50,9)[]{\footnotesize{$p$}}
\end{picture}}\hspace{.8cm}
\right\}=\frac{1}{2}p_{\mu}\frac{\partial}{\partial p_{\mu}}\left\{\parbox{10mm}{\begin{picture}(300,20)(0,0)
\SetScale{.4}
\Line(15,25)(45,25)
\Line(85,25)(115,25)
\GOval(65,25)(20,20)(0){.5}
\Text(2,9)[]{\footnotesize{$p$}}
\Text(50,9)[]{\footnotesize{$p$}}
\end{picture}}\hspace{.8cm}\right\}\hspace{1cm} \\ 
\hspace{-1cm} \nonumber
\end{eqnarray}

\vspace{.2cm}

\item \emph{3-point case.}

\begin{eqnarray}\label{INTRO:eq41}
\lefteqn{P^2\frac{\partial}{\partial P^2}\left\{\parbox{10mm}{\begin{picture}(300,20)(0,0)
\SetScale{.4}
\Line(25,50)(45,40)
\Line(25,0)(45,10)
\Line(80,25)(100,25)
\GOval(60,25)(20,20)(0){.5}
\Text(6,23)[]{\footnotesize{$p_{1}$}}
\Text(6,-3)[]{\footnotesize{$p_{2}$}}
\Text(45,10)[]{\footnotesize{$p_{3}$}}
\end{picture}}\hspace{.7cm}\right\}=}\nonumber\\ & &
\nonumber\\ & &
=\left[A\left(p_{1,\mu}\frac{\partial}{\partial p_{1,\mu}}+p_{2,\mu}\frac{\partial}{\partial p_{2,\mu}}\right)+B\left(p_{1,\mu}\frac{\partial}{\partial p_{2,\mu}}+p_{2,\mu}\frac{\partial}{\partial p_{1,\mu}}\right)\right]\left\{\parbox{10mm}{\begin{picture}(300,20)(0,0)
\SetScale{.4}
\Line(25,50)(45,40)
\Line(25,0)(45,10)
\Line(80,25)(100,25)
\GOval(60,25)(20,20)(0){.5}
\Text(6,23)[]{\footnotesize{$p_{1}$}}
\Text(6,-3)[]{\footnotesize{$p_{2}$}}
\Text(45,10)[]{\footnotesize{$p_{3}$}}
\end{picture}}\hspace{.7cm}\right\},\nonumber\\ & &
\,\,\,
\end{eqnarray}

with $P=p_{1}+p_{2}$ and $A,B$ rational coefficients.

\vspace{.3cm}

\item \emph{4-point case.}

\begin{eqnarray}\label{INTRO:eq42}
 &\, & P^2 \frac{\partial}{\partial P^2}\left\{\parbox{10mm}{\begin{picture}(300,20)(0,0)
\SetScale{.4}
\Line(25,50)(45,40)
\Line(25,0)(45,10)
\Line(75,40)(95,50)
\Line(75,10)(95,0)
\GOval(60,25)(20,20)(0){.5}
\Text(6,23)[]{\footnotesize{$p_{1}$}}
\Text(6,-3)[]{\footnotesize{$p_{2}$}}
\Text(44,23)[]{\footnotesize{$p_{3}$}}
\Text(44,-3)[]{\footnotesize{$p_{4}$}}
\end{picture}}\hspace{.7cm}\right\} =  
\left[C\left(p_{1,\mu}\frac{\partial}{\partial p_{1,\mu}}-p_{3,\mu}\frac{\partial}{\partial p_{3,\mu}}\right) + D p_{2,\mu}\frac{\partial}{\partial p_{2,\mu}}+\right.\nonumber\\
\nonumber\\
\hspace{-2cm} & + &\left. E(p_{1,\mu}+p_{3,\mu})
 \left(\frac{\partial}{\partial p_{3,\mu}}-\frac{\partial}{\partial p_{1,\mu}}+\frac{\partial}{\partial p_{2,\mu}}\right)\right]\left\{\parbox{10mm}{\begin{picture}(300,20)(0,0)
\SetScale{.4}
\Line(25,50)(45,40)
\Line(25,0)(45,10)
\Line(75,40)(95,50)
\Line(75,10)(95,0)
\GOval(60,25)(20,20)(0){.5}
\Text(6,23)[]{\footnotesize{$p_{1}$}}
\Text(6,-3)[]{\footnotesize{$p_{2}$}}
\Text(44,23)[]{\footnotesize{$p_{3}$}}
\Text(44,-3)[]{\footnotesize{$p_{4}$}}
\end{picture}}\hspace{.7cm}\right\},
\end{eqnarray}

\begin{eqnarray}\label{INTRO:eq43}
 &\, & Q^2 \frac{\partial}{\partial Q^2}\left\{\parbox{10mm}{\begin{picture}(300,20)(0,0)
\SetScale{.4}
\Line(25,50)(45,40)
\Line(25,0)(45,10)
\Line(75,40)(95,50)
\Line(75,10)(95,0)
\GOval(60,25)(20,20)(0){.5}
\Text(6,23)[]{\footnotesize{$p_{1}$}}
\Text(6,-3)[]{\footnotesize{$p_{2}$}}
\Text(44,23)[]{\footnotesize{$p_{3}$}}
\Text(44,-3)[]{\footnotesize{$p_{4}$}}
\end{picture}}\hspace{.7cm}\right\} =  
\left[F\left(p_{1,\mu}\frac{\partial}{\partial p_{1,\mu}}-p_{2,\mu}\frac{\partial}{\partial p_{2,\mu}}\right) + G p_{2,\mu}\frac{\partial}{\partial p_{2,\mu}}+\right.\nonumber\\
\nonumber\\
\hspace{-2cm} & + &\left. H(p_{2,\mu}-p_{1,\mu})
 \left(\frac{\partial}{\partial p_{1,\mu}}+\frac{\partial}{\partial p_{2,\mu}}+\frac{\partial}{\partial p_{3,\mu}}\right)\right]\left\{\parbox{10mm}{\begin{picture}(300,20)(0,0)
\SetScale{.4}
\Line(25,50)(45,40)
\Line(25,0)(45,10)
\Line(75,40)(95,50)
\Line(75,10)(95,0)
\GOval(60,25)(20,20)(0){.5}
\Text(6,23)[]{\footnotesize{$p_{1}$}}
\Text(6,-3)[]{\footnotesize{$p_{2}$}}
\Text(44,23)[]{\footnotesize{$p_{3}$}}
\Text(44,-3)[]{\footnotesize{$p_{4}$}}
\end{picture}}\hspace{.7cm}\right\},
\end{eqnarray}
\\
with $P=p_{1}+p_{2}$, $Q=p_{1}-p_{3}$ and $C,D,E,F,G,H$ rational coefficients.
\end{itemize}

\noindent
Equation (\ref{INTRO:eq38}) holds for any function $M({\bf s})$.
In particular, let us assume that $M({\bf s})$ is a master integral. 
We can now substitute the expression of $M$  in the 
{\it r.h.s.} of one of the
Eqs.(\ref{INTRO:eq40}-\ref{INTRO:eq43}),
according to the case, and perform the direct 
differentiation of the integrand. It is clear that we obtain a 
combination of several 
integrals, all belonging to the same topology as $M$. Therefore, 
we can use the solutions of the IBP-id's, LI-id's and other
identities for that  
topology and express everything 
in terms of the MI's of the considered topology (and its subtopologies). 
If there is more than one MI, the procedure 
can be repeated for all of them as well. In so doing, one obtains
a system of linear differential equations in ${\bf s}$ for $M$
and for the other MI's (if any), 
expressing their ${\bf s}$-derivatives in terms of the
MI's of the considered topology and of its subtopologies.

The system is formed by a set of 
{\it first-order differential equations} (ODE), one for each MI, say $M_j$,
whose general structure reads like the following,
\begin{eqnarray} 
\frac{\partial}{\partial s_{\alpha}} M_j(D,{\bf s}) & = & 
\sum_k A_k(D,{\bf s}) \ M_k(D,{\bf s}) 
+ \sum_h B_h(D,{\bf s}) \ N_h(D,{\bf s}) \,\,\, 
\label{INTRO:eq44} 
\end{eqnarray}
where
$\alpha=1,\cdots,{\cal{N}}$,
is the label of the invariants,
and $N_k$ are MI's of the subtopologies.
Note that the above equations are exact in $D$, and the coefficients $A_k, 
B_k$ are rational factors whose singularities
represent the thresholds and the  
pseudothresholds of the solution.


\noindent
The system of equations (\ref{INTRO:eq44}) for $M_j$
is not homogeneous, as they may involve MI's of subtopologies. 
It is therefore natural to proceed
bottom-up, starting from the equations for the MI's of the simplest 
topologies (i.e. with less denominators), solving those equations and 
using the results within the equations for the MI's of the more 
complicated topologies with additional propagators, whose 
non-homogeneous part can then be considered as known.

\subsection{Boundary conditions \label{THEM:boundary}}

The coefficients of the differential equations 
(\ref{INTRO:eq44}) are in general singular at some 
kinematic points (thresholds
and pseudothresholds), and correspondingly, 
the solutions of the equations can show singular 
behaviours in those points, while the unknown integral might have not. 
The boundary conditions for the differential equations
are found by exploiting the known analytical properties of the MI's 
under consideration, imposing the regularity or the 
finiteness of the solution at the {\it pseudo-thresholds} of the MI. This 
qualitative information is sufficient for the quantitative 
determination of the otherwise arbitrary integration constants, which 
naturally arise when solving a system of differential equations.

\subsection{Laurent expansion around $D=4$
\label{THEM:Laurent}}

The system of differential equations (\ref{INTRO:eq44}) is exact in $D$, and, when possible, its
solutions could be found for arbitrary value of the dimensional parameter.
Very often, the result of the computation will have to be anyway expanded around some typical
value of the space-time dimension. Indeed, in what follows, 
we will discuss the Laurent expansion of the 
solutions around $D=4$, though the procedure can be applied equivalently 
for other values of $D$. 
In general, we use the ansatz 
\begin{equation} 
M_j(D,{\bf s}) = \sum_{k=-n_0}^{n} \ {(D-4)^{k}} \ M_j^{(k)}({\bf s}) 
+ {\mathcal O} \Big((D-4)^{(n+1)}\Big) \ , 
\label{INTRO:eq45} 
\end{equation} 
where $n_0$ (positive) corresponds to the highest pole, 
and $n$ is the required order in $(D-4)$.

When expanding systematically in $(D-4)$ all the MI's (including 
those appearing in the non-homogeneous part) and all the $D$-dependent 
terms of (\ref{INTRO:eq45}), one obtains a system of chained 
equations for the Laurent coefficients $M_j^{(k)}$ of (\ref{INTRO:eq45}).  
The first equation, corresponding to the highest pole involves only the coefficient 
$M_i^{(-n_0)}$ as unknown; 
the next equation, corresponding to the next pole in $(D-4)$, 
involves the $M_i^{(-n_0+1)}$ as unknown, but it may involve
$M_i^{(-n_0)}$ in the non-homogeneous term (if it appears as multiplied by
any power of $(D-4)$);  
but such a term, $M_i^{(-n_0)}$,
has to be considered known once the equation for the highest pole 
has been solved. For the subsequent equations we have the same
structure: 
at a given order $k$ in $(D-4)$, the equations involve $M_j^{(k)}$ as unknown, 
and previous coefficients $M_j^{(\ell)} \ (-n_0 \le \ell < k)$ as known 
non-homogeneous terms. 

Let us note that the homogeneous part of all the equations arising 
from the $D\to4$ expansion of (\ref{INTRO:eq45}) is always the
same and obviously identical to the homogeneous part of
Eq.(\ref{INTRO:eq44}) read at
$D=4$. It is, therefore, natural to look for the solutions of the
chained non-homogeneous equations by means of Euler's method of the {\it variation of the
constants}, using repeatedly the solutions of the homogeneous equation as
integration kernel, as we will see in the following chapters. 

General algorithms for the solution of the homogeneous equations 
are not available; it turns out however that in all the considered cases 
the homogeneous equations at $D=4$ have (almost trivial) solutions, 
so that Euler's formula can immediately be written. 
With suitable changes of variable, according to the kinematics
of the problem, all integrations can further be carried out in closed analytic form, by exploiting the shuffle algebra induced by integration-by-parts
on the nested integral representation of the solution.

\section{Laplace Transform of Difference Equations
\label{sec:DifferenceEqn}}

\emph{Difference Equations} are functional relations among values of functions
shifted by integers, and can be considered a ''discretization'' 
of differential equations.
The set of identities used to derive Differential Equations in the external
invariants of Feynman integral, can be used as well to derive Difference 
Equation in one of the denominator powers, as discussed by 
Laporta\cite{Laporta:2001dd,Laporta:2003jz}.
Alternatively, the shifted dimension relations among scalar integrals
of Tarasov\cite{Tarasov:1996br} naturally yield equations where the integer
variable is the dimensional parameter $D$.
We briefly describe the former kind of difference equation and
their link to differential equations.

Let us consider the difference equation
\begin{equation}
\sum_{k=0}^N p_k(n) \ U(n+k) = 0 \ ,
\label{DiffenceEqn:generic}
\end{equation}
where $N$ is the equation order, and $p_k(n)$ are polynomial in $n$ of 
maximum degree $P$,
whose generic structure can be parametrised as follows,
\begin{equation}
p_k(n) = A_{k0} + \sum_{i=1}^P A_{ki} \prod_{j=0}^{i-1} (n + k + j) \ .
\end{equation}
The Laplace transform method consists in the substitution 
\begin{equation}
U(n) = \int_{\gamma} dt \ t^{n-1} \ v(t) \ ,
\label{DiffenceEqn:Udef}
\end{equation}
where $\gamma$ is a suitable integration path, and 
whose effect can be shown to be \cite{Laporta:2001dd}
the translation of the difference equation (\ref{DiffenceEqn:generic}),
into a differential equation for $v(t)$,
\begin{equation}
\sum_{i=0}^P \Phi_i(t) \ (-t)^i \ {d^{(i)} \over dt^i} v(t) = 0 \ ,
\label{DiffenceEqn:DiffEq4v}
\end{equation}
with
\begin{equation}
\Phi_i(t) = \sum_{k=0}^N A_{ki} \ t^k \ .
\end{equation}
The solution $v(t)$ of Eq.(\ref{DiffenceEqn:DiffEq4v}), still carry
unknown integration constants which will be present as well
in $U(n)$, reconstructed by integrating Eq.(\ref{DiffenceEqn:Udef}).
Finally, the value of the yet undetermined integration constants
can be fixed by imposing boundary conditions in the large-$n$ regime
of $U(n)$,
easily found by direct inspection of its original representation as
loop integral\footnote{
In the rest of the paper, for computational convenience, 
we adopt two different definitions of the loop measure:
{\it (i)} $d^D k/(2 \pi)^{D-2}$, for integrals treated with the Differential
Equations method;
{\it (ii)} $d^D k/(\pi)^{D/2}$, in case of use of Difference Equations method.
}.

{\footnotesize{
\subsubsection{Example: one-loop Tadpole}

\begin{equation}\label{DiffEq1}
U(n)=\pi^{-\frac{D}{2}}\int\frac{d^D k}{(k^2+1)^n}=\hspace{-.5cm}\parbox{20mm}{\begin{picture}(300,20)(0,0)
\SetScale{0.1}
\SetWidth{2.5}
\GOval(300,200)(100,100)(0){1.0}
\Line(200,100)(400,100)
\Vertex(300,300){10}
\Text(30,35)[]{\footnotesize{$n$}}
\end{picture}}.
\end{equation}
By means IBP Id's, one can write the following relationship between the integrals of family (\ref{DiffEq1})\\
\\
\begin{equation}\label{DiffEq2}
-\left(n-1-\frac{D}{2}\right)\hspace{-.5cm}
\parbox{20mm}{
\begin{picture}(300,20)(0,0)
\SetScale{0.1}
\SetWidth{2.5}
\GOval(300,200)(100,100)(0){1.0}
\Line(200,100)(400,100)
\Vertex(300,300){10}
\Text(30,35)[]{\footnotesize{$n-1$}}
\end{picture}
}\hspace{-.5cm}
+ (n-1)\hspace{-.5cm}
\parbox{20mm}{\begin{picture}(300,20)(0,0)
\SetScale{0.1}
\SetWidth{2.5}
\GOval(300,200)(100,100)(0){1.0}
\Line(200,100)(400,100)
\Vertex(300,300){10}
\Text(30,35)[]{\footnotesize{$n$}}
\end{picture}}
\hspace{-.5cm}
=0,
\end{equation}
which, after renaming $n \to n+1$, can be put in the form
(\ref{DiffenceEqn:generic}), and read as a first-order 
difference equation for the function $U(n)$,
\begin{equation}\label{DiffEq2bis}
- \bigg( n - {D \over 2} \bigg) U(n) + n \ U(n+1) = 0.
\end{equation}
The boundary condition is determined by the asymptotic behaviour ($n\to\infty$) 
of the integral (\ref{DiffEq1}), which reads
\begin{equation}\label{DiffEq5}
\pi^{-\frac{D}{2}}\int\frac{d^D k}{(k^2+1)^n}\thickapprox \pi^{-\frac{D}{2}}\int d^D k e^{-nk^2}=n^{-\frac{D}{2}} .
\end{equation}
One way to solve equation (\ref{DiffEq2bis}) is to 
seek for the solution as a \emph{factorial series} 
(see \cite{Laporta:2001dd,Laporta:2003jz} for details).
Equivalently, one can convert the difference equation  
into a differential equation, by means of the Laplace transform of $U(n)$.
Accordingly, with the ansatz
\begin{equation}
U(n) = \int_0^1 dt \ t^{n-1} \ v(t) \ ,
\label{DiffnceEqn:Un}
\end{equation}
and 
\begin{eqnarray}
A_{00} = {D \over 2} \ , & \qquad & A_{01} = -1 \ , \nonumber\\
A_{10} = {-1} \ , & \qquad & A_{11} = 1 \ , 
\end{eqnarray}
equation (\ref{DiffEq2bis}) becomes a differential equation in the form
of (\ref{DiffenceEqn:DiffEq4v})
for the function $v(t)$,
\begin{equation}
-t (t-1) v'(t) + \bigg({D \over 2} - t \bigg) v(t) = 0. 
\end{equation}
The solution is, 
\begin{equation}
v(t) = v_0 \ t^{-D/2} \ (1 - t)^{D/2 - 1} \ ,
\end{equation}
where $v_0$ is the yet unknown integration constant.
With the above solution, Eq.(\ref{DiffnceEqn:Un}) can be integrated and $U(n)$ reads,
\begin{equation}
U(n) = v_0 \frac{\Gamma \left(\frac{D}{2}\right) 
                 \Gamma \left(n-\frac{D}{2}\right)}{\Gamma (n)} \ .
\end{equation}
By comparing its large-$n$ expansion,
\begin{equation}
U(n) \stackrel{n\to\infty}{=} 
v_0 \ 
n^{-D/2} \bigg( \Gamma \left(\frac{D}{2}\right)
               + {\cal O}\bigg( \frac{1}{n} \bigg) \bigg)
\end{equation}
with the asymptotic limit explicitly computed in Eq.(\ref{DiffEq5}),
one finally determines the value of the constant,
\begin{equation}
v_0 = \frac{1}{\Gamma \left(\frac{D}{2}\right) } \ .
\end{equation} 
}} 

\begin{section}{One-Loop Vacuum Polarization in QED}

\begin{figure}[ht]
\begin{center}
\begin{picture}(150,50)(0,0)
\SetWidth{1.0}
\SetScale{0.7}
\Photon(25,25)(50,25){2}{3}    
\Photon(100,25)(125,25){2}{3}
\ArrowArcn(75,25)(25,0,180)
\ArrowArcn(75,25)(25,180,360)
\Text(53,41)[]{\small{$p$}}
\Text(53,-6)[]{\small{$p-k$}}
\Text(12,18.5)[]{\small{$k$}}
%
\end{picture}
\end{center}
\caption{
1-loop vacuum polarization diagram.
}
\label{fig:VP1L}
\end{figure}
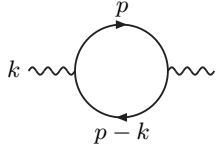

As an application, we calculate the one-loop correction to the 
photon propagator: the so called Vacuum Polarization tensor.
The only contributing diagram is shown in Fig. \ref{fig:VP1L}, and 
corresponds to the expression

\begin{equation}\label{VP1L:eq1}
i\left(\frac{\alpha}{\pi}\right)\Pi_{\mu\nu}(k)=(-ie)^2(-1)\int\frac{d^{D}p}{(2\pi)^{D}}{\textrm{Tr}}\left\{\frac{\gamma_{\mu}(-ip\hspace*{-0.2 cm}/+m)\gamma_{\nu}[-i(p\hspace*{-0.2 cm}/-k\hspace*{-0.2 cm}/)+m]}{(p^2+m^2)[(p-k)^2+m^2]}\right\}.
\end{equation}

$\Pi_{\mu\nu}$ is a tensor which depends only on the external momentum $k_{\mu}$ and the metric tensor $\delta_{\mu\nu}$ and can be decomposed as a sum of two contributions

\begin{equation}\label{VP1L:eq2}
\Pi_{\mu\nu}(k)=\Pi(D,k^2)k_{\mu}k_{\nu}+\Delta(D,k^2)\delta_{\mu\nu},
\end{equation}
where $\Pi(D,k^2)$ and $\Delta(D,k^2)$ are scalar functions.
The Ward identity (current conservation) tells us that $k_{\mu}\Pi_{\mu\nu}(k)=0$, hence

\begin{equation}\label{VP1L:eq3}
k_{\mu}\Pi_{\mu\nu}(k)=0\Longrightarrow k_{\nu}k^2\Pi(D,k^2)+k_{\nu}\Delta(D,k^2)=0\Longrightarrow \Delta(D,k^2)=-k^2\Pi(D,k^2), 
\end{equation}

and we can rewrite Eq. (\ref{VP1L:eq2}) as

\begin{equation}\label{VP1L:eq4}
\Pi_{\mu\nu}(k)=
\Pi(D,k^2)(k_{\mu}k_{\nu}-k^2\delta_{\mu\nu}) \ .
\end{equation}
Let us extract the scalar function $\Pi(D,k^2)$ taking the trace of the (\ref{VP1L:eq4})

\begin{equation}\label{VP1L:eq5}
\Pi_{\mu\mu}(D,k)=\Pi(D,k^2)(k^2-k^2 D)\Longrightarrow \Pi(D,k^2)=\frac{1}{k^2(1-D)}\Pi_{\mu\mu}(k).
\end{equation}

From (\ref{VP1L:eq1}), setting $\alpha=\frac{e^2}{4\pi}$, we have

\begin{equation}\label{VP1L:eq6}
i\Pi_{\mu\nu}(k)=\int\frac{d^{D}p}{(2\pi)^{D-2}}\frac{{\textrm{Tr}}(-\gamma_{\mu}p\hspace*{-0.2 cm}/\gamma_{\nu}q\hspace*{-0.2 cm}/+m^2\gamma_{\mu}\gamma_{\nu})}{(p^2+m^2)[(p-k)^2+m^2]};
\end{equation}

\noindent
after some operations within the Dirac algebra, our expression becomes

\begin{eqnarray}\label{VP1L:eq7}
\lefteqn{{\textrm{Tr}}(-\gamma_{\mu}p\hspace*{-0.2 cm}/\gamma_{\nu}q\hspace*{-0.2 cm}/+m^2\gamma_{\mu}\gamma_{\nu})=}\nonumber\\ & &
=-p_{\alpha}q_{\beta}{\textrm{Tr}}(\gamma_{\mu}\gamma_{\alpha}\gamma_{\nu}\gamma_{\beta})+m^2{\textrm{Tr}}(\gamma_{\mu}\gamma_{\nu})=\nonumber\\ & &
=-{\textrm{Tr}}(\mathbf{I})_{D} \ p_{\alpha}q_{\beta}(\delta_{\mu\alpha}\delta_{\nu\beta}-\delta_{\mu\nu}\delta_{\alpha\beta}+\delta_{\mu\beta}\delta_{\nu\alpha})+{\textrm{Tr}}(\mathbf{I})_{D} \ m^2\delta_{\mu\nu}\nonumber\\ & &
=-{\textrm{Tr}}(\mathbf{I})_{D} \ (p_{\mu}q_{\nu}-p\cdot q\delta_{\mu\nu}+p_{\nu}q_{\mu}-m^2\delta_{\mu\nu})=\nonumber\\ & & 
=-{\textrm{Tr}}(\mathbf{I})_{D} \ [2p_{\mu}p_{\nu}-p_{\mu}k_{\nu}-p_{\nu}k_{\mu}-(p^2-p\cdot k+m^2)\delta_{\mu\nu}],
\end{eqnarray}
where ${\textrm{Tr}}(\mathbf{I})_{D}$ is the trace of the unit matrix in $D$ dimensions (this is, in general, different from $D$). However, $\lim_{D\to 4}{\textrm{Tr}}(\mathbf{I})_{D} \ =4$. Substituting (\ref{VP1L:eq7}) in (\ref{VP1L:eq6}) and putting $\mu=\nu$ we obtain

\begin{eqnarray}\label{VP1L:eq8}
\lefteqn{\Pi_{\mu\mu}(k)=}
\nonumber\\ & &
={\textrm{Tr}}(\mathbf{I})_{D} \ i\int\frac{d^{D}p}{(2\pi)^{D-2}}\frac{2p^2-2p\cdot k-D(p^2-p\cdot k+m^2)}{(p^2+m^2)[(p-k)^2+m^2]}=\nonumber\\ & &
={\textrm{Tr}}(\mathbf{I})_{D} \ i\int\frac{d^{D}p}{(2\pi)^{D-2}}\frac{(D-2)(p\cdot k)-(D-2)p^2-D\,m^2}{(p^2+m^2)[(p-k)^2+m^2]}.
\end{eqnarray} 
We can define,
\begin{displaymath}
\begin{array}{l}{\mathcal{D}}_{1}=p^2+m^2
\\
{\mathcal{D}}_{2}=(p-k)^2+m^2
\end{array} 
\end{displaymath} 
so that after the trivial tensor reduction, according to
\begin{eqnarray}
p^2 &=& D_1^2 - m^2 \\
p\cdot k&=&\frac{1}{2}({\mathcal{D}}_{1}-{\mathcal{D}}_{2}+k^2)
\end{eqnarray}
Eq. (\ref{VP1L:eq8}) becomes

\begin{eqnarray}\label{VP1L:eq9}
\lefteqn{\Pi_{\mu\mu}(k)=}\nonumber\\ & &
={\textrm{Tr}}(\mathbf{I})_{D} \ i\int\frac{d^{D}p}{(2\pi)^{D-2}}\frac{(D-2)(p\cdot k)-(D-2)p^2-D\,m^2}{{\mathcal{D}}_{1}{\mathcal{D}}_{2}}=\nonumber\\ & &
={\textrm{Tr}}(\mathbf{I})_{D} \ i\int\frac{d^{D}p}{(2\pi)^{D-2}}\frac{(D-2)[({\mathcal{D}}_{1}-{\mathcal{D}}_{2}+k^2)/2]-(D-2)({\mathcal{D}}_{1}-m^2)-D\,m^2}{{\mathcal{D}}_{1}{\mathcal{D}}_{2}}=\nonumber\\ & &
={\textrm{Tr}}(\mathbf{I})_{D} \ i\left\{\frac{(2-D)}{2}\int\frac{d^{D}p}{(2\pi)^{D-2}}\left(\frac{1}{{\mathcal{D}}_{1}}+\frac{1}{{\mathcal{D}}_{2}}\right)+\left[\frac{k^2(D-2)}{2}-2m^2\right]\int\frac{d^{D}p}{(2\pi)^{D-2}}\frac{1}{{\mathcal{D}}_{1}{\mathcal{D}}_{2}}\right\}=\nonumber\\ & &
={\textrm{Tr}}(\mathbf{I})_{D} \ i\left\{\frac{(2-D)}{2}(J_{10}+J_{01})+\left[\frac{k^2(D-2)}{2}-2m^2\right]J_{11}\right\},
\end{eqnarray}
where all the integrals can be seen as belonging to the class,
\begin{equation}
J_{m_{1}m_{2}}=\int\frac{d^{D}p}{(2\pi)^{D-2}}\frac{1}{{\mathcal{D}}_{1}^{m_{1}}{\mathcal{D}}_{2}^{m_{2}}} \ .
\end{equation}

\noindent
Finally, the scalar vacuum polarization function can be written as
\begin{eqnarray}\label{VP1L:eq10}
\lefteqn{\Pi(D,k^2)=}\nonumber\\ & &=\frac{{\textrm{Tr}}(\mathbf{I})_{D} \ i}{2 k^2(1-D)}
\left\{(2-D)(J_{10}+J_{01})+[k^2(D-2)-4m^2]J_{11}\right\}=\nonumber\\ & & 
=\frac{{\textrm{Tr}}(\mathbf{I})_{D} \ i}{k^2(1-D)}\left\{(2-D)T(D,m^2)+\frac{1}{2}[k^2(D-2)-4m^2]J(D,k^2)\right\}.\nonumber\\ & &
\end{eqnarray}
One may notice the result is expressed in terms of two integrals: 
the \emph{massive Tadpole} \cite{Remiddi:1997ny}

\begin{equation}\label{VP1L:eq11}
T(D,m^2)=J_{10}=J_{01}=\parbox{20mm}{\begin{picture}(300,20)(0,0)
\SetScale{0.1}
\SetWidth{2.5}
\GOval(300,200)(100,100)(0){1.0}
\Photon(200,100)(400,100){6}{3}
\end{picture}}=\int\frac{d^D p}{(2\pi)^{D-2}}\frac{1}{{\cal{D}}_{1}}
= C(D) {m^{D-2} \over (D-2) (D-4)} \ ,
\end{equation}
and the \emph{1-loop 2-point function}
\begin{equation}\label{VP1L:eq12}
J(D,k^2)=J_{11}=\,\parbox{20mm}{\begin{picture}(300,20)(0,0)
\SetScale{0.1}
\SetWidth{2.5}
\Photon(100,100)(200,100){6}{2}    
\Photon(400,100)(500,100){6}{2}      
%
\GOval(300,100)(100,100)(0){1.0}
\end{picture}}=\int\frac{d^D p}{(2\pi)^{D-2}}\frac{1}{{\cal{D}}_{1}{\cal{D}}_{2}} \ ,
\end{equation}
where
\begin{equation}
C(D) = (4 \pi)^{4-D \over 2} \Gamma(3 - D/2) \ ,
\end{equation}
is an overall loop factor whose value in 4 dimension is $C(4) = 1$.
In the following pages, when not ambiguous, we will give 
as understood a factor $C(D)$ for every loop.

\begin{subsection}{Reduction to master integrals}

We have seen that the scalar vacuum polarization function $\Pi(D,k^2)$ is reducible to the sum of a Tadpole and an integral belonging to the 1-loop 2-point topology, 
the most general integral of this class being 

\begin{equation}\label{VP1L:eq13}
J_{m_{1}m_{2}}=\int\frac{d^D p}{(2\pi)^{D-2}}\frac{1}{{\cal{D}}_{1}^{m_{1}}{\cal{D}}_{2}^{m_{2}}} \ .
\end{equation}

By means of IBP id's, we can reduce every integral of the type
(\ref{VP1L:eq13}) to 

\begin{equation}\label{VP1L:eq14}
J_{11}=J(D,k^2)=\int\frac{d^D p}{(2\pi)^{D-2}}\frac{1}{{\cal{D}}_{1}{\cal{D}}_{2}},
\end{equation} 
which is, actually, the only MI of this topology, and to the tadpole $J_{10}$,
which is the MI of the only possible subtopology.
\\
%
%
%
%
%
Let us write explicitly some useful IBP id's for the integral (\ref{VP1L:eq14}). we have

\begin{equation}\label{VP1L:eq15}
\left\{\begin{array}{l}\int\frac{d^D p}{(2\pi)^{D-2}}\frac{\partial}{\partial p_{\mu}}\left(\frac{k_{\mu}}{{\mathcal{D}}_{1}{\mathcal{D}}_{2}}\right)=0\\
\int\frac{d^D p}{(2\pi)^{D-2}}\frac{\partial}{\partial p_{\mu}}\left(\frac{p_{\mu}}{{\mathcal{D}}_{1}{\mathcal{D}}_{2}}\right)=0
\end{array}\right.,
\end{equation}
where ${\cal{D}}_{1}=p^2+m^2$ and ${\cal{D}}_{2}=(p-k)^2+m^2$.
After the trivial tensor reduction we obtain the identities

\begin{equation}\label{VP1L:eq16}
\left\{\begin{array}{l}\parbox{10mm}{\begin{picture}(300,20)(0,0)
\SetScale{0.1}
\SetWidth{2.5}
\Photon(100,100)(200,100){6}{2}    
\Photon(400,100)(500,100){6}{2}      
%
\GOval(300,100)(100,100)(0){1.0}
\Vertex(300,200){10}
\end{picture}}\hspace{1cm}=\parbox{10mm}{\begin{picture}(300,20)(0,0)
\SetScale{0.1}
\SetWidth{2.5}
\Photon(100,100)(200,100){6}{2}    
\Photon(400,100)(500,100){6}{2}      
%
\GOval(300,100)(100,100)(0){1.0}
\Vertex(300,0){10}
\end{picture}}\\
\\
\parbox{15mm}{\begin{picture}(300,20)(0,0)
\SetScale{0.1}
\SetWidth{2.5}
\Photon(100,100)(200,100){6}{2}    
\Photon(400,100)(500,100){6}{2}      
%
\GOval(300,100)(100,100)(0){1.0}
\Vertex(300,200){10}
\end{picture}}\,\,\,\,\,\,\,\,\,\,\,=-\frac{(D-3)}{(4m^2+k^2)}\,\,\parbox{10mm}{\begin{picture}(300,20)(0,0)
\SetScale{0.1}
\SetWidth{2.5}
\Photon(100,100)(200,100){6}{2}    
\Photon(400,100)(500,100){6}{2}      
%
\GOval(300,100)(100,100)(0){1.0}
\end{picture}}\hspace{1cm}+\frac{1}{(4m^2+k^2)}\hspace{-.5cm}\parbox{10mm}{\begin{picture}(300,20)(0,0)
\SetScale{0.1}
\SetWidth{2.5}
\GOval(300,200)(100,100)(0){1.0}
\Photon(200,100)(400,100){6}{3}
\Vertex(300,300){10}
\end{picture}}
\end{array}\right.\hspace{.5cm},
\end{equation}

\noindent
with

\begin{eqnarray}
\parbox{10mm}{\begin{picture}(300,20)(0,0)
\SetScale{0.1}
\SetWidth{2.5}
\Photon(100,100)(200,100){6}{2}    
\Photon(400,100)(500,100){6}{2}      
%
\GOval(300,100)(100,100)(0){1.0}
\end{picture}}\hspace{1cm}&=&\int\frac{d^D p}{(2\pi)^{D-2}}\frac{1}{{\mathcal{D}}_{1}{\mathcal{D}}_{2}}\,\,, \\
&& \nonumber \\
&& \nonumber \\
\hspace{.3cm}\parbox{10mm}{\begin{picture}(300,20)(0,0)
\SetScale{0.1}
\SetWidth{2.5}
\Photon(100,100)(200,100){6}{2}    
\Photon(400,100)(500,100){6}{2}      
%
\GOval(300,100)(100,100)(0){1.0}
\Vertex(300,200){10}
\end{picture}}\hspace{1cm}&=&\int\frac{d^D p}{(2\pi)^{D-2}}\frac{1}{{\mathcal{D}}_{1}^2{\mathcal{D}}_{2}} \\
&& \nonumber \\
&& \nonumber \\
\parbox{10mm}{\begin{picture}(300,20)(0,0)
\SetScale{0.1}
\SetWidth{2.5}
\Photon(100,100)(200,100){6}{2}    
\Photon(400,100)(500,100){6}{2}      
%
\GOval(300,100)(100,100)(0){1.0}
\Vertex(300,0){10}
\end{picture}}\hspace{1cm}&=&\int\frac{d^D p}{(2\pi)^{D-2}}\frac{1}{{\mathcal{D}}_{1}{\mathcal{D}}_{2}^2}\,\,, \\
&& \nonumber \\
&& \nonumber \\
\hspace{.3cm}\parbox{10mm}{
\begin{picture}(300,20)(0,0)
\SetScale{0.1}
\SetWidth{2.5}
\GOval(300,200)(100,100)(0){1.0}
\Vertex(300,300){10}
\Photon(200,100)(400,100){6}{3}
\end{picture}}\hspace{.7cm}&=&\int\frac{d^D p}{(2\pi)^{D-2}}\frac{1}{{\mathcal{D}}_{1}^2} = 
-{ (D-2) \over 2 \ m^2}
\hspace*{-0.5cm}
\parbox{10mm}{
\begin{picture}(300,20)(0,0)
\SetScale{0.1}
\SetWidth{2.5}
\GOval(300,200)(100,100)(0){1.0}
\Photon(200,100)(400,100){6}{3}
\end{picture}}
\label{VP1L:tadpolIBPid}
\end{eqnarray}
where the last equation has been obtained as IBP identity for the
tadpole \cite{Remiddi:1997ny}.

\end{subsection}

\begin{subsection}{Differential equation for $J(D,k^2)$}

The master integral  $J(D,k^2)$ is an analytic function of the argument $k^2$ and it can be viewed as the solution of a suitable differential equation. Let us see how to build and solve such an equation.  For $J(D,k^2)$ the following trivial identity holds,

\begin{equation}\label{VP1L:eq17}
\frac{\partial J}{\partial k_{\mu}}=\frac{\partial J}{\partial k^2}\frac{\partial k^2}{\partial k_{\mu}}=2k_{\mu}\frac{\partial J}{\partial k^2}.
\end{equation}

By contracting (\ref{VP1L:eq17}) with the vector $k_{\mu}$ we have

\begin{equation}\label{VP1L:eq18}
k_{\mu}\frac{\partial J}{\partial k_{\mu}}=2k^2\frac{\partial J}{\partial k^2}.
\end{equation}

On the other hand 

\begin{equation}\label{VP1L:eq19}
\frac{\partial J}{\partial k_{\mu}}=\int\frac{d^D p}{(2\pi)^{D-2}}\frac{\partial }{\partial k_{\mu}}\left(\frac{1}{{\mathcal{D}}_{1}{\mathcal{D}}_{2}}\right)=\int\frac{d^D p}{(2\pi)^{D-2}}\frac{2(p_{\mu}-k_{\mu})}{{\mathcal{D}}_{1}{\mathcal{D}}_{2}^2},
\end{equation}

so

\begin{eqnarray}\label{VP1L:eq20}
\lefteqn{k_{\mu}\frac{\partial J}{\partial k_{\mu}}=
\int\frac{d^D p}{(2\pi)^{D-2}}\frac{2(p\cdot k-k^2)}{{\mathcal{D}}_{1}{\mathcal{D}}_{2}^2}\underbrace{=}_{2p\cdot k={\mathcal{D}}_{1}-{\mathcal{D}}_{2}+k^2}}\nonumber\\ & &
=\int\frac{d^D p}{(2\pi)^{D-2}}\frac{1}{{\mathcal{D}}_{2}^2}-\int\frac{d^D p}{(2\pi)^{D-2}}\frac{1}{{\mathcal{D}}_{1}{\mathcal{D}}_{2}}-\int\frac{d^D p}{(2\pi)^{D-2}}\frac{k^2}{{\mathcal{D}}_{1}{\mathcal{D}}_{2}^2}=\nonumber\\ & &
=\parbox{15mm}{ \begin{picture}(300,20)(0,0)
\SetScale{0.1}
\SetWidth{2.5}
\GOval(300,100)(100,100)(0){1.0}
\Vertex(300,200){10}
\Photon(200,0)(400,0){6}{3}
\end{picture}}\,-\parbox{10mm}{\begin{picture}(300,20)(0,0)
\SetScale{0.1}
\SetWidth{2.5}
\Photon(100,100)(200,100){6}{2}    
\Photon(400,100)(500,100){6}{2}      
%
\GOval(300,100)(100,100)(0){1.0}
\end{picture}}\hspace{1cm}-k^2\parbox{15mm}{\begin{picture}(300,20)(0,0)
\SetScale{0.1}
\SetWidth{2.5}
\Photon(100,100)(200,100){6}{2}    
\Photon(400,100)(500,100){6}{2}      
%
\GOval(300,100)(100,100)(0){1.0}
\Vertex(300,0){10}
\end{picture}}
\end{eqnarray}

By substituting Eq. (\ref{VP1L:eq20}) in Eq. (\ref{VP1L:eq18}) we have

\begin{equation}\label{VP1L:eq21}
\frac{d}{dk^2}\parbox{10mm}{\begin{picture}(300,20)(0,0)
\SetScale{0.1}
\SetWidth{2.5}
\Photon(100,100)(200,100){6}{2}    
\Photon(400,100)(500,100){6}{2}      
%
\GOval(300,100)(100,100)(0){1.0}
\end{picture}}\hspace{1cm}=\frac{1}{2k^2}\hspace{-.5cm}\parbox{10mm}{\begin{picture}(300,20)(0,0)
\SetScale{0.1}
\SetWidth{2.5}
\GOval(300,100)(100,100)(0){1.0}
\Vertex(300,200){10}
\Photon(200,0)(400,0){6}{3}
\end{picture}}\hspace{.5cm}-\frac{1}{2k^2}\parbox{10mm}{\begin{picture}(300,20)(0,0)
\SetScale{0.1}
\SetWidth{2.5}
\Photon(100,100)(200,100){6}{2}    
\Photon(400,100)(500,100){6}{2}      
%
\GOval(300,100)(100,100)(0){1.0}
\end{picture}}\hspace{1cm}-\frac{1}{2}\,\parbox{10mm}{\begin{picture}(300,20)(0,0)
\SetScale{0.1}
\SetWidth{2.5}
\Photon(100,100)(200,100){6}{2}    
\Photon(400,100)(500,100){6}{2}      
%
\GOval(300,100)(100,100)(0){1.0}
\Vertex(300,0){10}
\end{picture}}\hspace{1cm},
\end{equation}
which is rewritten, thanks to the second identity of the (\ref{VP1L:eq16})
and to (\ref{VP1L:tadpolIBPid}), 
as a non-homogeneous first-order differential equation for $J(D,k^2)$
\begin{eqnarray}\label{VP1L:eq22}
\lefteqn{\frac{d}{dk^2}\,\,\parbox{10mm}{\begin{picture}(300,20)(0,0)
\SetScale{0.1}
\SetWidth{2.5}
\Photon(100,100)(200,100){6}{2}    
\Photon(400,100)(500,100){6}{2}      
%
\GOval(300,100)(100,100)(0){1.0}
\end{picture}}\hspace{1cm}+\frac{1}{2}\left[\frac{1}{k^2}-\frac{(D-3)}{(k^2+4m^2)}\right]\parbox{10mm}{\begin{picture}(300,20)(0,0)
\SetScale{0.1}
\SetWidth{2.5}
\Photon(100,100)(200,100){6}{2}    
\Photon(400,100)(500,100){6}{2}      
%
\GOval(300,100)(100,100)(0){1.0}
\end{picture}
}}\nonumber\\ & &
=-\frac{(D-2)}{4m^2}\left[\frac{1}{k^2}-\frac{1}{(k^2+4m^2)}\right]\hspace{-.5cm}\parbox{10mm}{\begin{picture}(300,20)(0,0)
\SetScale{0.1}
\SetWidth{2.5}
\GOval(300,100)(100,100)(0){1.0}
\Photon(200,0)(400,0){6}{3}
\end{picture}}\hspace{1cm}.
\end{eqnarray}
Eq. (\ref{VP1L:eq22}) contains the boundary condition for the solution. In fact, thanks to the analytic properties of Feynman integrals, we know that $J(D,k^2)$ must be a regular function in $k^2=0$, that is 

\begin{equation}\label{VP1L:eq23}
\lim_{k^2\rightarrow 0}k^2\,\frac{dJ}{dk^2}=0.
\end{equation}

By multiplying Eq. (\ref{VP1L:eq22}) by $k^2$ and taking the limit $k^2\to 0$, we have

\begin{eqnarray}\label{VP1L:eq24} 
\lefteqn{\hspace{-.5cm}\lim_{k^2\rightarrow 0}k^2\frac{d}{dk^2}\parbox{20mm}{\begin{picture}(300,20)(0,0)
\SetScale{0.1}
\SetWidth{2.5}
\Photon(100,100)(200,100){6}{2}    
\Photon(400,100)(500,100){6}{2}      
%
\GOval(300,100)(100,100)(0){1.0}
\end{picture}}+\lim_{k^2\rightarrow 0}\frac{1}{2}\left[1-\frac{(D-3)k^2}{(k^2+4m^2)}\right]\hspace{-.4cm}\parbox{20mm}{\begin{picture}(300,20)(0,0)
\SetScale{0.1}
\SetWidth{2.5}
\Photon(100,100)(200,100){6}{2}    
\Photon(400,100)(500,100){6}{2}      
%
\GOval(300,100)(100,100)(0){1.0}
\end{picture}}\hspace{-.3cm}=}\nonumber\\ & &
=-\lim_{k^2\rightarrow 0}\frac{(D-2)}{4m^2}\left[1-\frac{k^2}{(k^2+4m^2)}\right]\hspace{-.4cm}\parbox{10mm}{\begin{picture}(300,20)(0,0)
\SetScale{0.1}
\SetWidth{2.5}
\GOval(300,100)(100,100)(0){1.0}
\Photon(200,0)(400,0){6}{3}
\end{picture}}
\hspace{.8cm},
\end{eqnarray}

out of which one has,
\begin{equation}\label{VP1L:eq25}
\lim_{k^2\rightarrow 0}J(D,k^2)=J(D,0)=-\frac{(D-2)}{2m^2}\hspace{-.4cm}\parbox{10mm}{\begin{picture}(300,20)(0,0)
\SetScale{0.1}
\SetWidth{2.5}
\GOval(300,100)(100,100)(0){1.0}
\Photon(200,0)(400,0){6}{3}
\end{picture}}\hspace{.5cm}
\end{equation} 

Eq. (\ref{VP1L:eq22}) with the condition (\ref{VP1L:eq25}) constitutes 
the Cauchy problem we have to solve.

\end{subsection}

\begin{subsection}{Exact solution in $D$ dimensions}

Equation (\ref{VP1L:eq22}) can be solved exactly in $D$ dimensions. Let us introduce the dimensionless variable $x=\frac{k^2}{4m^2}$, obviously $\frac{d}{dk^2}=\frac{dx}{dk^2}\frac{d}{dx}\rightarrow \frac{d}{dk^2}=\frac{1}{4m^2}\frac{d}{dx}$, and the equation (\ref{VP1L:eq22}) can be rewritten as follows

\begin{equation}\label{VP1L:eq26}
 \frac{d}{dx}\hspace{-.2cm}\parbox{10mm}{\begin{picture}(300,20)(0,0)
\SetScale{0.1}
\SetWidth{2.5}
\Photon(100,100)(200,100){6}{2}    
\Photon(400,100)(500,100){6}{2}      
%
\GOval(300,100)(100,100)(0){1.0}
\end{picture}}\hspace{.8cm}+\frac{1}{2}\left[\frac{1}{x}-\frac{(D-3)}{(1+x)}\right]\hspace{-.3cm}\parbox{10mm}{\begin{picture}(300,20)(0,0)
\SetScale{0.1}
\SetWidth{2.5}
\Photon(100,100)(200,100){6}{2}    
\Photon(400,100)(500,100){6}{2}      
%
\GOval(300,100)(100,100)(0){1.0}
\end{picture}}
\hspace{.7cm}=-\frac{(D-2)}{4m^2}\left[\frac{1}{x}-\frac{1}{(1+x)}\right]\hspace{-.5cm}\parbox{10mm}{\begin{picture}(300,20)(0,0)
\SetScale{0.1}
\SetWidth{2.5}
\GOval(300,100)(100,100)(0){1.0}
\Photon(200,0)(400,0){6}{3}
\end{picture}}\hspace{1cm}.
\end{equation}

The general solution of Eq. (\ref{VP1L:eq26}) is the sum of solution of the homogeneous 
equation, say $J_{0}$,
 and particular solution of the complete equation, say $J^*$.
The homogeneous equation is

\begin{equation}\label{VP1L:eq27}
\frac{dJ_{0}}{dx}=-\frac{1}{2}\left[\frac{1}{x}-\frac{(D-3)}{(1+x)}\right]J_{0}
\end{equation}
with solution,
\begin{equation}\label{VP1L:eq28}
J_{0}(D,x)=Ax^{-\frac{1}{2}}(1+x)^{\frac{D-3}{2}}.
\end{equation}

To find a particular solution, $J^{*}(D,x)$ of the complete equation, we use Euler's variation of constants method, and write 

\begin{equation}\label{VP1L:eq29}
J^{*}(D,x)=x^{-\frac{1}{2}}(1+x)^{\frac{(D-3)}{2}}\phi(x),
\end{equation} 

where $\phi(x)$ is an unknown function to be found by
imposing $J^*$ fulfills Eq.(\ref{VP1L:eq26}).
In so doing, we obtain the following equation for $\phi(x)$, 

\begin{equation}\label{VP1L:eq30}
\frac{d\phi}{dx}=-\frac{(D-2)}{4m^2}\hspace{-.4cm}\parbox{10mm}{\begin{picture}(300,20)(0,0)
\SetScale{0.1}
\SetWidth{2.5}
\GOval(300,100)(100,100)(0){1.0}
\Photon(200,0)(400,0){6}{3}
\end{picture}}\hspace{.6cm}[x^{-\frac{1}{2}}(1+x)^{\frac{3-D}{2}}-x^{\frac{1}{2}}(1+x)^{\frac{1-D}{2}}],
\end{equation}

hence

\begin{equation}\label{VP1L:eq31}
\phi(x)=-\frac{(D-2)}{4m^2}\hspace{-.4cm}\parbox{10mm}{\begin{picture}(300,20)(0,0)
\SetScale{0.1}
\SetWidth{2.5}
\GOval(300,100)(100,100)(0){1.0}
\Photon(200,0)(400,0){6}{3}
\end{picture}}\hspace{.5cm}\left[\int x^{-\frac{1}{2}}(1+x)^{-\frac{D-3}{2}}dx-\int x^{\frac{1}{2}}(1+x)^{-\frac{D-1}{2}}dx\right].
\end{equation}

The integrals in Eq.(\ref{VP1L:eq31}) are a representation of hypergeometric functions

\begin{equation}\label{VP1L:eq32}
\begin{array}{l}
\int x^{-\frac{1}{2}}(1+x)^{-\frac{D-3}{2}}dx=2x^{\frac{1}{2}}\,\, _{2}F_{1}\left(\frac{1}{2},\frac{D-3}{2};\frac{3}{2};-x\right)\\
\int x^{\frac{1}{2}}(1+x)^{-\frac{D-1}{2}}dx=\frac{2}{3}x^{\frac{3}{2}}\,\, _{2}F_{1}\left(\frac{3}{2},\frac{D-1}{2};\frac{5}{2};-x\right),
\end{array}
\end{equation}

hence

\begin{eqnarray}\label{VP1L:eq33}
\lefteqn{J(D,x)=J_{0}+J^{*}=}\nonumber\\ & &
=Ax^{-\frac{1}{2}}(1+x)^{\frac{(D-3)}{2}}
-\frac{(D-2)}{4m^2}\hspace{-.4cm}\parbox{10mm}{\begin{picture}(300,20)(0,0)
\SetScale{0.1}
\SetWidth{2.5}
\GOval(300,100)(100,100)(0){1.0}
\Photon(200,0)(400,0){6}{3}
\end{picture}}\hspace{.6cm}(1+x)^{\frac{D-3}{2}}\times\nonumber\\ & &
\times\left[2\,_{2}F_{1}\left(\frac{1}{2},\frac{D-3}{2};\frac{3}{2};-x\right)-\frac{2}{3}x\,_{2}F_{1}\left(\frac{3}{2},\frac{D-1}{2};\frac{5}{2};-x\right)\right].
\end{eqnarray}

By imposing the boundary condition, we see that the constant $A=0$, because the term $x^{-\frac{1}{2}}(1+x)^{\frac{(D-3)}{2}}$ is singular for $x\to 0$, while the MI is regular by inspection.

Finally, the full solution of our Cauchy problem is

\begin{eqnarray}\label{VP1L:eq34}
\lefteqn{J(D,x)=-\frac{(D-2)}{2m^2}\hspace{-.4cm}\parbox{10mm}{\begin{picture}(300,20)(0,0)
\SetScale{0.1}
\SetWidth{2.5}
\GOval(300,100)(100,100)(0){1.0}
\Photon(200,0)(400,0){6}{3}
\end{picture}}\hspace{.6cm}(1+x)^{\frac{D-3}{2}}\times}\nonumber\\ & &
\times\left[\,_{2}F_{1}\left(\frac{1}{2},\frac{D-3}{2};\frac{3}{2};-x\right)-\frac{x}{3}\,_{2}F_{1}\left(\frac{3}{2},\frac{D-1}{2};\frac{5}{2};-x\right)\right]=\nonumber\\ & &
=-\frac{(D-2)}{2m^2}\hspace{-.4cm}\parbox{10mm}{\begin{picture}(300,20)(0,0)
\SetScale{0.1}
\SetWidth{2.5}
\GOval(300,100)(100,100)(0){1.0}
\Photon(200,0)(400,0){6}{3}
\end{picture}}\hspace{.6cm}(1+x)^{\frac{D-3}{2}}\,\,_{2}F_{1}\left(\frac{D-1}{2},\frac{1}{2};\frac{3}{2};-x\right)=\nonumber\\ & &
=-\frac{(D-2)}{2m^2}\hspace{-.4cm}\parbox{10mm}{\begin{picture}(300,20)(0,0)
\SetScale{0.1}
\SetWidth{2.5}
\GOval(300,100)(100,100)(0){1.0}
\Photon(200,0)(400,0){6}{3}
\end{picture}}\hspace{.6cm}\,_{2}F_{1}\left(\frac{4-D}{2},1;\frac{3}{2};-x\right)
\end{eqnarray}

\begin{subsection}{Renormalized vacuum polarization function}

Now we can write the exact $D$-dimensional expression for the renormalized vacuum polarization function. This is

\begin{equation}\label{VP1L:eq35}
\Pi(D,k^2)_{R}=\Pi(D,k^2)-\Pi(D,0),
\end{equation} 

where $\Pi(D,0)$ is the function evaluated at zero momentum transfer 

\begin{equation}\label{VP1L:eq36}
\Pi(D,0) =
\frac{{\textrm{Tr}}(\mathbf{I})_{D} \ i}{(1-D)}\frac{(2-D)}{4m^2}\hspace{-.4cm}\parbox{10mm}{\begin{picture}(300,20)(0,0)
\SetScale{0.1}
\SetWidth{2.5}
\GOval(300,100)(100,100)(0){1.0}
\Photon(200,0)(400,0){6}{3}
\end{picture}}\hspace{.6cm}\lim_{x\rightarrow 0}\frac{1}{x}
+\frac{D-2}{2}J(D,0)
-\frac{1}{2}\lim_{x\rightarrow 0}\frac{J(D,x)}{x}.
\end{equation}

We see that

\begin{equation}\label{VP1L:eq37}
\lim_{x\rightarrow 0}\frac{J(D,x)}{x}
 =J(D,0)\lim_{x\rightarrow 0}\frac{1}{x}
 +J'(D,0)=\frac{(2-D)}{2m^2}\hspace{-.4cm}\parbox{10mm}{\begin{picture}(300,20)(0,0)
\SetScale{0.1}
\SetWidth{2.5}
\GOval(300,100)(100,100)(0){1.0}
\Photon(200,0)(400,0){6}{3}
\end{picture}}\hspace{.6cm}\lim_{x\rightarrow 0}\frac{1}{x}
 +J'(D,0) \ ,
\end{equation}

hence

\begin{equation}\label{VP1L:eq38}
\Pi(D,0)=\frac{{\textrm{Tr}}(\mathbf{I})_{D} \ i}{(1-D)}\left\{\frac{D-2}{2}J(D,0)
-\frac{1}{2}J'(D,0)\right\}.
\end{equation}

Finally, one has the renormalization counterterm,

\begin{equation}\label{VP1L:eq39}
\Pi(D,0)={\textrm{Tr}}(\mathbf{I})_{D} \ i \ 
         \frac{(D-2)}{6 m^2}\hspace{-.4cm}
\parbox{10mm}{\begin{picture}(300,20)(0,0)
\SetScale{0.1}
\SetWidth{2.5}
\GOval(300,100)(100,100)(0){1.0}
\Photon(200,0)(400,0){6}{3}
\end{picture}}\hspace{.6cm}
\end{equation}

yielding the renormalised form factor,

\begin{eqnarray}\label{VP1L:eq40}
\lefteqn{\Pi_{R}(D,k^2)=\Pi(D,k^2)-\Pi(D,0)=
\frac{{\textrm{Tr}}(\mathbf{I})_{D} \ i}{(1-D)}\frac{(2-D)}{4m^2}\hspace{-.4cm}\parbox{10mm}{\begin{picture}(300,20)(0,0)
\SetScale{0.1}
\SetWidth{2.5}
\GOval(300,100)(100,100)(0){1.0}
\Photon(200,0)(400,0){6}{3}
\end{picture}}\hspace{.6cm}\times}\nonumber\\ & &
\times\left\{\frac{1}{x}+\left[(D-2)-\frac{1}{x}\right]\,_{2}F_{1}\left(\frac{4-D}{2},1;\frac{3}{2};-x\right)+\frac{2}{3}(1-D)\right\}.\nonumber\\
\end{eqnarray}
which can be ultimately expanded around $D=4$ up to the finite order
in terms of HPL's
\begin{eqnarray}\label{VP1L:eq41}
\Pi_{R}(D,k^2) &=&
       i \  \Bigg(
  \frac{5}{9} 
- \frac{4}{3\,{\left( 1 - z \right) }^2}  
+  \frac{4}{3\,\left( 1 - z \right) } \nonumber \\ 
&&
+
  \frac{\left( 1 + z \right) \,
     \left( 1 - 4\,z + z^2 \right) \,H(0,z)} {3\,
     {\left( 1 - z \right) }^3} 
\Bigg)
               + {\cal O}((D-4)).
\end{eqnarray}
with the spacelike variable $z$ defined as
\begin{equation}\label{VP1L:eq42}
z = {\sqrt{-x + 1} - \sqrt{-x} \over \sqrt{-x + 1} + \sqrt{-x}}
  = {\sqrt{-k^2 + 4 m^2} - \sqrt{-k^2} \over 
     \sqrt{-k^2 + 4 m^2} + \sqrt{-k^2}}
\end{equation}
 
\end{subsection}
\end{subsection}
\end{section}

\begin{section}{Two-Loop Vacuum Polarization in QED \label{2LVacPol}}

The two-loop corrections to the vacuum polarization function was first calculated in 1955 by K\"allen and Sabry\cite{KS}. At that time, it was one of the first applications of perturbative QED, soon after the theory was definitely laid out. The exact expression in dimensional regularization was calculated in 1993 by Broadhurst, Fleischer and Tarasov\cite{BroFleTar}. The result was expressed in terms of generalized Hypergeometric functions $_{2}F_{1}\left(\frac{k^2}{4m^2}\right)$ and $_{3}F_{2}\left(\frac{k^2}{4m^2}\right)$, where $k$ is the momentum of the external photon and $m$ the mass of the internal fermion.\\
In the following sections we will derive the same result by means of the method of differential equations. First of all, we will show that the use of IBP allows one to reduce the whole problem to five MI's, out of which three
are product of one-loop integrals (known from the previous section), and
two are actual two-loop integrals, as yet unknown. 
Then, we will write for them a system of two first order differential equations in the square of the external momentum. 
We will see that the system is equivalent to 
a pair of second order differential equations, one for each MI. These equations can be solved exactly in $D$ dimensions and the solution can be written in terms of generalized Hypergeometric functions. 

\begin{subsection}{Diagrams}
The \emph{one particle irreducible} contributions to the two-loop vacuum polarization are showed in Eq. (\ref{VP2L:eq1}). 
They consist of three bare diagrams, plus the corresponding
counterterms, namely two fermion self-energy counterterms 
and two vertex corrections counterterms.
Due to Ward identities of QED, some of these counterterms cancel each other:
the vertex subtractions cancel exactly the wavefunction renormalization counterterms; so the only effective contribution comes from the 
fermion mass counterterm, as shown in Eq. (\ref{VP2L:eq1}).
The amplitude reads  
\begin{eqnarray}\label{VP2L:eq1}
\lefteqn{i\left(\frac{\alpha}{\pi}\right)^2\Pi_{\mu\nu}(k)=}\nonumber\\ & &
\nonumber\\ & & 
=
\hspace{-.3cm}\parbox{10mm}{\begin{picture}(300,20)(0,0)
\SetScale{0.11}
\SetWidth{2.5}
\Photon(100,100)(200,100){6}{3}    
\Photon(400,100)(500,100){6}{3}      
%
\GOval(300,100)(100,100)(0){0.5}
\end{picture}}\hspace{1cm}=\hspace{-.3cm}\parbox{10mm}{\begin{picture}(300,20)(0,0)
\SetScale{0.11}
\SetWidth{2.5}
\Photon(100,100)(200,100){6}{3}    
\Photon(400,100)(500,100){6}{3}      
%
\GOval(300,100)(100,100)(0){1.0}
\Photon(300,200)(300,0){6}{4}
\end{picture}}\hspace{1cm}+\hspace{-.3cm}\parbox{10mm}{\begin{picture}(300,20)(0,0)
\SetScale{0.11}
\SetWidth{2.5}
\Photon(100,100)(200,100){6}{3}    
\Photon(400,100)(500,100){6}{3}      
%
\GOval(300,100)(100,100)(0){1.0}
\Photon(235,175)(365,175){6}{3}
\end{picture}}\hspace{1cm}+\hspace{-.3cm}\parbox{10mm}{\begin{picture}(300,20)(0,0)
\SetScale{0.11}
\SetWidth{2.5}
\Photon(100,100)(200,100){6}{3}    
\Photon(400,100)(500,100){6}{3}      
%
\GOval(300,100)(100,100)(0){1.0}
\Photon(235,35)(365,35){6}{3}
\end{picture}}\nonumber\\ & &
\nonumber\\ & &
\nonumber\\ & &
\hspace{2.3cm}+\parbox{10mm}{
\begin{picture}(300,20)(0,0)
\SetScale{0.11}
\SetWidth{2.5}
\Photon(100,100)(200,100){6}{3}    
\Photon(400,100)(500,100){6}{3}      
%
\GOval(300,100)(100,100)(0){1.0}
\Text(33,0.5)[]{$\times$}
\Text(33,-6)[]{\footnotesize{$i\delta m$}}
\end{picture}}\hspace{1cm}+\hspace{-.3cm}\parbox{10mm}{\begin{picture}(300,20)(0,0)
\SetScale{0.11}
\SetWidth{2.5}
\Photon(100,100)(200,100){6}{3}    
\Photon(400,100)(500,100){6}{3}      
%
\GOval(300,100)(100,100)(0){1.0}
\Text(33,22)[]{$\times$}
\Text(33,28)[]{\footnotesize{$i\delta m$}}
\end{picture}}
\hspace{1cm}
%
%
%
%
%
%
%
%
%
%
\end{eqnarray}

\end{subsection}

\begin{subsection}{Topologies and Master Integrals}

From the above Feynman diagrams, one can identify seven
\emph{independent} topologies, which cannot be 
related to each other by a transformation of the internal momenta, 
as shown in the second column of Tab. \ref{VP2L:table1}.\\ 

By the systematic application of the reduction algorithm, one
can express all the needed integrals as combination of just five
MI's, depicted in the last column of Tab. \ref{VP2L:table1}:
one MI with four denominators; three with three; and one with two.

\begin{table}[!h]
\caption{Independent topologies for the two-loop vacuum polarization}
\label{VP2L:table1}
\begin{center}
\begin{tabular}{||c|c|c||}\hline
$denominators$ & $Independent\,\,topologies$ & ${\textrm{MI}}$\\ 
\hline

$5$ & $\parbox{10mm}{\begin{picture}(300,20)(0,0)
\SetScale{0.08}
\SetWidth{2.5}
\Photon(100,100)(200,100){6}{3}    
\Photon(400,100)(500,100){6}{3}      
%
\GOval(300,100)(100,100)(0){1.0}
\Photon(300,200)(300,0){6}{4}
\end{picture}}$ & ${\textrm{completely reducible}}$\\ 
\hline

$4$ & $\parbox{12mm}{\begin{picture}(300,20)(0,0)
\SetScale{0.08}
\SetWidth{2.5}
\Photon(100,100)(200,100){6}{3}    
\Photon(400,100)(500,100){6}{3}
%
\GOval(300,100)(100,100)(0){1.0}
\Photon(300,200)(400,100){6}{3}
\end{picture}}\hspace{.3cm}\parbox{12mm}{\begin{picture}(300,20)(0,0)
\SetScale{0.08}
\SetWidth{2.5}
\Photon(0,100)(100,100){6}{3}    
\Photon(500,100)(600,100){6}{3}      
%
\GOval(200,100)(100,100)(0){1.0}
\GOval(400,100)(100,100)(0){1.0}
\end{picture}}$ & $
\parbox{12mm}{\begin{picture}(300,20)(0,0)
\SetScale{0.08}
\SetWidth{2.5}
\Photon(0,100)(100,100){6}{3}    
\Photon(500,100)(600,100){6}{3}      
%
\GOval(200,100)(100,100)(0){1.0}
\GOval(400,100)(100,100)(0){1.0}
\end{picture}}$\\
 \hline

$3$ & $\parbox{12mm}{\begin{picture}(300,20)(0,0)
\SetScale{0.08}
\SetWidth{2.5}
\Photon(100,100)(200,100){6}{3}    
\Photon(400,100)(500,100){6}{3}

\GOval(300,100)(100,100)(0){1.0}
\Photon(200,100)(400,100){6}{3}
\end{picture}}
\parbox{12mm}{\begin{picture}(300,20)(0,0)
\SetScale{0.08}
\SetWidth{2.5}
\Photon(100,100)(200,100){6}{3}    
\Photon(400,100)(400,0){6}{3}      
%
\GOval(300,100)(100,100)(0){1.0}
\GOval(500,100)(100,100)(0){1.0}
\end{picture}}
\parbox{12mm}{\begin{picture}(300,20)(0,0)
\SetScale{0.08}
\SetWidth{2.5}
%
%
\GOval(300,100)(100,100)(0){1.0}
\Photon(200,100)(400,100){6}{4}
\end{picture}}$ & 
\hspace{0.5cm}$\parbox{12mm}{\begin{picture}(300,20)(0,0)
\SetScale{0.08}
\SetWidth{2.5}
\Photon(100,100)(200,100){6}{3}    
\Photon(400,100)(500,100){6}{3}
%
\GOval(300,100)(100,100)(0){1.0}
\Photon(200,100)(400,100){6}{4}
\end{picture}}\,\,\parbox{10mm}{\begin{picture}(300,20)(0,0)
\SetScale{0.08}
\SetWidth{2.5}
\Photon(100,100)(200,100){6}{3}    
\Photon(400,100)(500,100){6}{3}
%
\GOval(300,100)(100,100)(0){1.0}
\Photon(200,100)(400,100){6}{4}
\Text(43,15)[]{{\tiny $(p_{1}\cdot p_{2})$}}
\end{picture}}
\hspace*{0.5cm}
\parbox{12mm}{\begin{picture}(300,20)(0,0)
\SetScale{0.08}
\SetWidth{2.5}
\Photon(100,100)(200,100){6}{3}    
\Photon(400,100)(400,0){6}{3}      
%
\GOval(300,100)(100,100)(0){1.0}
\GOval(500,100)(100,100)(0){1.0}
\end{picture}}
$ \hspace*{0.5cm} 
\\ \hline 
$2$ & $\parbox{10mm}{
\begin{picture}(300,20)(0,0)
\SetScale{0.08}
\SetWidth{2.5}
%
\GOval(300,100)(100,100)(0){1.0}
\GOval(500,100)(100,100)(0){1.0}
\end{picture}
}$ & $\parbox{10mm}{
\begin{picture}(300,20)(0,0)
\SetScale{0.08}
\SetWidth{2.5}
%
\GOval(300,100)(100,100)(0){1.0}
\GOval(500,100)(100,100)(0){1.0}
\end{picture}
}$\\ 
\hline
\end{tabular}
\end{center}
\end{table}

By giving a close look at them, one soon realizes that three of them 
are indeed product of the one-loop MI's we met in the previous section,
namely,
\begin{eqnarray}
\parbox{10mm}{
\begin{picture}(300,20)(0,0)
\SetScale{0.08}
\SetWidth{2.5}
%
\GOval(300,100)(100,100)(0){1.0}
\GOval(500,100)(100,100)(0){1.0}
\end{picture}
} \hspace*{0.8cm}
&=& T^2(D,m^2) \ , \\
\parbox{12mm}{
\begin{picture}(300,20)(0,0)
\SetScale{0.08}
\SetWidth{2.5}
\Photon(100,100)(200,100){6}{3}    
\Photon(400,100)(400,0){6}{3}      
%
\GOval(300,100)(100,100)(0){1.0}
\GOval(500,100)(100,100)(0){1.0}
\end{picture}
} \hspace*{0.8cm}
&=& J(D, k^2) \times T(D,m^2) \ , \\
\parbox{12mm}{\begin{picture}(300,20)(0,0)
\SetScale{0.08}
\SetWidth{2.5}
\Photon(0,100)(100,100){6}{3}    
\Photon(500,100)(600,100){6}{3}      
%
\GOval(200,100)(100,100)(0){1.0}
\GOval(400,100)(100,100)(0){1.0}
\end{picture}} \hspace*{0.8cm}
&=& 
J^2(D,k^2) \ ,
\end{eqnarray}
where $T(D,m^2)$ and $J(D,k^2)$ were given respectively in
Eqs. (\ref{VP1L:eq11}, \ref{VP1L:eq34}).

The two yet unknown MI's, both belonging to the same topology,
are,
\begin{equation}\label{VP2L:eq2} 
J_{1}(k^2)=\hspace{-.3cm}
\parbox{15mm}{\begin{picture}(300,20)(0,0)
\SetScale{0.1}
\SetWidth{2.5}
\Photon(100,100)(200,100){6}{2}    
\Photon(400,100)(500,100){6}{2}
\GOval(300,100)(100,100)(0){1.0}
\Photon(200,100)(400,100){6}{4}
\end{picture}}\hspace{.2cm}=\int\frac{d^D p_{1}}{(2\pi)^{D-2}}\int\frac{d^D p_{2}}{(2\pi)^{D-2}}\frac{1}{{\cal{D}}_{1}{\cal{D}}_{2}{\cal{D}}_{3}} 
\end{equation} 

\begin{equation}\label{VP2L:eq3} 
J_{2}(k^2)=\hspace{-.3cm}\parbox{15mm}{\begin{picture}(300,20)(0,0)
\SetScale{0.1}
\SetWidth{2.5}
\Photon(100,100)(200,100){6}{2}    
\Photon(400,100)(500,100){6}{2}
\GOval(300,100)(100,100)(0){1.0}
\Photon(200,100)(400,100){6}{4}
\Text(53,15)[]{{\tiny $(p_{1}\cdot p_{2})$}}
\end{picture}}\hspace{1cm}=\int\frac{d^D p_{1}}{(2\pi)^{D-2}}\int\frac{d^D p_{2}}{(2\pi)^{D-2}}\frac{(p_{1}\cdot p_{2})}{{\cal{D}}_{1}{\cal{D}}_{2}{\cal{D}}_{3}},
\end{equation}
\\
where ${\cal{D}}_{1}=p_{1}^2+m^2, {\cal{D}}_{2}=p_{2}^2, {\cal{D}}_{3}=(k-p_{1}-p_{2})^2+m^2$, $p_{1}$ and $p_{2}$ being the two loop momenta and $k$ being the external momentum. 
Let's discuss their evaluation.

\end{subsection}

\begin{subsection}{Differential equations for $J_{1}(k^2)$ and $J_{2}(k^2)$.}
The functions $J_{1}(k^2)$ and $J_{2}(k^2)$
are found to fulfill
a system of first-order differential 
equations in the external momentum squared that reads as 
 \begin{equation}\label{VP2L:eq4} 
\left\{\begin{array}{l}\frac{dJ_{1}}{dk^2}=\left[\frac{(D-3)}{2k^2}+\frac{(3D-7)}{2(k^2+4m^2)}\right]J_{1}-\frac{3(D-2)}{2m^2}\left[\frac{1}{k^2}-\frac{1}{k^2+4m^2}\right]J_{2}-\frac{(D-2)}{2m^2}\left[\frac{1}{k^2}-\frac{1}{k^2+4m^2}\right]T^2 
\\ 
\\ 
\frac{dJ_{2}}{dk^2}=\frac{(D-2)}{4}\,\,J_{1}-\frac{(D-2)}{2k^2}J_{2}-\frac{(D-2)}{4k^2}\,\,T^2 \end{array}\right. 
\end{equation}
where $T(D,m^2)$ is the massive tadpole.\\
Such a system is equivalent to a second-order differential 
for $J_{1}$   
\begin{eqnarray}\label{VP2L:eq5} 
 & \eta^2 & (\eta+4m^2)\frac{d^2}{d\eta^2}\hspace{-.3cm}\parbox{15mm}{\begin{picture}(300,20)(0,0)
\SetScale{0.1}
\SetWidth{2.5}
\Photon(100,100)(200,100){6}{2}    
\Photon(400,100)(500,100){6}{2}

\GOval(300,100)(100,100)(0){1.0}
\Photon(200,100)(400,100){6}{4}
\end{picture}}\hspace{.2cm}-\eta\left[\frac{3}{2}\left(D-4\right)\eta-6m^2\right]\frac{d}{d\eta}\hspace{-.3cm}\parbox{15mm}{\begin{picture}(300,20)(0,0)
\SetScale{0.1}
\SetWidth{2.5}
\Photon(100,100)(200,100){6}{2}    
\Photon(400,100)(500,100){6}{2}

\GOval(300,100)(100,100)(0){1.0}
\Photon(200,100)(400,100){6}{4}
\end{picture}}\,\,\,\,+\nonumber\\ 
& + &(D-3)\left[\eta\frac{(D-4)}{2}-(D-2)m^2\right]\hspace{-.3cm}\parbox{15mm}{\begin{picture}(300,20)(0,0)
\SetScale{0.1}
\SetWidth{2.5}
\Photon(100,100)(200,100){6}{2}    
\Photon(400,100)(500,100){6}{2}

\GOval(300,100)(100,100)(0){1.0}
\Photon(200,100)(400,100){6}{4}
\end{picture}}\hspace{.2cm}
-\frac{(D-2)^2}{2}\hspace{-.3cm}\parbox{15mm}{\begin{picture}(300,20)(0,0)
\SetScale{0.1}
\SetWidth{2.5}
%
%
%
\GOval(200,100)(100,100)(0){1.0}
\GOval(400,100)(100,100)(0){1.0}
\end{picture}}\hspace{.3cm}=0, 
\end{eqnarray}

where $\eta=k^2$. 

Differentiating once more with respect to $\eta$,
we obtain a third order differential equation for $J_{1}$

\begin{eqnarray}\label{VP2L:eq7} 
& \, &\eta^2(\eta+4m^2)\frac{d^3J_{1}}{d\eta^3}-\eta\left[\frac{3}{2}\left(D-6\right)\eta-14m^2\right]\frac{d^2J_{1}}{d\eta^2}
+\nonumber\\
& + &\left[\frac{(D-4)(D-9)}{2}\eta-D(D-5)m^2\right]\frac{dJ_{1}}{d\eta} + \frac{(3-D)(4-D)}{2}J_{1}=0, 
\end{eqnarray} 

which, introducing the dimensionless variable $x=\frac{\eta}{4m^2}$, becomes

\begin{eqnarray}\label{VP2L:eq8} 
\lefteqn{x^2(1+x)\frac{d^3J_{1}}{dx^3}-x\left[\frac{3(D-6)}{2}x-\frac{7}{2}\right]\frac{d^2J_{1}}{dx^2}+}\nonumber\\ & & 
+\left[\frac{(D-4)(D-9)}{2}x-\frac{D(D-5)}{4}\right]\frac{dJ_{1}}{dx}+\frac{(3-D)(4-D)}{2}J_{1}=0.\nonumber\\ 
\end{eqnarray} 

\end{subsection}

\begin{subsection}{Exact solution in $D$ dimensions.}

Quite in general, a differential equation like

\begin{eqnarray}\label{VP2L:eq9} 
\lefteqn{x^2(1+x)\frac{d^3\psi}{dx^3}+x[x(a_{1}+a_{2}+a_{3}+3)+(b_{1}+b_{2}+1)]\frac{d^2\psi}{dx^2}+}\nonumber \\ & & 
+[x(a_{1}a_{2}+a_{1}a_{3}+a_{2}a_{3}+a_{1}+a_{2}+a_{3}+1)+b_{1}b_{2}]\frac{d\psi}{dx}+\nonumber\\ & & 
+(a_{1}a_{2}a_{3})\psi=0, 
\end{eqnarray}

where $a_{1},a_{2},a_{3},b_{1},b_{2}$ are parameters, is classified as 
hypergeometric and its solution space is spanned by the functions

\begin{equation}\label{VP2L:eq10} 
\begin{array}{l}\psi_{1}(x)=\,_{3}F_{2}\left[\begin{array}{c}a_{1},a_{2},a_{3} \\ b_{1},b_{2} \end{array}; -x \right]\\ 
\\ 
\psi_{2}(x)=(-x)^{1-b_{1}}\,_{3}F_{2}\left[\begin{array}{c}1+a_{1}-b_{1},1+a_{2}-b_{1},1+a_{3}-b_{1} \\ 2-b_{1},1+b_{2}-b_{1} \end{array}; -x \right]\\ 
\\ 
\psi_{3}(x)=(-x)^{1-b_{2}}\,_{3}F_{2}\left[\begin{array}{c}1+a_{1}-b_{2},1+a_{2}-b_{2},1+a_{3}-b_{2} \\ 2-b_{2},1+b_{1}-b_{2} \end{array}; -x \right].\end{array} 
\end{equation} 

We can easily see that Eq. (\ref{VP2L:eq8}) is hypergeometric if we identify the parameters as follows: 



\begin{equation}\label{VP2L:eq12} 
\left\{\begin{array}{l}a_{1}=1 \\ 
a_{2}=3-D\\ 
a_{3}=(4-D)/2\\ 
b_{1}=D/2\\ 
b_{2}=(5-D)/2\,.\end{array}\right. 
\end{equation} 

With the above choice, the three independent solutions of Eq. (\ref{VP2L:eq8}) become

\begin{equation}\label{VP2L:eq13} 
\begin{array}{l} 
\psi_{1}(x)=\,_{3}F_{2}\left[\begin{array}{c}(3-D),(4-D)/2,1\\ 
D/2,(5-D)/2 
\end{array}; -x\right]\\ 
\psi_{2}(x)=(-x)^{-\frac{(D-2)}{2}}\,_{3}F_{2}\left[\begin{array}{c}(8-3D)/2,(3-D),(4-D)/2\\ 
(4-D)/2,(7-2D)/2 
\end{array}; -x\right]\\ 
\psi_{3}(x)=(-x)^{\frac{(D-3)}{2}}\,_{3}F_{2}\left[\begin{array}{c}(3-D),1/2,(D-1)/2\\ 
(2D-3)/2,(D-1)/2 
\end{array}; -x\right] 
\end{array} 
\end{equation}

Finally, the general solution of Eq. (\ref{VP2L:eq8}) can be written as a linear combination of  $\psi_{1}(x),\psi_{2}(x)$ e $\psi_{3}(x)$ 
 
\begin{equation}\label{VP2L:eq14} 
J_{1}(x)=A\psi_{1}(x)+B\psi_{2}(x)+C\psi_{3}(x). 
\end{equation}

Now we can choose the value of the three integration constants $A$, $B$ and $C$ by means of suitable boundary conditions. By inspection we see that 
for $x\to 0$ the MI $J_{1}$ becomes 
\begin{eqnarray}\label{VP2L:eq15} 
J_{1}(0)& = &\hspace{-.3cm}\parbox{15mm}{\begin{picture}(300,20)(0,0)
\SetScale{0.1}
\SetWidth{2.5}
\GOval(300,100)(100,100)(0){1.0}
\Photon(200,100)(400,100){6}{4}
\end{picture}}\,\,=\nonumber\\ 
& = &\int\frac{d^D p_{1}}{(2\pi)^{D-2}}\int\frac{d^D p_{2}}{(2\pi)^{D-2}}\frac{1}{(p_{1}^2+m^2)p_{2}^2[(p_{1}-p_{2})^2+m^2)}=\nonumber\\  
& = &-\frac{(D-2)}{2m^2(D-3)}\parbox{15mm}{\begin{picture}(300,20)(0,0)
\SetScale{0.1}
\SetWidth{2.5}
%
%
%
\GOval(200,100)(100,100)(0){1.0}
\GOval(400,100)(100,100)(0){1.0}
\end{picture}}\hspace{.2cm} \ ,
\end{eqnarray} 
where the last line has been obtained by means of IBP-id's.
If we study the behaviour of (\ref{VP2L:eq15}) in the UV and IR limits, we see that it is regular for $2<D<4$. This means that $B=C=0$. If this was not the case, in fact, terms such as $(-x)^{-\frac{(D-2)}{2}}$ and $(-x)^{\frac{(D-3)}{2}}$ would give rise, in the abovementioned range of $D$, to a divergent behaviour, and this would not be compatible with the finite result of (\ref{VP2L:eq15}).  

Furthermore, (\ref{VP2L:eq15}) allows one to choose once and for all the constant $A$   

\begin{equation}\label{VP2L:eq16} 
J_{1}(0)=A\psi_{1}(0)=A, 
\end{equation} 

because, like all Hypergeometric functions, $\psi_{1}(0)=1$. Comparing (\ref{VP2L:eq16}) with (\ref{VP2L:eq15}), we have 

\begin{displaymath} 
A=\hspace{-.5cm}\parbox{15mm}{\begin{picture}(300,20)(0,0)
\SetScale{0.1}
\SetWidth{2.5}
%
%
\GOval(300,100)(100,100)(0){1.0}
\Photon(200,100)(400,100){6}{4}
\end{picture}} \ ,
\end{displaymath} 
hence
\begin{equation}\label{VP2L:eq17} 
J_{1}(x)=-\frac{(D-2)}{2m^2(D-3)}\hspace{-.3cm}\parbox{15mm}{\begin{picture}(300,20)(0,0)
\SetScale{0.1}
\SetWidth{2.5}
%
%
%
\GOval(200,100)(100,100)(0){1.0}
\GOval(400,100)(100,100)(0){1.0}
\end{picture}}\hspace{.5cm}_{3}F_{2}\left[\begin{array}{c}(3-D),(4-D)/2,1\\ 
D/2,(5-D)/2 
\end{array}; -x\right]. 
\end{equation} 

Now we can obtain the expression for $J_{2}(x)$ by simply substituting $J_{1}(x)$ and $\frac{dJ_{1}}{dx}$ in the first equation of (\ref{VP2L:eq4}). To do so, we have to know the expression for the first order derivative of the Hypergeometric function  $_{3}F_{2}(x)$.\\
Quite in general, a function like     

\begin{displaymath} 
_{3}F_{2}\left[\begin{array}{c} a_{1},a_{2},a_{3} \\ b_{1},b_{2}\end{array}; -x\right] 
\end{displaymath} 

has the following series representation 

\begin{equation}\label{VP2L:eq18} 
_{3}F_{2}\left[\begin{array}{c} a_{1},a_{2},a_{3} \\ b_{1},b_{2}\end{array}; x\right]=\sum_{n=0}^{+\infty}\frac{(a_{1})_{n}(a_{2})_{n}(a_{3})_{n}}{(b_{1})_{n}(b_{2})_{n}}\frac{(-x)^n}{n!}, 
\end{equation}
 
where $(\xi)_{n}$ is the \emph{Pochhammer symbol}, defined as 
\begin{displaymath} 
(\xi)_{n}=\frac{\Gamma(\xi+n)}{\Gamma(\xi)}. 
\end{displaymath} 

By differentiating Eq. (\ref{VP2L:eq18}) with respect to $x$, we obtain 

\begin{eqnarray}\label{VP2L:eq19} 
\lefteqn{\frac{d}{dx}\,_{3}F_{2}\left[\begin{array}{c} a_{1},a_{2},a_{3} \\ b_{1},b_{2}\end{array}; x\right]=}\nonumber\\ & & 
=\frac{d}{dx}\sum_{n=0}^{+\infty}\frac{(a_{1})_{n}(a_{2})_{n}(a_{3})_{n}}{(b_{1})_{n}(b_{2})_{n}}\frac{(-x)^n}{n!}=-\sum_{n=0}^{+\infty}\frac{(a_{1})_{n}(a_{2})_{n}(a_{3})_{n}}{(b_{1})_{n}(b_{2})_{n}}n\frac{(-x)^{n-1}}{n!}=\nonumber\\ & & 
=-\sum_{n=1}^{+\infty}\frac{(a_{1})_{n}(a_{2})_{n}(a_{3})_{n}}{(b_{1})_{n}(b_{2})_{n}}\frac{(-x)^{n-1}}{(n-1)!}=\nonumber\\ & & 
=-\sum_{k=0}^{+\infty}\frac{(a_{1})_{k+1}(a_{2})_{k+1}(a_{3})_{k+1}}{(b_{1})_{k+1}(b_{2})_{k+1}}\frac{(-x)^{k}}{k!}. 
\end{eqnarray} 

Now, we see that 

\begin{displaymath} 
(\xi)_{k+1}=\frac{\Gamma(\xi+k+1)}{\Gamma(\xi)}=x\frac{\Gamma(\xi+1+k)}{
\Gamma(\xi+1)}=\xi(\xi+1)_{k}, 
\end{displaymath} 

hence

\begin{eqnarray}\label{VP2L:eq20} 
\lefteqn{\frac{d}{dx}\,_{3}F_{2}\left[\begin{array}{c} a_{1},a_{2},a_{3} \\ b_{1},b_{2}\end{array}; -x\right]=}\nonumber\\ & & 
=-\frac{(a_{1})(a_{2})(a_{3})}{(b_{1})(b_{2})}\sum_{k=0}^{+\infty}\frac{(a_{1}+1)_{k}(a_{2}+1)_{k}(a_{3}+1)_{k}}{(b_{1}+1)_{k}(b_{2}+1)_{k}}\frac{(-x)^{k}}{k!}=\nonumber\\ & & 
=-\frac{(a_{1})(a_{2})(a_{3})}{(b_{1})(b_{2})}\,_{3}F_{2}\left[\begin{array}{c} a_{1}+1,a_{2}+1,a_{3}+1 \\ b_{1}+1,b_{2}+1\end{array}; -x\right] 
\end{eqnarray}
 
Solving (\ref{VP2L:eq4}) with respect to $J_{2}$ we have 

\begin{eqnarray}\label{VP2L:eq21} 
J_{2}(x) & = & \frac{a}{3(D-2)}[2(2D-5)x+(D-3)]J_{1}(x)+\nonumber\\  
& - &\frac{2m^2}{3(D-2)}x(x+1)\frac{dJ_{1}(x)}{dx}-\frac{1}{3}T^2(D,m^2). 
\end{eqnarray}
 
Substituting (\ref{VP2L:eq17}) in (\ref{VP2L:eq21}) and because of 

\begin{equation}\label{VP2L:eq22} 
\frac{d}{dx}\,_{3}F_{2}\left[\begin{array}{c}(3-D),(4-D)/2,1\\ 
D/2,(5-D)/2 
\end{array}; -x\right]
=\frac{2(D-3)(D-4)}{D(D-5)}\,_{3}F_{2}\left[\begin{array}{c}(4-D),(6-D)/2,2\\ 
(D+2)/2,(7-D)/2 
\end{array}; -x\right], 
\end{equation} 

we can write down finally the exact expression in $D$ for 
$J_{2}(x)$  

\begin{eqnarray}\label{VP2L:eq23} 
J_{2}(z) & = &
\Bigg\{
- \frac{\Big[2(2D-5)x+(D-3)\Big]}{6(D-3)}
\,_{3}F_{2}\left[\begin{array}{c}(3-D),(4-D)/2,1\\ 
D/2,(5-D)/2 
\end{array}; -x\right]+\nonumber\\  
& & \quad +\frac{2 (D-4)}{3 D (D-5)}x(x+1)
     _{3}F_{2}\left[\begin{array}{c}(4-D),(6-D)/2,2\\ 
(D+2)/2,(7-D)/2 
\end{array}; -x\right]+\nonumber\\
& & \quad - \frac{1}{3}
\Bigg\}
\parbox{15mm}{\begin{picture}(300,20)(0,0)
\SetScale{0.1}
\SetWidth{2.5}
%
%
%
\GOval(200,100)(100,100)(0){1.0}
\GOval(400,100)(100,100)(0){1.0}
\end{picture}}
\end{eqnarray}

\subsection{Renormalized vacuum polarization function}

By means of the integrals' reduction, 
the two-loop $\Pi_{2R}$, with the one-loop subdivergences already subtracted
off according to Eq.(\ref{VP2L:eq1}), admits the following
decomposition in terms of MI's,

\begin{eqnarray}\label{VP2L:eq24}
\Pi_{2L,1R}(D,k^2) &=&          
c_1 \hspace{-.3cm}
\parbox{15mm}{\begin{picture}(300,20)(0,0)
\SetScale{0.1}
\SetWidth{2.5}
\Photon(100,100)(200,100){6}{2}    
\Photon(400,100)(500,100){6}{2}
\GOval(300,100)(100,100)(0){1.0}
\Photon(200,100)(400,100){6}{4}
\end{picture}}\hspace{0.5cm} 
+ c_2 \hspace{-.3cm}
\parbox{15mm}{\begin{picture}(300,20)(0,0)
\SetScale{0.1}
\SetWidth{2.5}
\Photon(100,100)(200,100){6}{2}    
\Photon(400,100)(500,100){6}{2}
\GOval(300,100)(100,100)(0){1.0}
\Photon(200,100)(400,100){6}{4}
\Text(53,15)[]{{\tiny $(p_{1}\cdot p_{2})$}}
\end{picture}}\hspace{1cm} 
+ c_3 
\parbox{12mm}{
\begin{picture}(300,20)(0,0)
\SetScale{0.08}
\SetWidth{2.5}
\Photon(100,100)(200,100){6}{3}    
\Photon(400,100)(400,0){6}{3}      
%
\GOval(300,100)(100,100)(0){1.0}
\GOval(500,100)(100,100)(0){1.0}
\end{picture}
} \hspace*{0.8cm}
\nonumber \\
&&                         
+ c_4 \ \parbox{12mm}{\begin{picture}(300,20)(0,0)
\SetScale{0.08}
\SetWidth{2.5}
\Photon(0,100)(100,100){6}{3}    
\Photon(500,100)(600,100){6}{3}      
%
\GOval(200,100)(100,100)(0){1.0}
\GOval(400,100)(100,100)(0){1.0}
\end{picture}} \hspace*{0.8cm}
+ c_5 \hspace{-.3cm}
\parbox{10mm}{
\begin{picture}(300,20)(0,0)
\SetScale{0.08}
\SetWidth{2.5}
%
\GOval(300,100)(100,100)(0){1.0}
\GOval(500,100)(100,100)(0){1.0}
\end{picture}
} \hspace*{1cm} \ ,
\end{eqnarray}
with
\begin{eqnarray}\label{VP2L:eq25-29}
\hspace{-1cm}c_1 &=& 
\frac{-2\left(-2+D\right)}{\left( -12 + 19D -
  8\,D^2 + D^3 \right) \,k^2\,m^2\,\left( k^2 + 4\,m^2 \right) }
\bigg(\left(-16+18D-7D^2+D^3\right)k^4
      \nonumber \\ && \hspace*{2cm}
     +4\left(-14+22D - 9D^2+D^3\right)k^2 m^2 + 16\left( -3 + D \right)^2m^4 \bigg) \ ;
\\
c_2 &=&
\frac{12\,{\left( -2 + D \right) }^2\,\left( \left( 8 - 5\,D + D^2 \right) \,k^2 + 4\,\left( 10 - 7\, D+D^2 \right) \,m^2 \right) }
  {\left( -12 + 19\,D - 8\,D^2 + D^3 \right) \,k^2\,m^2\,\left( k^2 + 4\,m^2
    \right) } \ ;
\\
c_3 &=&
\frac{-2\,\left( -2 + D \right) }{\left( -12 + 19\,D - 8\,D^2 +
  D^3 \right) \,k^2\,m^2\,\left( k^2 + 4\,m^2 \right) }
\bigg( \left( -16 + 18\,D - 7\,D^2 + D^3 \right) \,k^4 
\nonumber \\ && \hspace*{2cm}
+ 4\,{\left( -4 + D \right) }^2\,\left( -1 + D \right) \,k^2\,m^2 - 
      32\,\left( -3 + D \right) \,m^4 \bigg) \ ;
\\
c_4 &=&
\frac{1}
  {\left( 4 - 5\, D+ D^2 \right) \,k^2\,\left( k^2 + 4\,m^2 \right) }
\bigg(-2\,\left( -32 + 30\,D - 9\,D^2 + D^3 \right) \,k^4 
      \nonumber \\ && \hspace*{2cm}
   - 8\,\left( -40 + 38\,D - 11\,D^2 + D^3 \right) \,k^2\,m^2 
   + 64\,m^4 \bigg) \ ;
\\
c_5 &=&
\frac{6\,{\left( -2 + D \right) }^2\,\left( \left( 8 - 5\,D + D^2 \right) \,k^2 + 4\,\left( 10 - 7\,D + D^2 \right) \,m^2 \right) }
  {\left( -12 + 19\,D - 8\,D^2 + D^3 \right) \,k^2\,m^2\,\left( k^2 + 4\,m^2
    \right) }
\end{eqnarray}
and where we used the mass counterterm defined, in terms of the 
one-loop tadpole, as, 

\begin{eqnarray}
\delta m = { (D-1)(D-2) \over 2 \ m \ (D-3) }\hspace{-.5cm}
\parbox{20mm}{\begin{picture}(300,20)(0,0)
\SetScale{0.1}
\SetWidth{2.5}
\GOval(300,200)(100,100)(0){1.0}
\Photon(200,100)(400,100){6}{3}
\end{picture}}
\end{eqnarray}

The completion of the two-loop renormalization
procedure requires the subtraction of the value of $\Pi_{(2L,1R)}$
at zero momentum, which can be obtained from the above expression,
\begin{equation}\label{VP2L:eq30}
\Pi_{2L,1R}(D,0) = \frac{34 - D\,\left( 41 + \left( -12 + D \right) \,D \right) }
  {2\,\left( -5 + D \right) \,\left( -4 + D \right) \,\left( -3 + D \right) \,D} 
\end{equation}
The two-loop renormalized expression
\begin{equation}\label{VP2L:eq31}
\Pi_{2R}(D,k^2) = \Pi_{2L,1R}(D,k^2) - \Pi_{2L,1R}(D,0) 
\end{equation}
agrees with the result in literature \cite{BroFleTar}.


\end{subsection}
\end{section}


\section{System of Three Differential Equations}

After having shown in its completeness the calculation of the vacuum polarization at 1-
and 2-loop in QED, we proceed with the discussion of a less trivial case of differential
equations whose solutions \cite{Bonciani:2003hc} are required to compute a set of integrals 
which enter the 2-loop QCD corrections to the heavy-quark form factors \cite{Bernreuther:2004ih,Bernreuther:2004th,Bernreuther:2005rw,Bernreuther:2005gq,Bernreuther:2005gw} 
We are going to describe directly the evaluation of three master integrals 
belonging to the same topology.
In this case we give as understood
{\it i)} the reduction to the master integrals,
namely assuming that all the MI's have been already identified;
and {\it ii)} the knowledge of the MI belonging to the subtopologies 
which enters the non-homogeneous term of our differential equations.

We will see, in this example, the technical details related to the Laurent 
expansion of the system of equations around specific values of the dimensional parameter
($D\to4$), and to the choice of the boundary conditions.

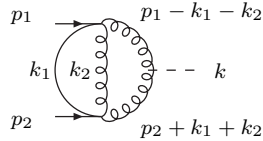
\begin{figure}[h]
$$
\begin{picture}(0,0)(0,0)
\SetScale{0.7}
\SetWidth{0.7}
\ArrowLine(-25,25)(0,25)
\CArc(0,0)(25,+90,-90)
\GlueArc(0,0)(25,-90,+90){3}{10}
\Gluon(0,25)(0,-25){3}{5}
\ArrowLine(-25,-25)(0,-25)
\DashLine(25,0)(50,0){5}
\Text( 45,0)[]{{\footnotesize $k$}}
\Text(-30, 20)[]{{\footnotesize $p_1$}}
\Text(-30,-20)[]{{\footnotesize $p_2$}}
\Text(-23, 0)[]{{\footnotesize $k_1$}}
\Text(-8, 0)[]{{\footnotesize $k_2$}}
\Text(38, 22)[]{{\footnotesize $p_1 - k_1 - k_2$}}
\Text(38,-22)[]{{\footnotesize $p_2 + k_1 + k_2$}}
\end{picture}
$$
\vspace*{0.5cm}
\caption{Two-loop vertex diagram: $p_1^2 = p_2^2 = -m^2; \ k=p_1+p_2 $; 
$k_{1,2}$ are loop variables; a curly line stands for massless 
propagator; a solid line, for propagator of mass $m$.}
\label{fig:nontrivialvertex}
\end{figure}

According to our assumptions, we have got the table of identities for reducing all 
the integrals belonging to the topology in Fig.\ref{fig:nontrivialvertex} and 
its subtopologies. 

The scalar integrals represented by the topology 
in Fig.\ref{fig:nontrivialvertex} are
\begin{equation}
{\cal{J}}(n_1,n_2,n_3;m_1,m_2,m_3,m_4)=
\int {d^Dk_1 \over (2 \pi)^{D-2}} {d^D k_2 \over (2 \pi)^{D-2}}
\frac{(k_1\cdot p_1)^{n_1}(k_2\cdot p_1)^{n_2}(k_2\cdot p_1)^{n_3}}{{\cal{D}}_{1}^{m_1}{\cal{D}}_{2}^{m_2}{\cal{D}}_{3}^{m_3}{\cal{D}}_{4}^{m_4}},
\end{equation}
where ${\cal{D}}_1=k_1^2+m^2,{\cal{D}}_2=k_2^2, {\cal{D}}_3=(p_1-k_1-k_2)^2, {\cal{D}}_4=(p_2+k_1+k_2)^2$.\\
At the end of the reduction, one is left over with three master integrals,
$\Phi_i (i=1,2,3)$, 
which we choose to be 

\begin{eqnarray}
\Phi_1(D,k^2) = \hspace*{0.5cm}
\parbox{2cm}{\begin{picture}(0,0)(0,0)
\SetScale{0.4}
\SetWidth{0.5}
\Line(-25,25)(0,25)
\CArc(0,0)(25,+90,-90)
\GlueArc(0,0)(25,-90,+90){3}{10}
\Gluon(0,25)(0,-25){3}{5}
\Line(-25,-25)(0,-25)
\DashLine(25,0)(50,0){5}
\end{picture}} \hspace*{-1.3cm} \ ;
%
\hspace*{0.2cm}
\Phi_2(D,k^2) = \hspace*{0.5cm}
\parbox{1.2cm}{\begin{picture}(0,0)(0,0)
\SetScale{0.4}
\SetWidth{0.5}
\Line(-25,25)(0,25)
\CArc(0,0)(25,+90,-90)
\GlueArc(0,0)(25,-90,+90){3}{10}
\Gluon(0,25)(0,-25){3}{5}
\Line(-25,-25)(0,-25)
\DashLine(25,0)(50,0){5}
\Text(30,10)[]{{\footnotesize $(k_1 \cdot p_1)$}}
\end{picture}}  \ ;
%
\hspace*{0.5cm}
\Phi_3(D,k^2) = \hspace*{0.5cm}
\parbox{2cm}{\begin{picture}(0,0)(0,0)
\SetScale{0.4}
\SetWidth{0.5}
\Line(-25,25)(0,25)
\CArc(0,0)(25,+90,-90)
\GlueArc(0,0)(25,-90,+90){3}{10}
\Gluon(0,25)(0,-25){3}{10}
\Line(-25,-25)(0,-25)
\DashLine(25,0)(50,0){5}
\Vertex(0,0){5}
\end{picture}} \hspace*{-1.0cm} \! .
\end{eqnarray}

These MI are found to fulfills the following system of first-order ODE, in the variable $k^2$, 
corresponding to the momentum transfer,


\begin{eqnarray}
\hspace*{-1.0cm} {d \over d k^2} \hspace*{0.5cm}
\parbox{2cm}{\begin{picture}(0,0)(0,0)
\SetScale{0.4}
\SetWidth{0.5}
\Line(-25,25)(0,25)
\CArc(0,0)(25,+90,-90)
\GlueArc(0,0)(25,-90,+90){3}{10}
\Gluon(0,25)(0,-25){3}{10}
\Line(-25,-25)(0,-25)
\DashLine(25,0)(50,0){5}
\end{picture}}
\hspace*{-1.2cm} &=& 
- 
  \frac{\left( 22 - 13\,D + 2\,D^2 \right) }{2\,k^2} \hspace*{0.5cm} 
\parbox{2cm}{\begin{picture}(0,0)(0,0)
\SetScale{0.4}
\SetWidth{0.5}
\Line(-25,25)(0,25)
\CArc(0,0)(25,+90,-90)
\GlueArc(0,0)(25,-90,+90){3}{10}
\Gluon(0,25)(0,-25){3}{10}
\Line(-25,-25)(0,-25)
\DashLine(25,0)(50,0){5}
\end{picture}}
\nonumber \\
&&
- 
  \frac{\left( -2 + D \right) \,\left( -7 + 2\,D \right) \,\left( -8 + 3\,D \right) }
   {\left( -3 + D \right) \,k^2\,\left( k^2 + 4\,m^2 \right) } \hspace*{0.5cm}
\parbox{1.2cm}{\begin{picture}(0,0)(0,0)
\SetScale{0.4}
\SetWidth{0.5}
\Line(-25,25)(0,25)
\CArc(0,0)(25,+90,-90)
\GlueArc(0,0)(25,-90,+90){3}{10}
\Gluon(0,25)(0,-25){3}{10}
\Line(-25,-25)(0,-25)
\DashLine(25,0)(50,0){5}
\Text(30,10)[]{{\footnotesize $(k_1 \cdot p_1)$}}
\end{picture}} 
\nonumber \\
&&
+ 
  \frac{\left( -4 + D \right) \,\left( -5 + 2\,D \right) \,m^2}
   {2\,\left( -3 + D \right) \,\left( k^2 + 4\,m^2 \right) }\ \hspace*{0.5cm}
\parbox{2cm}{\begin{picture}(0,0)(0,0)
\SetScale{0.4}
\SetWidth{0.5}
\Line(-25,25)(0,25)
\CArc(0,0)(25,+90,-90)
\GlueArc(0,0)(25,-90,+90){3}{10}
\Gluon(0,25)(0,-25){3}{10}
\Line(-25,-25)(0,-25)
\DashLine(25,0)(50,0){5}
\Vertex(0,0){5}
\end{picture}}
\nonumber \\
&&
+ 
 \frac{\,\left( -4 + D \right) \,\left( -2 + D \right) \,\left( -5 + 2\,D \right) }
   {2\,\left( -3 + D \right) \,k^2\,\left( k^2 + 4\,m^2 \right) } \hspace*{1.0cm}
\parbox{2cm}{\begin{picture}(0,0)(0,0)
\SetScale{0.4}
\SetWidth{0.5}
\DashLine(-50,0)(-25,0){5}
\GlueArc(0,0)(25,0,180){3}{10}
\CArc(0,0)(25,180,360)
\DashLine(25,0)(25,-30){5}
\CArc(50,0)(25,0,360)
\end{picture}}
\nonumber \\
&&
- 
  \frac{\left( -4 + D \right) \,\left( -5 + 2\,D \right) \,\left( -10 + 3\,D \right) \,\left( -8 + 3\,D \right) \,
    }{4\,\left( -3 + D \right) \,\left( -7 + 2\,D \right)
   \,k^2\,\left( k^2 + 4\,m^2 \right) } \hspace*{1.0cm}
\parbox{2cm}{\begin{picture}(0,0)(0,0)
\SetScale{0.4}
\SetWidth{0.5}
\Line(-50,0)(50,0)
\GlueArc(0,0)(25,0,360){3}{20}
\end{picture}}
\hspace*{-1.0cm} \ ; 
\label{vertici:dePhi1inD} 
\end{eqnarray}

\begin{eqnarray}
\hspace*{-1.0cm} {d \over d k^2} \hspace*{0.5cm}
\parbox{1.2cm}{\begin{picture}(0,0)(0,0)
\SetScale{0.4}
\SetWidth{0.5}
\Line(-25,25)(0,25)
\CArc(0,0)(25,+90,-90)
\GlueArc(0,0)(25,-90,+90){3}{10}
\Gluon(0,25)(0,-25){3}{10}
\Line(-25,-25)(0,-25)
\DashLine(25,0)(50,0){5}
\Text(30,10)[]{{\footnotesize $(k_1 \cdot p_1)$}}
\end{picture}} 
&=& 
  \frac{\left( -3 + D \right) \,\left( (-4 + D)\,k^2 + 4 (- 3 + D)\,m^2
    \right) }{8\,k^2} \hspace*{0.5cm} 
\parbox{2cm}{\begin{picture}(0,0)(0,0)
\SetScale{0.4}
\SetWidth{0.5}
\Line(-25,25)(0,25)
\CArc(0,0)(25,+90,-90)
\GlueArc(0,0)(25,-90,+90){3}{10}
\Gluon(0,25)(0,-25){3}{10}
\Line(-25,-25)(0,-25)
\DashLine(25,0)(50,0){5}
\end{picture}}
\nonumber \\
&&
+ 
  \frac{\left( (20 - 16\,D + 3\,D^2)\,k^2 + (56 - 56\,D + 12\,D^2)\,m^2 \right) }
   {4\,k^2\,\left( k^2 + 4\,m^2 \right) } \hspace*{0.5cm}
\parbox{1.2cm}{\begin{picture}(0,0)(0,0)
\SetScale{0.4}
\SetWidth{0.5}
\Line(-25,25)(0,25)
\CArc(0,0)(25,+90,-90)
\GlueArc(0,0)(25,-90,+90){3}{10}
\Gluon(0,25)(0,-25){3}{10}
\Line(-25,-25)(0,-25)
\DashLine(25,0)(50,0){5}
\Text(30,10)[]{{\footnotesize $(k_1 \cdot p_1)$}}
\end{picture}} 
\nonumber \\
&&
- 
\frac{\left( -4 + D \right) \,m^2}{8} \hspace*{0.5cm}
\parbox{2cm}{\begin{picture}(0,0)(0,0)
\SetScale{0.4}
\SetWidth{0.5}
\Line(-25,25)(0,25)
\CArc(0,0)(25,+90,-90)
\GlueArc(0,0)(25,-90,+90){3}{10}
\Gluon(0,25)(0,-25){3}{10}
\Line(-25,-25)(0,-25)
\DashLine(25,0)(50,0){5}
\Vertex(0,0){5}
\end{picture}}
\nonumber \\
&&
-
\frac{\left( 10 - 6\,D + D^2 \right)  
}{8\,k^2} \hspace*{1.0cm}
\parbox{2cm}{\begin{picture}(0,0)(0,0)
\SetScale{0.4}
\SetWidth{0.5}
\DashLine(-50,0)(-25,0){5}
\GlueArc(0,0)(25,0,180){3}{10}
\CArc(0,0)(25,180,360)
\DashLine(25,0)(25,-30){5}
\CArc(50,0)(25,0,360)
\end{picture}}
\nonumber \\
&&
+ 
  \frac{\left( -236 + 244\,D - 82\,D^2 + 9\,D^3 \right)
    }{16\,\left( -7 + 2\,D \right) \,k^2} \hspace*{1.0cm}
\parbox{2cm}{\begin{picture}(0,0)(0,0)
\SetScale{0.4}
\SetWidth{0.5}
\Line(-50,0)(50,0)
\GlueArc(0,0)(25,0,360){3}{20}
\end{picture}}
\hspace*{-1.0cm} \ ; 
\label{vertici:dePhi2inD} 
\end{eqnarray}

\begin{eqnarray}
\hspace*{-1.0cm} {d \over d k^2} \hspace*{0.5cm}
\parbox{2cm}{\begin{picture}(0,0)(0,0)
\SetScale{0.4}
\SetWidth{0.5}
\Line(-25,25)(0,25)
\CArc(0,0)(25,+90,-90)
\GlueArc(0,0)(25,-90,+90){3}{10}
\Gluon(0,25)(0,-25){3}{10}
\Line(-25,-25)(0,-25)
\DashLine(25,0)(50,0){5}
\Vertex(0,0){5}
\end{picture}}
\hspace*{-1.2cm} &=& 
- 
\frac{\left( -4 + D \right) \,\left( -3 + D \right) \,
     \left( \left( -2 + D \right) \,k^2 + 4\,\left( -3 + D \right) \,m^2
     \right) }{4\,k^4\,m^2} \hspace*{0.5cm} 
\parbox{2cm}{\begin{picture}(0,0)(0,0)
\SetScale{0.4}
\SetWidth{0.5}
\Line(-25,25)(0,25)
\CArc(0,0)(25,+90,-90)
\GlueArc(0,0)(25,-90,+90){3}{10}
\Gluon(0,25)(0,-25){3}{10}
\Line(-25,-25)(0,-25)
\DashLine(25,0)(50,0){5}
\end{picture}}
\nonumber \\
&&
- 
  \frac{\left( -4 + D \right) \,\left( -2 + D \right) \,\left( -8 + 3\,D
    \right) }{2\,k^4\,m^2} \hspace*{0.5cm}
\parbox{1.2cm}{\begin{picture}(0,0)(0,0)
\SetScale{0.4}
\SetWidth{0.5}
\Line(-25,25)(0,25)
\CArc(0,0)(25,+90,-90)
\GlueArc(0,0)(25,-90,+90){3}{10}
\Gluon(0,25)(0,-25){3}{10}
\Line(-25,-25)(0,-25)
\DashLine(25,0)(50,0){5}
\Text(30,10)[]{{\footnotesize $(k_1 \cdot p_1)$}}
\end{picture}} 
\nonumber \\
&&
+ 
  \frac{\left( \left( 4 - 6\,D + D^2 \right) \,k^2 + 4\,\left( -2 - 4\,D + D^2 \right) \,m^2 \right) }
   {4\,k^2\,\left( k^2 + 4\,m^2 \right) } \hspace*{0.5cm}
\parbox{2cm}{\begin{picture}(0,0)(0,0)
\SetScale{0.4}
\SetWidth{0.5}
\Line(-25,25)(0,25)
\CArc(0,0)(25,+90,-90)
\GlueArc(0,0)(25,-90,+90){3}{10}
\Gluon(0,25)(0,-25){3}{10}
\Line(-25,-25)(0,-25)
\DashLine(25,0)(50,0){5}
\Vertex(0,0){5}
\end{picture}}
\nonumber \\
&&
+
\frac{\left( -2 + D \right) \,\left( \left( 14 - 8\,D + D^2 \right) \,k^2 + 
       4\,\left( 8 - 6\,D + D^2 \right) \,m^2 \right) }{4\,k^4\,m^2\,\left(
  k^2 + 4\,m^2 \right) } \hspace*{1.0cm}
\parbox{2cm}{\begin{picture}(0,0)(0,0)
\SetScale{0.4}
\SetWidth{0.5}
\DashLine(-50,0)(-25,0){5}
\GlueArc(0,0)(25,0,180){3}{10}
\CArc(0,0)(25,180,360)
\DashLine(25,0)(25,-30){5}
\CArc(50,0)(25,0,360)
\end{picture}}
\nonumber \\
&&
- 
\frac{\left( -8 + 3\,D \right)}{8\,\left( -7 + 2\,D \right) \,k^4\,
     m^2\,\left( k^2 + 4\,m^2 \right) } 
  \bigg\{ \left( -164 + 136\,D - 36\,D^2 + 3\,D^3 \right) \,k^2 + \nonumber \\ && 
       + 4\,\left( -104 + 98\,D - 30\,D^2 + 3\,D^3 \right) \,m^2 
\bigg\} \hspace*{1.0cm}
\parbox{2cm}{\begin{picture}(0,0)(0,0)
\SetScale{0.4}
\SetWidth{0.5}
\Line(-50,0)(50,0)
\GlueArc(0,0)(25,0,360){3}{20}
\end{picture}}
\hspace*{-1.0cm} \ .
\label{vertici:dePhi3inD}
\end{eqnarray}
The non-homogeneous terms of the above equations contain the integrals,

\begin{eqnarray}
\parbox{2cm}{\begin{picture}(0,0)(0,0)
\SetScale{0.4}
\SetWidth{0.5}
\DashLine(-50,0)(-25,0){5}
\GlueArc(0,0)(25,0,180){3}{10}
\CArc(0,0)(25,180,360)
\DashLine(25,0)(25,-30){5}
\CArc(50,0)(25,0,360)
\end{picture}}
\hspace*{-0.8cm}
&=& 
J(D,k^2) \times T(D,m^2) \ , \\
&& \nonumber \\
\parbox{2cm}{\begin{picture}(0,0)(0,0)
\SetScale{0.4}
\SetWidth{0.5}
\Line(-50,0)(50,0)
\GlueArc(0,0)(25,0,360){3}{20}
\end{picture}}
\hspace*{-1.0cm}
&=&
- {(m^2)^{D-3} \over 8 \ (D-4)}
{ \Gamma( - D - 5 )
\Gamma^2((D-2)/2)
\Gamma(2 D - 5)
\over
\Gamma( - (D-6)/2)
\Gamma(D-2)
\Gamma((3D-9)/2)
} \ ,
\end{eqnarray} 
which are MI's of the non-vanishing subtopologies - any other subdiagrams
would contain a massless tadpole, vanishing in dimensional regularization.
The former is the product of
the massive tadpole $T(D,m^2)$, given in Eq. (\ref{VP1L:eq11}), and 
the 1-loop 2-point function $J(D,k^2)$, in Eq. (\ref{VP1L:eq34});
while the latter, known as 2-loop {\it sunrise}, could be easily 
evaluated by direct parametric integration.

%
%
%

%
%
%

In principle, solving a system of three first-order ODE's is
equivalent to solve a single third-order ODE for one of the
three MI, say $\Phi_1.$
But instead of writing directly the third-order ODE,
one can observe that: in Eqs.(\ref{vertici:dePhi1inD},\ref{vertici:dePhi2inD}), $\Phi_3$ appears
to be multiplied by $(D-4)$; and in Eq.(\ref{vertici:dePhi3inD}), $\Phi_1$ and
$\Phi_2$ are multiplied by $(D-4)$.
This features means that, after expanding around $D=4$, the original system of three
coupled equations, get simplified: it decouples in a system of two coupled
equations for $\Phi_1$ and $\Phi_2$, plus a single equation for $\Phi_3$.

It is important to remark that the shape of the differential equations depends strongly 
on the choice of the MI: 
a different choice for $\Phi_i (i=1,2,3)$ could lead to a 
system which does not get simplified after the Laurent expansion either.
In all the problems we studied, given the arbitrariness of the choice,
the practical criterion of having to deal with a simpler system 
of differential equations has determined which integrals had to be picked up as master ones
- though there is no apriori assurance to find any simplification at all.

\subsection{Laurent expansion in $(D-4)$}

Indeed, 
one rearranges the system of equations 
(\ref{vertici:dePhi1inD},\ref{vertici:dePhi2inD},\ref{vertici:dePhi3inD})
as follows.
\begin{itemize}
\item Writes a second-order ODE for $\Phi_1$, from
Eqs.(\ref{vertici:dePhi1inD},\ref{vertici:dePhi2inD}). 
\item Take the first-order ODE
Eq.(\ref{vertici:dePhi3inD}) for $\Phi_3$.
\item Perform the change of variable
\begin{equation} 
k^2 \to x = {  \sqrt{-k^2} - \sqrt{-k^2 + m^2}  \over 
       \sqrt{-k^2} + \sqrt{-k^2 + m^2} } \ .
\label{vertici:xdef}
\end{equation}
\item Finally,
expands both equations around $D=4$,
with a Laurent ansatz for the solutions,
\begin{eqnarray}
\Phi_1(D,x) &=& \sum_{k=-2}^\infty (D-4)^k \ \Phi_1^{(k)}(x) \\
\Phi_2(D,x) &=& \sum_{k=-2}^\infty (D-4)^k \ \Phi_2^{(k)}(x) \\
\Phi_3(D,x) &=& \sum_{k=-2}^\infty (D-4)^k \ \Phi_3^{(k)}(x) \ .
\label{vertici:LaurentPhi}
\end{eqnarray}
\end{itemize}
After the Laurents expansion, the first-order ODE for $\Phi_3(D,x)$ will induce,
order-by-order in $(D-4)$,
a system of chained first-order ODE for the Laurent coeffiecients, $\Phi_3^{(k)}(x)$,
which at the first two orders, $k=-2,-1$, reads,
\begin{eqnarray}
0 &=& 
\Bigg\{ \frac{\left( 1 + x^2 \right)}{\left( -1 + x \right) \,x\,\left( 1 + x \right) } 
+ {d  \over dx} \Bigg\} \Phi_3^{(-2)}(x)
\ ; 
\label{vertici:dePhi3m2}
\\
&& \nonumber \\
0 &=&
\frac{1 + x}{8\,{\left( -1 + x \right) }^3}
+ \frac{H(0,x)}{8\,\left( -1 + x \right) \,\left( 1 + x \right) } + 
  \frac{H(1,x)}{4\,\left( -1 + x \right) \,\left( 1 + x \right) } +
\nonumber \\
&& + 
  \frac{\left( 1 + x \right) \,\left( 1 + x^2 \right)
    \,\Phi_1^{(-2)}(x)}{2\,{\left( -1 + x \right) }^3\,x} 
+ \frac{4\,\left( 1 + x \right) \,\Phi_2^{(-2)}(x)}{{\left( -1 + x \right) }^3} 
- \frac{\left( 1 + 6\,x + x^2 \right) \,\Phi_3^{(-2)}(x)}{2\,\left( -1 + x
    \right) \,x\,\left( 1 + x \right) } 
+
\nonumber \\
&& + 
\Bigg\{ \frac{\left( 1 + x^2 \right)}{\left( -1 + x \right) \,x\,\left( 1 + x \right) } 
+ {d  \over dx} \Bigg\} \Phi_3^{(-1)}(x)
\ .
\label{vertici:dePhi3m1}
\end{eqnarray}
We remark that the non-homogeneous term of (\ref{vertici:dePhi3m1}), 
the equation for $\Phi_3^{(-1)}(x)$, demands for the knowledge of
the previous order solutions,
$\Phi_1^{(-2)}(x)$, $\Phi_2^{(-2)}(x)$, and $\Phi_3^{(-2)}(x)$.

Analogously, for the second-order ODE for $\Phi_1(D,x)$, 
the system of chained second-order ODE just for the first two coefficients
of the Laurent series of $\Phi_1(D,x)$, for $k=-2,-1$ reads,
\begin{eqnarray}
0 &=& 
\frac{1}{4\,{\left( -1 + x \right) }^2\,x} - 
\Bigg\{
  \frac{2\,}{{\left( -1 + x \right) }^2\,x} 
- \frac{2\,}{1 + x}{d \over dx} 
- {d^2 \over dx^2} 
\Bigg\} \Phi_1^{(-2)}(x)
\ ;
\label{vertici:dePhi1m2}
\\
&& \nonumber \\
0 &=&
\frac{-1}{8\,{\left( -1 + x \right) }^2\,x}
- \frac{H(0,x)}{4\,{\left( -1 + x \right) }^2\,x} - 
  \frac{H(1,x)}{2\,{\left( -1 + x \right) }^2\,x} 
+ \frac{\Phi_3^{(-2)}(x) }{2\,x^2}
+
\nonumber \\
&&  
- \Bigg[ 
  \frac{\left( 1 + x^2 \right)}{{\left( -1 + x \right) }^2\,x^2} 
- \frac{\left( 1 + 6\,x + x^2 \right)}{2\,x - 2\,x^3} {d \over dx}
\Bigg] \Phi_1^{(-2)}(x)
+
\nonumber \\
&& 
- \Bigg\{
  \frac{2\,}{{\left( -1 + x \right) }^2\,x} 
- \frac{2\,}{1 + x}{d \over dx} 
- {d^2 \over dx^2} 
\Bigg\} \Phi_1^{(-1)}(x)
\ .
\label{vertici:dePhi1m1}
\end{eqnarray}
In this case, the non-homogeneous term of (\ref{vertici:dePhi1m1}), 
the equation for $\Phi_1^{(-1)}(x)$, requires the knowledge of
the previous order solutions,
$\Phi_1^{(-2)}(x)$ and $\Phi_3^{(-2)}(x)$.

Therefore,
by looking at the Eqs.(\ref{vertici:dePhi3m2},\ref{vertici:dePhi1m2}) 
for $k=-2$,
and 
at the structure of the non-homogeneous term of 
Eq.(\ref{vertici:dePhi3m1},\ref{vertici:dePhi1m1}) for $k=-1$,
the computational strategy is soon outlined:
\begin{itemize}
\item $k=-2$. 
      \begin{enumerate}
       \item solve Eq.(\ref{vertici:dePhi3m2}) to find $\Phi_3^{(-2)}(x)$;
       \item solve Eq.(\ref{vertici:dePhi1m2}) to find $\Phi_1^{(-2)}(x)$;
       \item invert the $D\to4$ expansion Eq.(\ref{vertici:dePhi1inD}),
             and substitute the expressions 
             of $\Phi_3^{(-2)}(x)$ and $\Phi_1^{(-2)}(x)$ in it, 
             to find $\Phi_2^{(-2)}(x)$. 
      \end{enumerate}

\item $k=-1$. 
      \begin{enumerate}
       \item plug the results of the previous order, 
             $\Phi_1^{(-2)}(x)$, $\Phi_2^{(-2)}(x)$ and
             $\Phi_3^{(-2)}(x)$, in the
             non-homogeneous term of Eqs.(\ref{vertici:dePhi3m1},\ref{vertici:dePhi1m1});
       \item solve Eq.(\ref{vertici:dePhi3m1}) to find $\Phi_3^{(-1)}(x)$;
       \item solve Eq.(\ref{vertici:dePhi1m1}) to find $\Phi_1^{(-1)}(x)$;
       \item invert the $D\to4$ expansion Eq.(\ref{vertici:dePhi1inD}),
             and substitute the expressions 
             of $\Phi_3^{(-1)}(x)$ and $\Phi_1^{(-1)}(x)$ in it,
             to find $\Phi_2^{(-1)}(x)$. 
      \end{enumerate}
\end{itemize}
The construction of Laurent coefficients for $k \ge 0$ goes on
by repeating the steps (1--4),

\begin{itemize}
\item $k=j, \ j \ge 0$. 
      \begin{enumerate}
       \item plug the results of the previous orders, 
             $\Phi_1^{(i)}(x)$, $\Phi_2^{(i)}(x)$ and
             $\Phi_3^{(i)}(x)$ $(-2 \le i < j)$, in the
             non-homogeneous term of the equations;
       \item solve the first order equation for $\Phi_3^{(j)}(x)$;
       \item solve the second order equation for  $\Phi_1^{(j)}(x)$;
       \item invert Eq.(\ref{vertici:dePhi1inD}) to find $\Phi_2^{(j)}(x)$. 
      \end{enumerate}
\end{itemize}

\noindent
Let us see in detail the case $k=-2$.

\subsection{Homogeneous differential equations}

The homogeneous differential equation is the same at
every order in $(D-4)$, as one can realize by looking at 
the operator in the curly brakets in 
Eqs.(\ref{vertici:dePhi3m2},\ref{vertici:dePhi3m1}) for $\Phi_3^{(k)}(x)$
and, correspondingly, at Eqs.(\ref{vertici:dePhi1m2},\ref{vertici:dePhi1m1}) for 
$\Phi_1^{(k)}(x)$. 

The solution $\phi_3(x)$ of the first-order homogeneous equation is
\begin{equation} 
\phi_3(x) = - \frac{x}{\left( 1 - x \right) \,\left( 1 + x \right) } 
\end{equation} 

The two solutions $\phi_{1,1}(x), \phi_{1,2}(x)$ of 
the $2^{\rm nd}$ order homogeneous equation are
\begin{eqnarray} 
\phi_{1,1}(x) &=& \frac{x}{\left( 1 - x \right) \,\left( 1 + x \right) }
\\
\phi_{1,2}(x) &=& 
\frac{x }{\left(1 - x \right) \,\left( 1 + x \right) }
 \left( -\frac{1}{x} + x - 2\,H(0,x) \right)
\end{eqnarray} 
whose Wronskian reads 
\begin{eqnarray}
W(x) =  \begin{array}{|c c|}
             \phi_1(x)  & \phi_2(x)  \\ 
               & \\
             \phi_1'(x)  & \phi_2'(x) 
        \end{array} =  -{\left( 1 + x \right) }^{-2} \ .
\label{vertex:W} 
\end{eqnarray}

\subsection{Solution of the chained equations}

Let us discuss the solutions of the system of equations at order $k=-2$, according
to the above strategy.

\begin{enumerate}

\item One easily solves Eq. (\ref{vertici:dePhi3m2}), 
finding
\begin{equation}
\Phi_3^{(-2)}(x) = - \frac{x}{1 - x^2} \Phi_3^{(1,-2)}
\label{vertici:solPhi3indet}
\end{equation}
where $\Phi_3^{(1,-2)}$ is an integration constant to be later fixed by
imposing the boundary conditions. 

\item Afterwards, one solves Eq. (\ref{vertici:dePhi1m2}) by Euler variation of constants,
\begin{eqnarray}
 \Phi_1^{(-2)}(x) &=&  \phi_{1,1}(x) 
                      \left[ \Phi_1^{(1,-2)} 
                             + \int_0^x \frac{dx'}{W(x')} \phi_{1,2}(x') K^{(-2)}(x') 
                      \right] \nonumber \\
                 && + \phi_{1,2}(x) 
                      \left[ \Phi_i^{(2,-2)} 
                             - \int_0^x \frac{dx'}{W(x')} \phi_{1,1}(x) K^{(-2)}(x') 
                      \right] 
\end{eqnarray}
with $\Phi_1^{(i,-2)}, (i=1,2)$ being integration constants, 
and $K^{(-2)}(x)$ the non-homogeneous term of Eq.(\ref{vertici:dePhi1m2}), thus
obtaining,
\begin{eqnarray}
\Phi_{1}^{(-2)}(x) &=& 
\frac{x \, (1 + H(0,x)) }{4\,\left( -1 + x \right) \,\left( 1 + x \right) } 
- \frac{x}{\left( -1 + x \right) \,\left( 1 + x\right) } \Phi_1^{(1,-2)}
+ 
\nonumber \\
&&
- \frac{\left( -1 + x^2 - 2\,x\,H(0,x) \right)}
       {\left( -1 + x \right) \,\left( 1 + x \right) } \Phi_1^{(2,-2)}
\label{vertici:solPhi1indet}
\end{eqnarray}

\item Then, the knowledge of $\Phi_3^{(-2)}(x)$ and $\Phi_1^{(-2)}(x)$, and of theirs
derivatives, allows the determination of the
the coefficient of Laurent expansion of the third MI,
by inverting the $D\to4$ expansion of Eq.(\ref{vertici:dePhi1inD}),
\begin{eqnarray}
\Phi_{2}^{(-2)}(x) &=& 
\frac{-\left( 1 + x \right) (1 + H(0,x)) }{32\,\left( -1 + x \right) }
+ \frac{\left( 1 + x \right)}{8\,\left( -1 + x \right) } \Phi_1^{(1,-2)}
+ 
\nonumber \\
&&
+ \frac{\left( 1 + x \right) \,\left( -1 + x^2 - 2\,x\,H(0,x) \right) }
       {8\,\left( -1 + x \right) \,x}\Phi_1^{(2,-2)} \ .
\label{vertici:solPhi2indet}
\end{eqnarray}

\end{enumerate}

\noindent
At the end of this three steps, one knows the expressions of the
$1/(D-4)^2$-coefficient of the three MI, $\Phi_i(D,x), (i=1,3)$
up to the determination of the {\it real} integration constants, 
$\Phi_3^{(1,-2)}, \Phi_1^{(1,-2)}, \Phi_1^{(2,-2)}$.

\subsection{Boundary conditions}
Usually boundary conditions for Feynman integrals can be read at special
values of the kinematic variables, and they do correspond to integrals
belonging to subdiagrams, therefore to simpler functions. 
That is true when the limit to such a specific values is smooth, as it happens
around the pseudo-thresholds of the corresponding diagrams. 
In our case the value of $x=-1$, meaning $k^2=4m^2$, is a pseudo-threshold.
To reach the point $x=-1$, being $x$ a space-like variable, as defined in
Eq.(\ref{vertici:xdef}),
one need an analytic continuation to the region, $x \to y=-x + i \epsilon$,
where the MI's develop an imaginary part,

\begin{eqnarray}
\Phi_{3}^{(-2)}(y) &=&
- \frac{y}{\left( -1 + y \right) \,\left( 1
  + y \right) } \Phi_3^{(1,-2)}
\\
\Phi_{1}^{(-2)}(y) &=& 
\frac{-y (1 + H(0,y))}{4\,\left( -1 + y \right) \,\left( 1 + y \right) } 
+ \frac{y}{\left( -1 + y \right) \,\left( 1 + y
    \right) } \Phi_1^{(1,-2)}
+ 
\nonumber \\
&&
- \frac{\left( -1 + y^2 + 2\,y\,H(0,y) \right)}{\left( -1 + y \right) \,\left(
  1 + y \right) }\Phi_1^{(2,-2)}
- i \pi
\frac{ y\,\left( 1 + 8 \Phi_1^{(2,-2)} \right)  
}{4\,\left( -1 + y \right) \,\left( 1 + y \right) }
\end{eqnarray}

Their expansion around $y=1$ reads,
\begin{eqnarray}
\Phi_{3}^{(-2)}(y) &=&
\frac{1}{2\,\left( 1 - y \right) } \Phi_3^{(1,-2)} 
 + {\cal O}\Big( (1-y)^0 \Big) 
\\
&& \nonumber \\
\Phi_{3}^{(-2)}(y) &=&
\frac{1}{(1 - y)}
\Bigg(
\frac{1}{8} - \frac{\Phi_1^{(1,-2)}}{2} 
+ i \,\pi  \Bigg( \frac{1 }{8} + \Phi_1^{(2,-2)} \Bigg)
\Bigg)
 + {\cal O}\Big( (1-y)^0 \Big) 
\end{eqnarray}
Finally, the three conditions needed to fix the value of the arbitrary
constants, order by order in $D-4$, are:
{\it i)} the regularity of $\Phi_3^{(k)}(y)$ as $y \to 1$;
{\it ii)} the regularity of $\Phi_1^{(k)}(y)$ as $y \to 1$,
 meaning the vanishing of the $1/(1-y)$-coefficients in both cases; 
{\it iii)} the realness of the constants, 
meaning that real and imaginary part of the 
$1/(1-y)$-coefficient must vanish separately.
From the above equations, that translates to,
\begin{eqnarray}
\Phi_3^{(1,-2)} = 0 ; \quad 
\Phi_1^{(1,-2)} = {1 \over 4} ; \quad 
\Phi_1^{(2,-2)} = - {1 \over 8} .
\end{eqnarray}

\noindent
We have all the ingredient to go up in the chain of Laurent coefficients, 
and repeat the previous steps for the case $k=-1$. 
The iterative structure of the method yields a bottom-up reconstruction 
of the three master integrals, $\Phi_1(D,x)$, $\Phi_2(D,x)$, and $\Phi_3(D,x)$, 
around $D=4$.

\section{System of Four Differential Equations}
\label{SUN4sec:int} 

\begin{figure}[h]
$$
\begin{picture}(0,0)(0,0)
\SetScale{1.3}
\SetWidth{0.7}
\CArc(0,0)(25,0,360)
\PhotonArc(0,-13)(27.5,+30,+150){1.5}{5}
\PhotonArc(0,+13)(27.5,210,330){1.5}{5}
\DashLine(-40,0)(40,0){3}
\Text(-60,0)[]{{\footnotesize $P$}}
\Text(-5, 25)[]{{\footnotesize $k_1$}}
\Text(5,-25)[]{{\footnotesize $k_2$}}
\Text( 5, 40)[]{{\footnotesize $k_3$}}
\Text(-5,-40)[]{{\footnotesize $k_4$}}
\Text(0,-7)[]{{\footnotesize $P - \sum_i k_i$}}
\end{picture}
$$
\vspace*{1.0cm}
\caption{Four-loop sunrise diagram: $P^2 = -m^2$; 
$k_{1,2,3,4}$ are loop variables; a wavy line stands for massless 
propagator; a solid line, for propagator of mass $M$; 
a dashed line, for propagator of mass $m$; 
}
\label{SUN4SUN4:fig1} 
\end{figure}
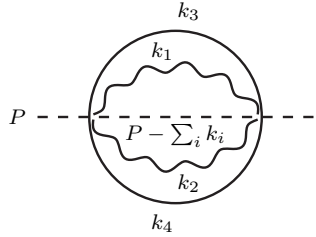

This section is devoted to the analytic evaluation of the MI's 
associated to the 4-loop sunrise graph with two massless lines, 
two massive lines of equal mass $M$, 
another massive line of mass $m$, with $m \ne M$, and the external invariant 
timelike and equal to $m^2$, as depicted in Fig.~\ref{SUN4SUN4:fig1}
\cite{Laporta:2003xa}. \par 

\vspace*{0.5cm}

In this case, the heart of the analytic 
calculation is the study of a homogeneous fourth order differential 
equation, whose solutions turned out to be, in a remarkably simple way, 
either a rational fraction 
or repeated quadratures of rational fractions. The required four-loop 
integral could then be obtained almost mechanically by repeated 
quadratures in terms of HPL's. \par 
Following the reduction algorithm
- by now sounding familiar to the reader -,
we identify the MI's of the problem; 
write the system of differential equations 
in $x=m/M$ satisfied by the MI's; convert it into a higher order 
differential equation for a single MI; Laurent-expand it around $D=4$; 
solve the associated homogeneous equation at $D=4$ 
and then use recursively Euler's method of 
the variation of the constants for obtaining the coefficient of the 
$(D-4)$-expansion in closed analytic form. The result involves 
HPL's of argument $x$ and weight increasing with the order in $(D-4)$. 
The integration constants are fixed at $x=0\ $.
After solving the differential equations for arbitrary $x$, we will 
show how to compute, independently,
the numerical value of the solution at $x=1$ by using 
the Finite Difference method of Laporta ~\cite{Laporta:2001dd}, discussed
in Sec.\ref{sec:DifferenceEqn},
to show the relation among Differential and Difference Equations for Feynman
integrals.

\par

\subsection{Master Integrals and Differential Equations} 
\label{SUN4sec:MI} 
We find that the problem has 5 MI's, which we choose to be 
\begin{equation} 
  F_i(D) = \int \frac{d^Dk_1 \ldots d^Dk_4}{(2\pi)^{4(D-2)}} 
         \frac{ \ N_i} 
      {{\cal{D}}_1{\cal{D}}_2{\cal{D}}_3{\cal{D}}_4{\cal{D}}_5} \ , 
\label{SUN4eq:defFi} 
\end{equation} 
where the numerators $N_i$ are 
$(M^2, k_1 \cdot k_3, p \cdot k_3, k_1 \cdot k_2, p \cdot k_2)$ and the denominators ${\cal{D}}_1=k_1^2,{\cal{D}}_2=k_2^2, {\cal{D}}_3=k_3^2+M^2, {\cal{D}}_4=k_4^2+M^2, {\cal{D}}_5=(P-k_1-k_2-k_3-k_4)+m^2$ . 
In terms of the dimensionless variable $ x = m/M $ and putting 
$P=Mp$ one can introduce 5 dimensionless functions $\Phi_i(d,x)$ through 
\begin{equation} 
     F_i(D) = M^{4D-8} \ C^4(D)\ \Phi_i(D,x) \ , 
\label{SUN4eq:FtoPhi} 
\end{equation} 
where $C(D)= (4\pi)^\frac{4-D}{2}\Gamma(3-D/2) $ is an overall loop 
normalization factor, with the limiting value $C(4)=1$ at $D=4$. 
\par 
The derivation of the system of differential equations is 
straightforward; the derivatives of the MI's, \ie of the 5 functions 
$\Phi_i(D,x)$, with respect to $x$ are easily 
carried out in their representation as loop-integrals Eq.(\ref{SUN4eq:defFi}); 
when the result is in turn expressed in terms of the same MI's, one 
obtains the following linear system of first order differential equations 
in $x$ 
\begin{eqnarray}
\frac{d \Phi_1(D,x)}{d x}
 & = &
        \Bigg\{ 
                 \frac{(3D - 7)}{x} 
               + \frac{3 \ (D-2)}{2 \ (1-x)}
               - \frac{3 \ (D-2)}{2 \ (1+x)}
        \Bigg\} \Phi_1(D,x) 
      - \Bigg\{ 
                 \frac{3(D - 2)}{x} 
               + \frac{3 \ (D-2)}{2 \ (1-x)}
        \nonumber \\
        & &
               - \frac{3 \ (D-2)}{2 \ (1+x)}
        \Bigg\}
        \bigg(
                 3 \Phi_2(D,x) 
               - 3 \Phi_3(D,x) 
               +   \Phi_4(D,x)
               - 3 \Phi_5(D,x)
        \bigg) \ ,
\label{SUN41stDEPhi1} \\
 & & \nonumber \\
\frac{d \Phi_2(D,x)}{d x}
 & = & - \frac{(D-2)}{x} \bigg(
                                  \Phi_2(D,x)
                                - 2 \Phi_3(D,x)
                                + \Phi_4(D,x)
                                - 2 \Phi_5(D,x)
                         \bigg) \ ,
\label{SUN41stdEPhi2} \\
 & & \nonumber \\
\frac{d \Phi_3(D,x)}{d x}
& = &
       - \Bigg\{ 
                 \frac{3 \ (D-2)}{2 \ (1-x)}
               - \frac{3 \ (D-2)}{2 \ (1+x)}
         \Bigg\} \bigg(
                         \Phi_1(D,x)
                       - 3 \Phi_2(D,x)
                       - \Phi_4(D,x)
                 \bigg) 
         \nonumber \\
         & &
       - \Bigg\{ 
                 \frac{3(D - 2)}{x} 
               + \frac{9 \ (D-2)}{2 \ (1-x)}
               - \frac{9 \ (D-2)}{2 \ (1+x)}
         \Bigg\} \bigg(
                         \Phi_3(D,x)
                       + \Phi_5(D,x)
                 \bigg) \ ,
\label{SUN41stDEPhi3} \\
 & & \nonumber \\
\frac{d \Phi_4(D,x)}{d x}
 & = & \frac{2(D-2)}{x} \bigg(
                                  \Phi_2(D,x)
                                + \Phi_4(D,x)
                         \bigg) \ ,
\label{SUN41stDEPhi4} \\
 & & \nonumber \\
\frac{d \Phi_5(D,x)}{d x}
 & = & \frac{2(D-2)}{x} \bigg(
                                  \Phi_3(D,x)
                                + \Phi_5(D,x)
                         \bigg) \ ,
\label{SUN41stDEPhi5}
\end{eqnarray}

The system is homogeneous: indeed, quite in general the non homo\-ge\-neous 
terms are given by the MI's of the subtopologies of the considered graph, 
obtained by shrinking to a point any of its propagator lines. When that 
is done for the parent topology with five propagators of 
Fig.(\ref{SUN4SUN4:fig1}), 
one obtains the product of 4 tadpoles; but as the considered graph has 
two massless propagators, at least one massless tadpole is always 
present in the product; as in the $D$-dimensional regularization 
massless tadpoles vanish, the product of the 4 tadpoles is always equal 
to zero -- and therefore the differential equations are homogeneous. 

By inspection, one sees that $\Phi_3(D,x), \Phi_5(D,x)$ appear in the 
r.h.s. of Eq.s(\ref{SUN41stDEPhi1}-\ref{SUN41stDEPhi5}) only in the
combination  
\begin{equation} 
 \Psi_3(D,x) = \Phi_3(D,x) + \Phi_5(D,x) \ ; 
\label{SUN4defPsi3} 
\end{equation} 
the other linearly independent combination of the two MIs, say 
\begin{equation} 
 \Psi_5(D,x) = \Phi_3(D,x) - \Phi_5(D,x) \ , 
\label{SUN4defPsi5} 
\end{equation} 
decouples and can be expressed in terms of the other integrals by means 
of the trivial 1st order differential equation 
\begin{eqnarray} 
\frac{d \Psi_5(D,x)}{d x} 
 & = & 
       - \Bigg\{ 
                 \frac{3 \ (D-2)}{2 \ (1-x)} 
               - \frac{3 \ (D-2)}{2 \ (1+x)} 
         \Bigg\} \bigg( 
                         \Phi_1(D,x) 
                       - 3 \Phi_2(D,x) 
                       - \Phi_4(D,x) 
                 \bigg) 
         \nonumber \\ 
         & & 
       - \Bigg\{ 
            \frac{5 (D-2)}{x} 
          + \frac{9 \ (D-2)}{2 \ (1-x)} 
          - \frac{9 \ (D-2)}{2 \ (1+x)} 
         \Bigg\} \Psi_3(D,x) 
\label{SUN4Psi5eq} 
\end{eqnarray} 

\noindent
As $\Psi_5(D,x)$ does not enter in the r.h.s. of 
Eq.s(\ref{SUN41stDEPhi1}-\ref{SUN41stDEPhi5}), the 4 linear 
equations for $\Phi_1(D,x)$, $\Phi_2(D,x)$, $\Psi_3(D,x)$, and 
$\Phi_4(D,x)$ can be written as a fourth order equation for 
$\Phi_1(x)$, which will be called simply $\Phi(D,x)$ from 
now on, and which is therefore equal to 
\begin{equation} 
  \Phi(D,x) = \frac{C^{-4}(D)}{(2\pi)^{4(D-2)}} \int 
     \frac{d^Dk_1\  d^Dk_2 \ d^Dk_3 \ d^Dk_4 } 
      {{\cal{D}}_1{\cal{D}}_2{\cal{D}}_3{\cal{D}}_4{\cal{D}}_5} \ , \ (p^2=-x^2) \ . 
\label{SUN4eq:defPhi} 
\end{equation}

\noindent
One obtains for $\Phi(D,x)$ the following fourth-order ODE
\begin{eqnarray} 
  x^3 (1-x^2) \frac{d^4 \Phi(D,x)}{d x^4}
+ x^2 \bigg\{ 1 + 5 x^2 - 3 (D-4) (1 - 3 x^2)
      \bigg\} \frac{d^3 \Phi(D,x)}{d x^3}
&& \nonumber\\ 
- x \bigg\{
            12
          + 6 x^2
          + (D-4) (13 + 32 x^2)
          + (D-4)^2 (1 + 26 x^2)
    \bigg\} \frac{d^2 \Phi(D,x)}{d x^2} 
&& \nonumber\\ 
+  \bigg\{
            12
          - 18 x^2
          + (D-4) (25 - 2 x^2)
          + 8 (D-4)^2 (2 + 5 x^2)+ \hspace*{2cm}&& \nonumber \\
          + 3 (D-4)^3 (1 + 8 x^2)
   \bigg\} \frac{d \Phi(D,x)}{d x}
&& \nonumber\\ 
+  4 x \bigg\{ 
          + 12 
          + 29 (D-4)
          + 23 (D-4)^2
          + 6 (D-4)^3
     \bigg\} \Phi(D,x) 
&& {\kern-5pt} =  0   \nonumber\\. 
\label{SUN4eq:4thordeq} 
\end{eqnarray}

\subsection{Behaviour of $\Phi(D,x)$ in the limit $x\to 0$} 

\label{SUN4sec:xto0p} 

By inspection, one finds that the most general solution of 
Eq.(\ref{SUN4eq:4thordeq}) can be expanded for $x\to 0$ in the form 
\begin{equation} 
  \Phi(D,x) = \sum_{i=1}^4 x^{\alpha_i} 
              \left( \sum_{n=0}^\infty A_n^{(i)}(D) x^{2n} \right) \ , 
\label{SUN4eq:xexpPhi} 
\end{equation} 
where the values of the 4 exponent $\alpha_i$ are 
\begin{eqnarray} 
   \alpha_1 &=& 0 \ , \nonumber\\ 
   \alpha_2 &=& (D-2) \ , \nonumber\\ 
   \alpha_3 &=& -(D-2) \ , \nonumber\\ 
   \alpha_4 &=& (3D-7) \ ; 
\label{SUN4eq:alphaval} 
\end{eqnarray} 
the $A_0^{(i)}(D)$ are the 4 integration constants, and all the other 
coefficients $A_n^{(i)}(D)$ for $n>0$ are determined by the differential 
equation Eq.(\ref{SUN4eq:4thordeq}), once the integration constants are fixed. 
\par 
A qualitative inspection of the integrals which one tries to evaluate by 
means of the differential equation Eq.(\ref{SUN4eq:defPhi}) shows 
that it is finite (just finite, not analytic!) for $x\to 0^+$ 
and $(D-2)>0$; that is sufficient to rule out from their expression as 
solutions of the differential equation the terms 
with the behaviour of the third and the fourth exponent (which is 
negative when $D$ is just above 2). 
\par
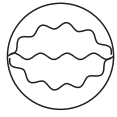
\begin{figure}[t]
$$
\begin{picture}(0,0)(0,0)
\SetScale{0.8}
\SetWidth{0.7}
\CArc(0,0)(25,0,360)
\PhotonArc(0,-13)(27.5,+30,+150){1.5}{5}
\PhotonArc(0,+13)(27.5,210,330){1.5}{5}
\Photon(-25,0)(25,0){1.5}{4}
\end{picture}
$$
\vspace*{0.3cm}
\caption{Four-loop watermelon diagram: 
a wavy line stands for massless 
propagator; a solid line, for propagator of mass $M$. 
}
\label{SUN4fig:3} 
\end{figure}

 
In the current case, as the equation for $\Phi(D,x)$ is homogeneous, the 
only information one gets out 
is that $A_0^{(3)}(D)$ and $A_0^{(4)}(D)$ are both equal 
to zero, due to the finiteness for $x\to0^+$; by substituting the 
{\it ansatz} Eq.(\ref{SUN4eq:xexpPhi}) in Eq.(\ref{SUN4eq:4thordeq}) and dropping 
$A_0^{(3)}(D), A_0^{(4)}(D),\ $ one finds for $\Phi(D,x)$ Eq.(\ref{SUN4eq:defPhi}) 
the $x\to 0$ expansion 
\begin{eqnarray} 
\Phi(D,x) &=& 
            A_0^{(1)}(D) 
              \left(   1 
                     - \frac{2 (2D-5) (3D-8)}{3 D (D-4)} \ x^2 
                     + O(x^4) 
              \right) \nonumber\\ 
          &+& A_0^{(2)}(D)\ x^{D-2} 
              \left(   1 
                     + \frac{(D-3)(D-4)(3D-8)}{2 D (2D-7)} \ x^2 
              + O(x^4) \right) \ . 
\label{SUN4eq:Phiexpexpl} 
\end{eqnarray} 
The expansion depends on the two as yet unspecified integrations 
constants $A_0^{(1)}(D), A_0^{(2)}(D)$.
To fix them, one has to provide some independent information, such as 
the value of the required Feynman integral and of its first derivative at
$x=0$.
Those value can be provided by an explicit conventional calculation, 
say in parameter space, which is in any case much easier than a 
calculation for non-zero values of the variable $x$
\footnote{In the present case the 
knowledge of the regularity of the solution at $x=1$ does not provide 
any additional information.}. 
That is done explicitly in \cite{Laporta:2003xa}, and the results are
%
\begin{eqnarray} 
 A_0^{(1)} &=& -\ \frac{3D-11}{ 8(D-2)(D-3)(D-4)^3(2D-5)(2D-7)(3D-8)(3D-10) } 
                                             \nonumber\\ 
  &\times& \frac{ \Gamma(1-(D-4))\Gamma(1-2(D-4)) 
           \Gamma^2\left(1+\frac{1}{2}(D-4)\right) 
           \Gamma^2\left(1-\frac{3}{2}(D-4)\right) } 
         { \Gamma^4\left(1-\frac{1}{2}(D-4)\right) 
           \Gamma(1-3(D-4)) }                \nonumber\\ 
 A_0^{(2)} &=& -\ \frac{ 2\ (2D-7) } { 3(D-2)^2(D-3)(D-4)^4(3D-8)(3D-10) } 
                                             \nonumber\\ 
  &\times& \frac{ \Gamma\left(1+\frac{1}{2}(D-4)\right) 
                  \Gamma\left(1-\frac{3}{2}(D-4)\right) 
                  \Gamma^2(1-(D-4)) } 
                { \Gamma^2\left(1-\frac{1}{2}(D-4)\right) 
                  \Gamma(1-2(D-4)) } \ ,
\label{SUN4eq:Aval} 
\end{eqnarray} 
where the term $A_0^{(1)}$ is the value of the vacuum graph
in Fig.\ref{SUN4fig:3}.

\vspace*{0.5cm}

\subsection{Expansion around $D=4$ and homogeneous equation} 
\label{SUN4sec:expD-4} 

The Laurent's expansion in $(D-4)$ of $\Phi(D,x)$ Eq.(\ref{SUN4eq:defPhi}) is 
\begin{equation} 
   \Phi(D,x) = \sum_{n=-4}^\infty (D-4)^n \Phi^{(n)}(x) \ , 
\label{SUN4eq:dexpPhi} 
\end{equation} 
as it is known on general grounds that it develops at most a fourth 
order pole in $(D-4)$. By substituting in Eq.(\ref{SUN4eq:4thordeq}) 
one obtains a system of inhomogeneous, chained equations for the 
coefficients $\Phi^{(n)}(x)$ of the expansion in $(D-4)$; the 
generic equation reads 

\begin{equation}
{\cal D} \Phi^{(n)}(x) = K^{(n)}(x) \ , 
\label{SUN4eq:eqPhin}
\end{equation}
with
\begin{equation} 
{\cal D} = \left[ 
  x^3 (1-x^2) \frac{d^4}{d x^4}
+ x^2 (1 + 5 x^2) \frac{d^3}{d x^3}
- 6 x (2 + x^2) \frac{d^2}{d x^2}
+ 6 (2 - 3 x^2) \frac{d}{d x}
+  48 x 
 \right] \ ,  
\end{equation} 
and
\begin{eqnarray} 
  K^{(n)}(x) &=&
          \bigg\{
            24 x
          + \big( 3 + 24 x^2 \big) \frac{d}{dx}
        \bigg\} \Phi^{(n-3)}(x) 
        \nonumber \\
& &
        + \bigg\{
            92 x
          + \big( 16 + 40 x^2 \big) \frac{d}{dx}
          - \big( x + 26 x^3 \big) \frac{d^2}{dx^2}
        \bigg\} \Phi^{(n-2)}(x)
        \nonumber \\
& &
        + \bigg\{
            116 x
          + \big( 25 - 2 x^2 \big) \frac{d}{dx}
          - \big( 13 x + 32 x^3 \big) \frac{d^2}{dx^2}
          - \big( 3 x^2 - 9 x^4 \big) \frac{d^3}{dx^3}
        \bigg\} \Phi^{(n-1)}(x)\nonumber\\
\ , 
\label{SUN4eq:Kn} 
\end{eqnarray} 
which shows that the equation at a given order $n$ for $\Phi^{(n)}(x)$ 
involves in the inhomogeneous term the coefficients 
$\Phi^{(k)}(x)$ (and their derivatives) with $k<n$ 
(obviously $\Phi^{(k)}(x) = 0$ when $k < -4$). 
Such a structure calls for an algorithm of solution bottom-up, i.e. 
starting from the lowest value of $n$ (which is $n=-4$) and proceeding 
recursively to the next $n+1$ value up to the required order. 
\par 
The Eq.s(\ref{SUN4eq:eqPhin}) have all the same associated homogeneous 
equation, independent of $n$, 
\begin{equation} 
 \left[ 
  x^3 (1-x^2) \frac{d^4}{d x^4}
+ x^2 (1 + 5 x^2) \frac{d^3}{d x^3}
- 6 x (2 + x^2) \frac{d^2}{d x^2}
+ 6 (2 - 3 x^2) \frac{d}{d x}
+  48 x 
 \right] \phi(x) = 0 \ ; 
\label{SUN4eq:eqhomophi} 
\end{equation} 
once the solutions of Eq.(\ref{SUN4eq:eqhomophi}) are known, all the 
Eq.s(\ref{SUN4eq:eqPhin}) can be solved by the method of the 
variation of the constants of Euler. 
\par 
To our (pleasant) surprise, the solutions of Eq.(\ref{SUN4eq:eqhomophi}) are 
almost elementary. By trial and error, a first solution is found to be 
\begin{equation} 
 \phi_1(x) = x^2 \ . 
\label{SUN4eq:phi1} 
\end{equation} 
We then substitute $\phi(x) = \phi_1(x) \xi(x) $ in Eq.(\ref{SUN4eq:eqhomophi}), 
obtaining the following 3rd order equation for the derivative of
$\xi(x) $  
\begin{equation} 
 \left[ 
  x^3 (1 - x^2) \frac{d^3}{d x^3}
+ 3 x^2 (3 - x^2) \frac{d^2}{d x^2}
+ 6 x (1 + 2 x^2) \frac{d}{d x}
- 6 ( 5 + 2 x^2 ) 
 \right] \xi'(x) = 0\ , 
\label{SUN4eq:eqxi} 
\end{equation} 
and find that it admits the solution 
\begin{equation} 
 \xi'_2(x) = \frac{1}{x^3}(1-x^2+x^4) \ . 
\label{SUN4eq:xi2} 
\end{equation} 
Substituting $\xi'(x) = \xi'_2(x) \chi(x)$ in Eq.(\ref{SUN4eq:eqxi}) we 
obtain the following 2nd order equation for $\chi'(x)$
\begin{equation} 
 \left[ 
    x^2 (1 - x^2) (1-x^2+x^4) \frac{d^2}{d x^2}
+ 6 x^5 (2 - x^2) \frac{d}{d x}
- 6 ( 2 - 2 x^4 + x^6 ) 
 \right] \chi'(x) = 0\ , 
\label{SUN4eq:eqchi} 
\end{equation} 
which admits as solution 
\begin{equation} 
 \chi'_3(x) = \frac{1}{x^3}(1-x^2)^4\ \frac{5-2x^2+5x^4}{(1-x^2+x^4)^2} \ . 
\label{SUN4eq:chi3} 
\end{equation} 
Finally, substituting $ \chi'(x) = \chi'_3(x)\tau(x) $ in 
Eq.(\ref{SUN4eq:eqchi}), we obtain the equation 
\begin{eqnarray} 
 \bigg[ 
 x (1 - x^2)^5 (1-x^2+x^4) (5-2x^2+5x^4) \frac{d}{d x}
  \hspace*{1.5cm} & & \nonumber \\
-2 (1 - x^2)^4 (15 - 12x^2 + 11x^4 + 30x^6 - 24x^8 + 20x^{10}) 
 \bigg]& &  \tau'(x) = 0\ , 
\label{SUN4eq:eqtau} 
\end{eqnarray} 
which has the solution 
\begin{equation} 
 \tau'_4(x) = \frac{x^6}{(1-x^2)^5}\ \frac{1-x^2+x^4} {5-2x^2+5x^4}\ . 
\label{SUN4eq:tau4} 
\end{equation} 

By repeated quadratures in $x$ and multiplications by the previous 
solutions we obtain the explicit analytic expressions of the 4 solutions 
of Eq.(\ref{SUN4eq:eqhomophi}); the nasty denominators $(1-x^2+x^4)$ and 
$(5-2x^2+5x^4)$ disappear in the final results, while the repeated 
integrations of the terms with denominators $x, (1+x)$ and $(1-x)$ 
generate, almost by definition, HPL's 
of argument $x$ and weight up to 3.

\begin{eqnarray}
 \phi_2(x) &=&
       - \frac{1}{2} (1 - x^4)
       - H(0,x) x^2 \ ,  \\ 
      & & \nonumber \\
 \phi_3(x) &=&
      \frac{(5 + 18x^2 + 14x^6 + 5x^8)}{8 \ x^2}
       + \frac{1}{2} (12 + x^2 - 12 x^4 ) H(0;x)
       + 12 \ x^2 H(0,0;x) \ ,  \\ 
      & & \nonumber \\
 \phi_4(x) &=&
          \frac{(1 + x^2) (15 + 182 x^2 + 15 x^4)}{65536 \ x} 
        + \frac{3 (1-x^2)^2 (5 + x^2) (1 + 5x^2)}{131072 \ x^2} 
              \Big[H(-1;x) + H(1;x) \Big] \nonumber \\
      & &   - \frac{9 (1 - x^4)}{8192}  \Big[ H(0,-1;x) + H(0,1;x) \Big] 
            - \frac{9 x^2}{4096}  \Big[ H(0,0,-1;x) + H(0,0,1;x) \Big] \ .
\label{SUN4eq:phis} 
\end{eqnarray}
The corresponding Wronskian has the remarkably simple expression 
\begin{eqnarray}
W(x) =  \begin{array}{|c c c c|}
             \phi_1(x)  & \phi_2(x)  & \phi_3(x)  &  \phi_4(x) \\ 
               & & & \\
             \phi_1'(x)  & \phi_2'(x)  & \phi_3'(x)  &  \phi_4'(x) \\
               & & & \\
             \phi_1''(x)  & \phi_2''(x)  & \phi_3''(x)  &  \phi_4''(x) \\
               & & & \\
             \phi_1'''(x)  & \phi_2'''(x)  & \phi_3'''(x)  &  \phi_4'''(x)
        \end{array} = \frac{(1-x^2)^3}{x} \ ,
\label{SUN4eq:W} 
\end{eqnarray}
in agreement (of course) with the coefficients of the $4^{th}$ and
$3^{rd}$ $x$-derivative of $\phi(x)$ in Eq.(\ref{SUN4eq:eqhomophi}). 

\subsection{Solutions of the chained equations} 
\label{SUN4sec:solved-4} 

With the results established in the previous Section one can 
use Euler's method of the variation of the constants for solving 
Eq.s(\ref{SUN4eq:eqPhin}) recursively in $n$, starting from $n=-4$. 
We write Euler's formula as 
\begin{equation} 
 \Phi^{(n)}(x) = \sum_{i=1}^4 \ \phi_i(x) \left[ \Phi_i^{(n)} 
               + \int_0^x \frac{dx'}{W(x')} M_i(x') K^{(n)}(x') \right] \ , 
\label{SUN4eq:Euler} 
\end{equation} 
where the $\phi_i(x)$ are the solutions of the homogeneous equation 
given in Eq.s(\ref{SUN4eq:phi1},\ref{SUN4eq:phis}), the $ \Phi_i^{(n)} $ 
are the as yet undetermined integration constants, the 
Wronskian $W(x)$ can be read from Eq.(\ref{SUN4eq:W}), the $M_i(x)$ 
are the minors of the $\phi_i^{'''}(x) $ in the determinant 
Eq.(\ref{SUN4eq:W}), and the $K^{(n)}(x)$ 
are the inhomogeneous terms of Eq.(\ref{SUN4eq:Kn}). 
The constants $\Phi_i^{(n)}$ are then fixed by comparing the 
expansion in $x$ for $x\to 0$ of Eq.(\ref{SUN4eq:Euler}) with the 
expansion in $(D-4)$ for $D\to4$ of Eq.(\ref{SUN4eq:Phiexpexpl}). 
Explicitly, we find, for the first two coefficients of the Laurent expansion
in Eq.(\ref{SUN4eq:dexpPhi}),  
\begin{eqnarray} 
 \Phi^{(-4)}(x) &=& - \frac{1}{64} x^2 \ , 
\label{SUN4eq:Phi-4x} \\
 \Phi^{(-3)}(x) &=&  
       - \frac{1}{384}
       + \frac{9}{256} x^2
       - \frac{1}{192} x^4
       - \frac{1}{48} x^2 H(0;x)  \ .
\label{SUN4eq:Phi-3x} \\
\end{eqnarray} 
The full results become quickly too lengthy to be reported explicitly here.

\subsection{Value at $x=1$ from Difference Equations} 
\label{SUN4sec:numcal} 

In this section we will calculate $\Phi(D,x=1)$ 
by using the Difference Equation technique described by Laporta
in~\cite{Laporta:2001dd}, 
which is a formidable way to get numerical results with 
high accuracy.
We will see in this example, the link between Differential Equations and 
Difference Equations for Feynman integral.

Having set ${\mathbb{D}}_1=(k_1^2+1),{\mathbb{D}}_2=((k_2-k_1)^2+1), {\mathbb{D}}_3=((k_3-k_2)^2+1), {\mathbb{D}}_4=(k_4-k_3)^2, {\mathbb{D}}_5=(p-k_4)^2$, we define 
\begin{equation} 
\label{SUN4eq:defI5} 
I_{5}(D,n)= \pi^{-2D} \int \frac{d^Dk_1\  d^Dk_2 \ d^Dk_3 \ d^Dk_4 } 
      {{\mathbb{D}}_1^n{\mathbb{D}}_2{\mathbb{D}}_3{\mathbb{D}}_4{\mathbb{D}}_5}
      \ , \quad p^2=-1\ ,  
\end{equation} 
so that $I_5(D,1)$ is equal to $\Phi(D,x)$ of Eq.(\ref{SUN4eq:defPhi}) 
at $x=1$ up to a known multiplicative factor  
\begin{equation}\label{SUN4eq:normal}
I_{5}(D,1) = {\left[4\Gamma(1+\ep)\right]^4} \Phi(D=4-2\ep,x=1)\ .
\end{equation}
By combining identities obtained by integration by parts 
one finds that $I_{5}(D,n)$ satisfies the third-order difference equation 
\begin{eqnarray} 
 32(n-1)(n-2)(n-3)(n-3D+5)&I_{5}(D,n)  & \nonumber\\ 
 -4(n-2)(n-3)\biggl[15n^2+(39-50D)n +27D^2 +5D -54 
                \biggr] &I_{5}(D,n-1)  & \nonumber\\ 
 +2(n-3) \bigg[12n^3-(38D+24)n^2+(23D^2+133D-84)n
  && \nonumber\\  
  +9D^3-141D^2 +134D-24 \bigg] &I_{5}(D,n-2)  & \nonumber\\ 
 +(n-D-1)(n-2D+1)(2n-3D)(2n-5D+4)&I_{5}(D,n-3)& = 0 \hspace{.5cm} \;.
\label{SUN4eq:dif1} 
\end{eqnarray}
We will solve this difference equation by using the Laplace {\it ansatz} 
\begin{equation}\label{SUN4eq:ansatz5} 
I_{5}(D,n)=\int_0^1 t^{n-1} v_{5}(D,t) \; dt\ , 
\end{equation} 
giving for $v_{5}(D,t)$ the fourth-order differential equation
\begin{eqnarray} 
  4t^4(8t+1)(t-1)^2 &\frac{d^4}{dt^4}v_5(D,t)&     \nonumber\\ 
  +4t^3(t-1) \biggl[ 24(D+1)t^2 +(18 - 26D)t - 7D + 12 \biggr] 
                          &\frac{d^3}{dt^3}v_5(D,t)& \nonumber\\
  t^2\bigg[ (576 (D-1)t^3 + ( - 108D^2 - 420D +648   )t^2 && \nonumber\\
      +(46d^2  + 38D  - 144 )t  + 71D^2  - 284D   + 288 \bigg] 
                          &\frac{d^2}{dt^2}v_5(D,t)& \nonumber\\
 +t\bigg[(576D-960)t^3 +(- 216D^2+360D)t^2 &&\nonumber\\ 
  \phantom{t[} +(- 18D^3 + 190D^2 - 496D +384)t  
                 + 77D^3  - 533D^2   + 1236D - 960) \bigg] 
                                   &\frac{d}{dt}v_5(D,t)& \nonumber\\
 + (D-3)(2D-5)(3D-8)(5D-12) &v_5(D,t) =0 &\;. \hspace*{1.0cm} 
\label{SUN4eq:equd1} 
\end{eqnarray} 

We will look for the solution of Eq.(\ref{SUN4eq:equd1}) in the form of 
a power series expansions, which, inserted in Eq.(\ref{SUN4eq:ansatz5}) 
and integrated term by term, will provide very accurate values 
of $I_5(D,n)$. As the convergence is faster for larger $n$, we will 
consider large enough values of the index $n$ (see below); the repeated 
use top-down of Eq.(\ref{SUN4eq:dif1}) ({\it i.e.} using it for 
expressing $I_5(D,n-3)$ in terms of the $I_5(k)$ with $k=n,n-1,n-2$) 
will give the values corresponding to smaller indices, till $I_5(D,1)$ 
is eventually obtained. 
To go on with this program, initial conditions for $v_5(D,t)$ are needed. 

From the definition Eq.(\ref{SUN4eq:defI5}), and introducing spherical 
coordinates in $D$-dimension for the loop momentum $k_1$, 
$ d^Dk_1 = k_1^{D-1}dk_1 d\Omega(D,\hat k_1) $ one has 
\begin{eqnarray}
I_{5}(D,n)&=&\dfrac{1}{\Gamma\left(\frac{D}{2}\right)} 
    \int_0^\oo \dfrac{(k_1^2)^{D/2-1}\;dk_1^2}{(k_1^2+1)^n} 
      f_{5}(D,k_1^2) \;, \\
\label{SUN4eq:deff2}
f_{5}(D,k_1^2)&=& \int \dfrac{d\Omega(D,\hat k_1)}{\Omega(D)} 
                         I_{4}(D,1,(p-k_1)^2) \;,
\end{eqnarray} 
where $\Omega(D)$ is the $D$-dimensional solid angle,
and $I_{4}(D,1,(p-k_1)^2) $ is the 3-loop (off mass-shell) sunrise integral 
\begin{equation}\label{SUN4eq:defI4} 
I_{4}(D,n,(p-k_1)^2)= \pi^{-3D/2} \int \frac{d^Dk_2\  d^Dk_3 \ d^Dk_4} 
      {{\mathbb{D}}_2^n{\mathbb{D}}_3{\mathbb{D}}_4{\mathbb{D}}_5  }
      \ . 
\end{equation} 
By the change of variable $1/(k_1^2+1)=t,\ $ $\ k_1^2=(1-t)/t,\ $ one finds
\begin{equation}
  I_{5}(D,n)=\dfrac{1}{\Gamma\left(\frac{D}{2}\right)} 
             \int_0^1 t^{n-1} (1-t)^{\frac{D}{2}-1} 
             t^{-\frac{D}{2}} f_{5}\left(D,\frac{1-t}{t}\right) \; dt\;,
\end{equation}
from which one gets the relation between $v_{5}(D,t) $ and 
$f_{5}(D,(1-t)/t)$
\begin{equation}\label{SUN4eq:vvff}
v_{5}(D,t)=\dfrac{1}{\Gamma\left(\frac{D}{2}\right)} 
             (1-t)^{\frac{D}{2}-1} 
             t^{-\frac{D}{2}} f_{5}\left(D,\frac{1-t}{t}\right) \; dt\;. 
\end{equation} 
From that relation we see that we can derive boundary conditions for 
$v_5(D,t)$ in the $t\to1$ limit from the expansion of $f_5(D,k_1^2)$ 
in the $k_1\to 0 $ limit, which is easy to obtain. 
Indeed, only the first denominator of \eqref{SUN4eq:defI4} depends on $k_1$; 
expanding for small $k_1$ and performing the angular integration one gets
\begin{equation}
\begin{split}
\int \dfrac{d\Omega(D,\hat k_1)}{\Omega(D)} \dfrac{1}{(k_2-k_1)^2+1}
=& \int \dfrac{d\Omega(D,\hat k_1)}{\Omega(D)} \left(
\dfrac{1}{k_2^2+1} 
-\dfrac{k_1^2-2k_1\cdot k_2}{(k_2^2+1)^2} 
+\dfrac{(k_1^2-2k_1\cdot k_2)^2}{(k_2^2+1)^3} + {\ldots} \; \right)\\
=&\dfrac{1}{k_2^2+1} 
+k_1^2\left(-\dfrac{1}{(k_2^2+1)^2} +\dfrac{4}{D} 
                        \dfrac{k_2^2}{(k_2^2+1)^3}\right) + {\ldots} \;.
\end{split}
\end{equation}
The above result gives the expansion of $f_{5}(D,k_1^2)$ at $k_1^2=0$:
\begin{equation}\label{SUN4eq:svilf5}
\begin{split}
 f_{5}(D,k_1^2)\ =& \ f_5^{(0)}(D) + f_5^{(1)}(D)\ k_1^2 + O(k_1^4)\;, \\
   f_5^{(0)}(D)\ =& \ I_{4}(D,1,p^2) \\ 
   f_5^{(1)}(D)\ =& \ -I_{4}(D,2,p^2)
  +\dfrac{4}{D}\left[I_{4}(D,2,p^2) - I_{4}(D,3,p^2) \right] \ . \\ 
\end{split} 
\end{equation}
Note that $f_5(D,k_1^2)$ is regular in the origin. 

By inspecting the differential equation \itref{SUN4eq:equd1} 
one finds that the behaviour at $t=1$ of the 4 independent solution
is $\sim(1-t)^{\alpha_i}$, with $\alpha_1=D/2-1$, $\alpha_2=D/2$, 
$\alpha_3=0$, and $\alpha_4=1$; for comparison with \eqref{SUN4eq:vvff} 
the behaviours $\alpha_3=0$, and $\alpha_4=1$ are ruled out and the 
expansion reads 
\begin{equation}\label{SUN4eq:cond0} 
v_{5}(D,t)= (1-t)^{\frac{D}{2}-1}\left( v_5^{(0)}(D) 
                   + v_5^{(1)}(D)\ (1-t) +O(1-t)^2 \right) \;; 
\end{equation}
by comparison with \eqref{SUN4eq:svilf5} ($t=1$ corresponds to $k_1^2=0$), 
one obtains 
\begin{equation}\label{SUN4eq:cond1} 
\begin {split} 
v_5^{(0)}(D) = &\frac{1}{\Gamma\left(\frac{D}{2}\right)} f_5^{(0)}(D) \;, \\ 
v_5^{(1)}(D) = &\frac{1}{\Gamma\left(\frac{D}{2}\right)} 
         \left[ \frac{D}{2}f_5^{(0)}(D) + f_5^{(1)}(D) \right] \;.
\end{split}
\end{equation} 
The values $I_{4}(D,n)$ of $I_{4}(D,n,p^2)$ at $p^2=-1$ are therefore 
required 
\begin{equation}\label{SUN4eq:defI4b} 
I_{4}(D,n)\equiv
I_{4}(D,n,p^2)=
\pi^{-3D/2} \int \frac{d^Dk_2\  d^Dk_3 \ d^Dk_4} 
      {(k_2^2+1)^n{\mathbb{D}}_3{\mathbb{D}}_4{\mathbb{D}}_5 }, \quad p^2=-1
      \ . 
\end{equation} 
The problem of evaluating the $I_{4}(D,n)$ is similar to the 
original problem of evaluating the $I_{5}(D,n)$, but in fact it 
is much simpler, as the $I_{4}(D,n)$ involve one less loop and one 
less propagator. 
As above, by using integration-by-parts identities one finds that 
$I_{4}(D,n)$ satisfies the third-order difference equation 
\begin{eqnarray} 
 && 6(n-1)(n-2)(n-3)I_{4}(D,n) \nonumber\\ 
 && -(n-2)(n-3)(10n-7D-10)I_{4}(D,n-1) \nonumber\\ 
 && +(n-3)(2n^2+(2D-18)n-7D^2+29D-8)I_{4}(D,n-2) \nonumber\\ 
 && +(n-D-1)(n-2D+1)(2n-3D)I_{4}(D,n-3)=0\;.
\label{SUN4eq:dif2}
\end{eqnarray} 
We solve the difference equation by using again the Laplace {\it ansatz} 
\begin{equation}\label{SUN4eq:ansatz4} 
I_{4}(D,n)=\int_0^1 t^{n-1} v_{4}(D,t) \; dt\;,
\end{equation} 
where $v_{4}(D,t)$ satisfies the differential equation
\begin{eqnarray} 
 && 2t^3(3t+1)(t-1)^2  \frac{d^3}{dt^3}v_{4}(d,t) \nonumber\\ 
 && +t^2(t-1)(36t^2+(6-7D)t - 9D + 18 ) \frac{d^2}{dt^2}v_{4}(D,t) \nonumber\\
 && +t(36t^3-14Dt^2 +(-7D^2+33D-36)t + 13D^2- 61D + 72) 
                                        \frac{d}{dt}v_{4}(D,t) \nonumber\\
 && +(D-3)(2D-5)(3D-8) v_{4}(D,t) =0\;.
\label{SUN4eq:equd2} 
\end{eqnarray} 

Following the procedure used above, we write
\begin{eqnarray}
I_{4}(D,n)&=&\dfrac{1}{\Gamma\left(\frac{D}{2}\right)} 
  \int_0^\oo\dfrac{(k_2^2)^{D/2-1}\; dk_2^2}{(k_2^2+1)^n} f_{4}(k_2^2) \;, \\
\label{SUN4eq:deff3} 
f_{4}(k_2^2)&=&\int \dfrac{d\Omega(D,\hat k_2)}{\Omega(D)} I_{3}(D,(p-k_2)^2)
\;, \\
\label{SUN4eq:defI3}
I_{3}(D,(p-k_2)^2)&=& \pi^{-D} \int \frac{d^Dk_3 \ d^Dk_4} 
      { ((k_3-k_2)^2+1) (k_4-k_3)^2 (p-k_4)^2  }
      \ , \\
\label{SUN4eq:vvff2}
v_{4}(D,t)&=&\dfrac{1}{\Gamma\left(\frac{D}{2}\right)} 
           (1-t)^{\frac{D}{2}-1} t^{-\frac{D}{2}} 
            f_{4}\left(D,\frac{1-t}{t}\right) \; . 
\end{eqnarray} 
At variance with the previous case, the function $f_{4}(D,k_2^2)$ is 
\emph{not} regular for $k_2\to 0$, as at $k_2=0$ the value 
of the external momentum squared $(p-k_2)^2$ becomes 
the threshold of the 2-loop sunrise graph associated to $I_{3}(D,p^2)$. 
But it is not difficult to evaluate 
analytically $I_{3}(D,q^2)$ for generic off-shell $q^2$ by using 
Feynman parameters:
\begin{equation*}
I_{3}(D,q^2)
=\dfrac{2\Gamma(5-D)\Gamma\left(3-\frac{D}{2}\right) 
        \Gamma^2\left(\frac{D}{2}-1\right) } 
       {(D-4)^2(3-D)\Gamma\left(\frac{D}{2}\right) } 
    \,{}_2F_1\left(3-D,2-\frac{D}{2};\frac{D}{2};-q^2\right)\ ,
\end{equation*}
where ${}_2F_1$ is the Gauss hypergeometric function.
The expansion of $I_{3}(D,q^2)$ in $q^2=-1$ consists of the sum of two series, 
\begin{equation}\label{SUN4eq:I3svil} 
 I_{3}(D,q^2)=  a_0(D) \biggl[1+O(q^2+1)\biggr]
  \ \ \ +\ b_0(D)\ (q^2+1)^{2D-5} \biggl[1+O(q^2+1)\biggr] \ ,
\end{equation}
\begin{equation*}
\begin{split}
a_0(D)\ =\ I_{3}(D,-1)\ =&\ \frac{2\Gamma(5-D)\Gamma\left(3-\frac{D}{2}\right) 
                                \Gamma^2\left(\frac{D}{2}-1\right) 
                    \Gamma(2D-5)}{(4-D)^2(3-D)\Gamma\left(\frac{3}{2}D-3\right) 
                                  \Gamma(D-2)}\;,\\
b_0(D)\ =&\ \Gamma^2\left(\frac{D}{2}-1\right)\Gamma(5-2D)\ .
\end{split}
\end{equation*}
Inserting \eqref{SUN4eq:I3svil} into \eqref{SUN4eq:deff3}, setting
$q=p-k_2$ and  performing the angular integration over $\hat k_2$
by means of the formula (see Eq.(88) of Ref.\cite{Laporta:2001dd})
valid for $k_2\to 0 $
\begin{equation}\label{SUN4eq:hypsvil0}
\dfrac{1}{\Omega(D)}\int \dfrac{d\Omega(D,\hat k_2)}{((p-k_2)^2+1)^N} \approx 
(k_2^2)^{-\frac{N}{2}} 
\dfrac{\Gamma\left(\frac{D}{2}\right) 
       \Gamma\left(\frac{N}{2}\right)} 
     {2\Gamma(N)\Gamma\left(\frac{1}{2}(D-N)\right)}\ ,\quad k_2 \to 0 \ ; 
\end{equation}
with $N=-(2D-5)$, as in the term $(q^2+1)^{2D-5}$ of \eqref{SUN4eq:I3svil}, 
one obtains 
\begin{equation}\label{SUN4eq:f4svil}
f_{4}(D,k_2^2)= a_0(D)\ \biggl[1+O(k_2^2)\biggr] \ + \ 
  \dfrac{\Gamma\left(\frac{D}{2}\right) 
         \Gamma\left(\frac{5}{2}-D\right)} 
       {2\Gamma(5-2D)\Gamma\left(\frac{1}{2}(3D-5)\right)} \ 
       b_0(D)\ (k_2^2)^{\frac{1}{2}(2D-5)}
       \ \biggl[1+O(k_2^2)\biggr] \;.
\end{equation}
Using the variable $1/(k_2^2+1)=t$ in \eqref{SUN4eq:f4svil} and inserting it
in \eqref{SUN4eq:vvff2} one gets  
the initial condition for $v_{4}(D,t)$ at the singular point $t=1$
\begin{eqnarray}\label{SUN4eq:cond2} 
v_{4}(D,t)&=& \dfrac{a_0(D)}{\Gamma\left(\frac{D}{2}\right)} 
            (1-t)^{\frac{D}{2}-1} \biggl[1+O(1-t)\biggr] \nonumber \\
&&
 + \dfrac{\Gamma\left(\frac{5}{2}-D\right)} 
   {2\Gamma(5-2D)\Gamma\left(\frac{1}{2}(3D-5)\right)} b_0(D) 
   (1-t)^{\frac{1}{2}(3D-7)} \biggl[1+O(1-t)\biggr] 
\;.
\end{eqnarray}
By inspecting the equation \itref{SUN4eq:equd2} 
one gets that the behaviour at $t=0$ of $v_4(D,t)$  is 
\begin{equation}\label{SUN4eq:v4at0} 
v_{4}(D,t\to 0)\approx c_4^{(1)}(D)\ t^{-D+3} 
                      +c_4^{(2)}(D)\ t^{-2D+5} 
                      +c_4^{(3)}(D)\ t^{\frac{1}{2}(-3D+8)} \;,
\end{equation}
so that for $D\to 4$ the integral \itref{SUN4eq:ansatz4} is convergent 
for $n\ge4$. 

All the quantities depending on $D$ are then systematically expanded 
in $(D-4)$; and the series are truncated at some fixed number of 
terms.
We solve finally the differential equation \itref{SUN4eq:equd2} 
with the initial condition \itref{SUN4eq:cond2} 
by a first expansions in series at $t=1$; due to the presence 
in \eqref{SUN4eq:equd2} of a singular point at $t=-1/3$, 
in order to have fast convergence 
till $t=0$, we switch to the subsequent series expansions 
at the intermediate points $1/2$, $1/4$, $1/8$ and $0$; 
then we calculate the integral \itref{SUN4eq:ansatz4} 
for $n=4,5,6,7,8$ by integrating the series term by term 
By applying repeatedly top-down the recurrence relation \itref{SUN4eq:dif2} 
to $I_{4}(D,8)$, $I_{4}(D,7)$, $I_{4}(D,6)$,
we obtain $I_{4}(D,5)$ and $I_{4}(D,4)$, 
$I_{4}(D,3)$, $I_{4}(D,2)$ 
and $I_{4}(D,1)$. 
Those values of $I_4(D,n)$ are used to determine the initial condition
for $v_5(D,t)$, Eq.s(\ref{SUN4eq:cond0},\ref{SUN4eq:cond1},\ref{SUN4eq:svilf5}).
The game must be repeated again for $v_4(d,t)$.
In fact, we solve the differential equation \itref{SUN4eq:equd1}
by expansions in series centered in the points
$t=1$, $1/2$, $1/4$, $1/8$, $1/16$ and $0$
(as above, this subdivision is due to the presence of a singular point at $t=-1/8$).
By inspecting the equation \itref{SUN4eq:equd1}
one gets that the behaviour at $t=0$ of
$v_4(d,t)$  is 
\begin{equation}\label{SUN4eq:v5at0} 
v_{5}(D,t\to 0)\approx c_5^{(1)}(D) t^{-D+3} 
                      +c_5^{(2)}(D) t^{-2D+5} 
                      +c_5^{(3)}(D) t^{(-3D+8)/2}
                      +c_5^{(4)}(D) t^{(-5D+12)/2}\;,
\end{equation}
so that, when $D \to 4$, the integral \itref{SUN4eq:ansatz5} is surely convergent
for $n\ge5$; 
then we calculate the integral \itref{SUN4eq:ansatz5}
for $n=5,6,7,8,9$ by integrating the series term by term. 
By using repeatedly top-down the recurrence relation \itref{SUN4eq:dif2}
starting from $n=9$, we finally obtain $I_{5}(D,6)$, $I_{5}(D,5)$,
$I_5(D,4),{\ldots}$, $I_{5}(D,1)$. 
By taking into account the normalization \itref{SUN4eq:normal}
one finds complete agreement with the value at $x=1$ of 
solution of the differential equation computed in the previous section.

\section{Conclusions}

The evaluation of multiloop Feynman diagrams in the last 
years has received a strong boost, thanks to the ability of turning
generic properties of scalar integrals in dimensional regularization  
into tools for computing them.
Integration-by-parts, Lorentz invariance, and kinematic symmetries
have been exploited to establish infinite sets of relations
among integrals sharing (partially) common integrands.
The Laporta algorithm systematizes the by now standard reduction
to Master Integrals, that is the algebraic procedure 
for expressing any Feynman integral as a linear combination 
of few basic integrals with the simplest integrands.

The completion of the computational task,
consisting in the actual evaluation of 
the Master Integrals can be as well afforded by employing the same 
set of identities.
In fact, by combining the differentiation of Master Integrals 
with respect to their Mandelstam invariants, 
and the reduction of the new born integrals, it is possible to derive
a system of non-homogeneous first order differential equations 
fulfilled by the Master Integrals themselves.

Solving such a system of differential equation amounts 
to the evaluation of the Master Integrals, alternatively to 
their direct parametric integration.

We have reviewed the method of differential equations by
its direct application, trying to follow a didactical path. 
We discussed the reduction algorithm plus
the general derivation of differential equations for Feynman
integrals. Successively, 
the calculation of Master Integrals in the context of the evaluation of the
one- and two-loop corrections to the photon propagator in QED;
whereas, in the last two sections, we presented two cases of less trivial
differential equations, to show more technical aspects 
related to the solution of homogeneous equations 
and to the choice of the boundary conditions.

In general, solving a system of first order differential equations 
for more than one Master Integral is equivalent to solving a higher order 
equation for a single Master Integral.
Despite to the lack of a theoretical procedure for solving  
differential equations in the most general case, 
Euler's variation of constants offers a viable procedure.
Accordingly, 
the solution of the non-homogeneous equation is obtained 
by quadrature, using as a kernel the Wronskian of the 
associated homogeneous equation -- whose solution can be preliminarily 
found by constants' variation as well.

The main achievement is the integration of the differential equation for exact
value of the parameters (Mandelstam invariants and dimensional-parameter).
When that is not possible, one can Laurent expand the equation, which then 
becomes a chained system of equations for the Laurent coefficients of 
the solution, suitable for a bottom-up solving algorithm, starting from
the lowest Laurent coefficient.

As a natural feature of Euler's variation of constants, 
the solution manifests an 
analytic integral representation,
generic of transcendental functions: a flexible nested structure of multiple
integrations (or equivalently, iterative summations) which benefits of the
shuffle algebras induced by the integration-by-parts.
Within this framework, the actual efforts required by the computation are
the finding out of the homogeneous solutions, and the definition
of new occurring functions, ordered according to their increasing 
trascendentality -- as required by the iterative fulfilment 
of the non-homogeneous equation.

Boundary conditions are found by imposing the regularity or the 
finiteness of the solution at the pseudo-thresholds of the Master 
Integrals. This qualitative information is sufficient for the quantitative 
determination of the arbitrary integration constants.
At the diagrammatic level, boundary conditions usually correspond to integrals
related to simpler diagrams.

The use of Differential Equation in the external invariants is
a very powerful tool for computing Feynman (master) integrals.
Dimensional regularization was fundamental for the derivation of the 
differential equations we discussed in this review. \\

In principle differential identities for integral functions 
can be derived whenever
it is allowed by the algebra of the integral representation under use - 
as induced by integration-by-parts.
And their use is not limited to the perturbative description of  
Feynman diagrams.
Therefore we like to conclude by remarking that the use of 
Differential Equations
for integrals' evaluation, is not just a technique, but a 
{\it point of view} from which any integral is
seen under a new {\it perspective},
where there appear, explicitly exposed, its analytic structure, its
singularities which finally determine its value.

\section{Acknowledgment}

We wish to thank Roberto Bonciani, Thomas Gehrmann, and Ettore Remiddi,
for their careful reading and comments on the manuscript,
but especially for their invaluable collaboration during these years.
The research of PM was supported by Marie-Curie-EIF under the contract 
MEIF-CT-2006-024178. 
MA wishes to anknowledge the Institute for Theoretical Physics of Z\"urich
for the kind hospitality while part of this work was performed, with the 
support of the abovementioned Marie-Curie-EIF founds.


\begin{thebibliography}{000} 




\bibitem{Kotikov:1990kg}
  A.~V.~Kotikov,
  %
  Phys.\ Lett.\ B {\bf 254} (1991) 158.


\bibitem{Boos:1990rg}
  E.~E.~Boos and A.~I.~Davydychev,
  %
  Theor.\ Math.\ Phys.\  {\bf 89} (1991) 1052
  [Teor.\ Mat.\ Fiz.\  {\bf 89} (1991) 56].


\bibitem{Kotikov:1991hy}
  A.~V.~Kotikov,
  %
ITF-91-26E

\bibitem{Kotikov:1991hm}
  A.~V.~Kotikov,
  %
  Phys.\ Lett.\ B {\bf 259} (1991) 314.

\bibitem{Lunev:1994sz}
  F.~A.~Lunev,
  %
  Phys.\ Rev.\ D {\bf 50} (1994) 6589
  [arXiv:hep-th/9407174].

\bibitem{Scharf:1994}
 R. Scharf, PhD theis, Wurzburg 1994.

\bibitem{Remiddi:1997ny}
  E.~Remiddi,
  %
  Nuovo Cim.\ A {\bf 110} (1997) 1435
  [arXiv:hep-th/9711188].

\bibitem{Gehrmann:1999as}
  T.~Gehrmann and E.~Remiddi,
  Nucl.\ Phys.\ B {\bf 580}, 485 (2000)
  [arXiv:hep-ph/9912329].

\bibitem{Gehrmann:2000zt}
  T.~Gehrmann and E.~Remiddi,
  Nucl.\ Phys.\ B {\bf 601}, 248 (2001)
  [arXiv:hep-ph/0008287].

\bibitem{Gehrmann:2001ck}
  T.~Gehrmann and E.~Remiddi,
  Nucl.\ Phys.\ B {\bf 601}, 287 (2001)
  [arXiv:hep-ph/0101124].



\bibitem{Ford:1991hw}
  C.~Ford and D.~R.~T.~Jones, 
  %
  Phys.\ Lett.\ B {\bf 274} (1992) 409
  [Erratum-ibid.\ B {\bf 285} (1992) 399].


\bibitem{Caffo:1998yd}
  M.~Caffo, H.~Czyz, S.~Laporta and E.~Remiddi,
  %
  Acta Phys.\ Polon.\ B {\bf 29}, 2627 (1998)
  [arXiv:hep-th/9807119].

\bibitem{Caffo:1998du}
  M.~Caffo, H.~Czyz, S.~Laporta and E.~Remiddi,
  Nuovo Cim.\ A {\bf 111}, 365 (1998)
  [arXiv:hep-th/9805118].

\bibitem{Bonciani:1999mj}
  R.~Bonciani,
  %
  Acta Phys.\ Polon.\ B {\bf 30} (1999) 3463.


\bibitem{Kotikov:2000ye}
  A.~V.~Kotikov,
  arXiv:hep-ph/0102178.

\bibitem{Czachor:2001mv}
  M.~Czachor and H.~Czyz,
  Acta Phys.\ Polon.\ B {\bf 32}, 3823 (2001)
  [arXiv:hep-ph/0110351].

\bibitem{Caffo:2002we}
  M.~Caffo, H.~Czyz and E.~Remiddi,
  Nucl.\ Instrum.\ Meth.\ A {\bf 502}, 613 (2003)
  [arXiv:hep-ph/0211171].

\bibitem{Caffo:2002wm}
  M.~Caffo, H.~Czyz and E.~Remiddi,
  Nucl.\ Phys.\ Proc.\ Suppl.\  {\bf 116}, 422 (2003)
  [arXiv:hep-ph/0211178].

\bibitem{Mastrolia:2002gt}
  P.~Mastrolia and E.~Remiddi,
  Nucl.\ Phys.\ Proc.\ Suppl.\  {\bf 116}, 412 (2003)
  [arXiv:hep-ph/0211210].

\bibitem{Aglietti:2004vs}
  U.~Aglietti,
  arXiv:hep-ph/0408014.

\bibitem{Bonciani:2004dz}
  R.~Bonciani,
  Acta Phys.\ Polon.\ B {\bf 35}, 2587 (2004)
  [arXiv:hep-ph/0410210].


\bibitem{Smirnov:2006ry}
  V.~A.~Smirnov,
  ``Feynman integral calculus,''
{\it  Berlin, Germany: Springer (2006) 283 p}



\bibitem{Caffo:2003gh}
  M.~Caffo,
  Acta Phys.\ Polon.\ B {\bf 34}, 5357 (2003)
  [arXiv:hep-ph/0311052].

\bibitem{Caffo:2003ma}
  M.~Caffo, H.~Czyz, A.~Grzelinska and E.~Remiddi,
  Nucl.\ Phys.\ B {\bf 681}, 230 (2004)
  [arXiv:hep-ph/0312189].




\bibitem{Gracey:1992ew}
  J.~A.~Gracey,
  Phys.\ Lett.\ B {\bf 277} (1992) 469.

\bibitem{Tarasov:1996br}{
  O.~V.~Tarasov,
  Phys.\ Rev.\  D {\bf 54} (1996) 6479
  [arXiv:hep-th/9606018]; \\
  Nucl.\ Phys.\ Proc.\ Suppl.\  {\bf 89} (2000) 237
  [arXiv:hep-ph/0102271].
}

\bibitem{Tarasov:1997kx}
  O.~V.~Tarasov,
  Nucl.\ Phys.\ B {\bf 502} (1997) 455
  [arXiv:hep-ph/9703319].

\bibitem{Fleischer:1997bw}
  J.~Fleischer, A.~V.~Kotikov and O.~L.~Veretin,
  Phys.\ Lett.\ B {\bf 417} (1998) 163
  [arXiv:hep-ph/9707492].

\bibitem{Berends:1997vk}
  F.~A.~Berends, A.~I.~Davydychev and N.~I.~Ussyukina,
  Phys.\ Lett.\ B {\bf 426} (1998) 95
  [arXiv:hep-ph/9712209].

\bibitem{Chung:1998mz}
  J.~M.~Chung and B.~K.~Chung,
  Phys.\ Rev.\ D {\bf 59} (1999) 105014
  [arXiv:hep-ph/9805432].


\bibitem{Fleischer:1998nb}
  J.~Fleischer, A.~V.~Kotikov and O.~L.~Veretin,
  Nucl.\ Phys.\ B {\bf 547}, 343 (1999)
  [arXiv:hep-ph/9808242].


\bibitem{Davydychev:1999ic}
  A.~I.~Davydychev and V.~A.~Smirnov,
  Nucl.\ Phys.\ B {\bf 554} (1999) 391
  [arXiv:hep-ph/9903328].

\bibitem{Fleischer:1999aa}
  J.~Fleischer, M.~Y.~Kalmykov and A.~V.~Kotikov,
  arXiv:hep-ph/9905379.

\bibitem{Fleischer:1999hp}
  J.~Fleischer, M.~Y.~Kalmykov and A.~V.~Kotikov,
  Phys.\ Lett.\ B {\bf 462} (1999) 169
  [arXiv:hep-ph/9905249].

\bibitem{Fleischer:1999tu}
  J.~Fleischer and M.~Y.~Kalmykov,
  Comput.\ Phys.\ Commun.\  {\bf 128} (2000) 531
  [arXiv:hep-ph/9907431].


\bibitem{Fleischer:1999mp}
  J.~Fleischer and M.~Y.~Kalmykov,
  Phys.\ Lett.\ B {\bf 470} (1999) 168
  [arXiv:hep-ph/9910223].

\bibitem{Fleischer:2000vb}
  J.~Fleischer, O.~V.~Tarasov and M.~Tentyukov,
  Nucl.\ Phys.\ Proc.\ Suppl.\  {\bf 89} (2000) 112.


\bibitem{Smirnov:2002kq}
  V.~A.~Smirnov,
  arXiv:hep-ph/0209177.

\bibitem{Smirnov:2002je}
  V.~A.~Smirnov,
  Nucl.\ Phys.\ Proc.\ Suppl.\  {\bf 116}, 417 (2003)
  [arXiv:hep-ph/0209295].


\bibitem{Onishchenko:2002ri}
  A.~Onishchenko and O.~Veretin,
  Phys.\ Atom.\ Nucl.\  {\bf 68} (2005) 1405
  [Yad.\ Fiz.\  {\bf 68} (2005) 1461]
  [arXiv:hep-ph/0207091].


\bibitem{Martin:2003qz}
  S.~P.~Martin,
  Phys.\ Rev.\ D {\bf 68}, 075002 (2003)
  [arXiv:hep-ph/0307101].


\bibitem{Remiddi:2003ci}
  E.~Remiddi,
  %
  Acta Phys.\ Polon.\ B {\bf 34} (2003) 5311
  [arXiv:hep-ph/0310332].


\bibitem{Remiddi:2004kv}
  E.~Remiddi,
  %
  Nucl.\ Phys.\ Proc.\ Suppl.\  {\bf 135} (2004) 247.


\bibitem{Actis:2004bp}
  S.~Actis, A.~Ferroglia, G.~Passarino, M.~Passera and S.~Uccirati,
  Nucl.\ Phys.\ B {\bf 703}, 3 (2004)
  [arXiv:hep-ph/0402132].


\bibitem{Birthwright:2004kk}
  T.~G.~Birthwright, E.~W.~N.~Glover and P.~Marquard,
  JHEP {\bf 0409}, 042 (2004)
  [arXiv:hep-ph/0407343].


\bibitem{Kniehl:2005yc}
  B.~A.~Kniehl and A.~V.~Kotikov,
  arXiv:hep-ph/0508238.


\bibitem{Bogdan:2005rg}
  A.~V.~Bogdan and R.~N.~Lee,
  Nucl.\ Phys.\ B {\bf 732}, 169 (2006)
  [arXiv:hep-ph/0509181].


\bibitem{Kniehl:2005bc}
  B.~A.~Kniehl, A.~V.~Kotikov, A.~Onishchenko and O.~Veretin,
  Nucl.\ Phys.\ B {\bf 738}, 306 (2006)
  [arXiv:hep-ph/0510235].


\bibitem{Tarasov:2006nk}
  O.~V.~Tarasov,
  Phys.\ Lett.\ B {\bf 638} (2006) 195
  [arXiv:hep-ph/0603227].





\bibitem{Anastasiou:2000mf}
  C.~Anastasiou, T.~Gehrmann, C.~Oleari, E.~Remiddi and J.~B.~Tausk,
  Nucl.\ Phys.\ B {\bf 580}, 577 (2000)
  [arXiv:hep-ph/0003261].

\bibitem{Gehrmann:2000xj}
  T.~Gehrmann and E.~Remiddi,
  Nucl.\ Phys.\ Proc.\ Suppl.\  {\bf 89}, 251 (2000)
  [arXiv:hep-ph/0005232].


\bibitem{Smirnov:2000ie}
  V.~A.~Smirnov,
  Phys.\ Lett.\ B {\bf 500}, 330 (2001)
  [arXiv:hep-ph/0011056].


\bibitem{Gehrmann:2001ih}
  T.~Gehrmann and E.~Remiddi,
in {\it Proc. of the 5th International Symposium on Radiative Corrections (RADCOR 2000) } ed. Howard E. Haber,
  arXiv:hep-ph/0101147.

\bibitem{Gehrmann:2001ru}
  T.~Gehrmann,
  arXiv:hep-ph/0107090.


\bibitem{Giele:2002hx}
  W.~Giele {\it et al.},
  arXiv:hep-ph/0204316.

\bibitem{Frixione:2002kn}
  S.~Frixione,
  Nucl.\ Phys.\ Proc.\ Suppl.\  {\bf 117}, 222 (2003)
  [arXiv:hep-ph/0211434].


\bibitem{allonshell}{
J.~B.~Tausk,
Phys.\ Lett.\  B {\bf 469} (1999) 225
[arXiv:hep-ph/9909506];\\
Z.~Bern, L.J.~Dixon and A.~Ghinculov, Phys.\ Rev.\ D {\bf 63} (2001)
053007 [hep-ph/0010075];\\
C.\ Anastasiou, E.W.N.~Glover, C.\ Oleari and M.E.\ Tejeda-Yeomans,
Nucl.\ Phys.\ B~{\bf 601}~(2001) 318~[hep-ph/0010212];~{\bf 601}~(2001)~347 [hep-ph/0011094];
 {\bf 605} (2001) 486 [hep-ph/0101304];\\
E.W.N.~Glover, C.~Oleari and M.E.~Tejeda-Yeomans,
Nucl.\ Phys.\ {\bf 605} (2001) 467 [hep-ph/0102201];\\
C.~Anastasiou, E.W.N.~Glover and M.E.~Tejeda-Yeomans,
Nucl.\ Phys.\ B {\bf 629} (2002) 255 [hep-ph/0201274];\\
E.W.N.~Glover and M.E.~Tejeda-Yeomans,
JHEP {\bf 0306} (2003) 033
[hep-ph/0304169];\\
E.W.N.~Glover,
JHEP {\bf 0404} (2004) 021
[hep-ph/0401119];\\
Z.~Bern, A.~De Freitas and L.J.~Dixon,
JHEP {\bf 0109} (2001) 037 [hep-ph/0109078];
JHEP {\bf 0203} (2002) 018 [hep-ph/0201161];
JHEP {\bf 0306} (2003) 028
[hep-ph/0304168];\\
A.~De Freitas and Z.~Bern,
JHEP {\bf 0409} (2004) 039
[hep-ph/0409007];\\
Z.~Bern, A.~De Freitas, L.J.~Dixon, A.~Ghinculov and H.L.~Wong,
JHEP {\bf 0111} (2001) 031 [hep-ph/0109079];\\
T.~Binoth, E.W.N.~Glover, P.~Marquard and J.J.~van der Bij,
JHEP {\bf 0205} (2002) 060
[hep-ph/0202266];\\
}

\bibitem{oneoffshell}{
L.W.~Garland, T.~Gehrmann, E.W.N.~Glover, A.~Koukoutsakis and E.~Remiddi,
Nucl.\ Phys.\ B {\bf 627} (2002) 107 [hep-ph/0112081] and
{\bf 642} (2002) 227 [hep-ph/0206067].
T.~Gehrmann and E.~Remiddi,
Nucl.\ Phys.\ B {\bf 640} (2002) 379
[hep-ph/0207020].\\
}





\bibitem{Buttar:2006zd}
  C.~Buttar {\it et al.},
  arXiv:hep-ph/0604120.

\bibitem{Davydychev:1996fn}
  A.~I.~Davydychev,
  Acta Phys.\ Polon.\ B {\bf 28} (1997) 841
  [arXiv:hep-ph/9610510].

\bibitem{Anastasiou:1999ui}
  C.~Anastasiou, E.~W.~N.~Glover and C.~Oleari,
  Nucl.\ Phys.\ B {\bf 572} (2000) 307
  [arXiv:hep-ph/9907494].


\bibitem{Kotikov:2000yd}
  A.~V.~Kotikov,
  arXiv:hep-ph/0102177.



\bibitem{Passarino:2001wv}{
  G.~Passarino,
  Nucl.\ Phys.\ B {\bf 619}, 257 (2001)
  [arXiv:hep-ph/0108252]; \\
 G.~Passarino and S.~Uccirati,
 Nucl.\ Phys.\  B {\bf 629}, 97 (2002)
 [arXiv:hep-ph/0112004];\\
 A.~Ferroglia, M.~Passera, G.~Passarino and S.~Uccirati,
 Nucl.\ Phys.\  B {\bf 680}, 199 (2004)
 [arXiv:hep-ph/0311186]; \\
 S.~Actis, A.~Ferroglia, G.~Passarino, M.~Passera and S.~Uccirati,
 Nucl.\ Phys.\  B {\bf 703}, 3 (2004)
 [arXiv:hep-ph/0402132] \\
 G.~Passarino and S.~Uccirati,
 Nucl.\ Phys.\  B {\bf 747}, 113 (2006)
 [arXiv:hep-ph/0603121].
}

\bibitem{Kotikov:2001sd}
  A.~V.~Kotikov,
  arXiv:hep-ph/0112347.


\bibitem{Laporta:2001dd}
  S.~Laporta,
  Int.\ J.\ Mod.\ Phys.\ A {\bf 15}, 5087 (2000)
  [arXiv:hep-ph/0102033].

\bibitem{Laporta:2003jz}
  S.~Laporta, 
  Acta Phys.\ Polon.\  B {\bf 34} (2003) 5323
  [arXiv:hep-ph/0311065].


\bibitem{Laporta:2002pg}
  S.~Laporta,
  Phys.\ Lett.\ B {\bf 549}, 115 (2002)
  [arXiv:hep-ph/0210336].


\bibitem{Ferroglia:2002mz}
  A.~Ferroglia, M.~Passera, G.~Passarino and S.~Uccirati,
  Nucl.\ Phys.\ B {\bf 650} (2003) 162
  [arXiv:hep-ph/0209219].


\bibitem{Suzuki:2003jn}
  A.~T.~Suzuki, E.~S.~Santos and A.~G.~M.~Schmidt,
  J.\ Phys.\ A {\bf 36} (2003) 11859
  [arXiv:hep-ph/0309080].


\bibitem{Smirnov:2004ip}
  V.~A.~Smirnov,
  Nucl.\ Phys.\ Proc.\ Suppl.\  {\bf 135}, 252 (2004)
  [arXiv:hep-ph/0406052].


\bibitem{Binoth:2000ps}
  T.~Binoth and G.~Heinrich,
  Nucl.\ Phys.\  B {\bf 585} (2000) 741
  [arXiv:hep-ph/0004013].

\bibitem{Binoth:2003ak}
  T.~Binoth and G.~Heinrich,
  Nucl.\ Phys.\  B {\bf 680} (2004) 375
  [arXiv:hep-ph/0305234].

\bibitem{Duplancic:2003tv}
  G.~Duplancic and B.~Nizic,
  Eur.\ Phys.\ J.\  C {\bf 35}, 105 (2004)
  [arXiv:hep-ph/0303184].



\bibitem{Czakon:2005rk}
  M.~Czakon,
  arXiv:hep-ph/0511200.


\bibitem{Anastasiou:2005cb}
  C.~Anastasiou and A.~Daleo,
  arXiv:hep-ph/0511176.


\bibitem{Smirnov:2006vt}
  V.~A.~Smirnov,
  Nucl.\ Phys.\ Proc.\ Suppl.\  {\bf 157}, 131 (2006)
  [arXiv:hep-ph/0601268].


\bibitem{Weinzierl:2006qs}
  S.~Weinzierl,
  arXiv:hep-ph/0604068.

\bibitem{Grozin:2003ak}
  A.~G.~Grozin,
  Int.\ J.\ Mod.\ Phys.\  A {\bf 19} (2004) 473
  [arXiv:hep-ph/0307297].

\bibitem{Smirnov:2004ym}
  V.~A.~Smirnov,
  Springer Tracts Mod.\ Phys.\  {\bf 211} (2004) 1.



\bibitem{Bonciani:2003te}
  R.~Bonciani, P.~Mastrolia and E.~Remiddi,
  Nucl.\ Phys.\ B {\bf 661}, 289 (2003)
  [Erratum-ibid.\ B {\bf 702}, 359 (2004)]
  [arXiv:hep-ph/0301170].

\bibitem{Bonciani:2003ai}
  R.~Bonciani, P.~Mastrolia and E.~Remiddi,
  Nucl.\ Phys.\ B {\bf 676}, 399 (2004)
  [arXiv:hep-ph/0307295].

\bibitem{Bonciani:2004xp}
  R.~Bonciani,
  Nucl.\ Phys.\ Proc.\ Suppl.\  {\bf 152}, 168 (2006)
  [arXiv:hep-ph/0410092].




\bibitem{Bonciani:2003hc}
  R.~Bonciani, P.~Mastrolia and E.~Remiddi,
  Nucl.\ Phys.\ B {\bf 690}, 138 (2004)
  [arXiv:hep-ph/0311145].


\bibitem{Bernreuther:2004ih}
  W.~Bernreuther, R.~Bonciani, T.~Gehrmann, R.~Heinesch, T.~Leineweber, P.~Mastrolia and E.~Remiddi,
  Nucl.\ Phys.\ B {\bf 706}, 245 (2005)
  [arXiv:hep-ph/0406046].

\bibitem{Bernreuther:2004th}
  W.~Bernreuther, R.~Bonciani, T.~Gehrmann, R.~Heinesch, T.~Leineweber, P.~Mastrolia and E.~Remiddi,
  Nucl.\ Phys.\ B {\bf 712}, 229 (2005)
  [arXiv:hep-ph/0412259].


\bibitem{Bernreuther:2005rw}
  W.~Bernreuther, R.~Bonciani, T.~Gehrmann, R.~Heinesch, T.~Leineweber and E.~Remiddi,
  Nucl.\ Phys.\ B {\bf 723}, 91 (2005)
  [arXiv:hep-ph/0504190].


\bibitem{Bernreuther:2005gq}
  W.~Bernreuther, R.~Bonciani, T.~Gehrmann, R.~Heinesch, T.~Leineweber, P.~Mastrolia and E.~Remiddi,
  %
  Phys.\ Rev.\ Lett.\  {\bf 95} (2005) 261802
  [arXiv:hep-ph/0509341].





\bibitem{Kotikov:2001ct}
  A.~V.~Kotikov, A.~V.~Lipatov, G.~Parente and N.~P.~Zotov,
  Eur.\ Phys.\ J.\ C {\bf 26} (2002) 51
  [arXiv:hep-ph/0107135].


\bibitem{Kotikov:2002ab}
  A.~V.~Kotikov and L.~N.~Lipatov,
  Nucl.\ Phys.\ B {\bf 661} (2003) 19
  [Erratum-ibid.\ B {\bf 685} (2004) 405]
  [arXiv:hep-ph/0208220].


\bibitem{Jegerlehner:2003py}
  F.~Jegerlehner and M.~Y.~Kalmykov,
  Nucl.\ Phys.\ B {\bf 676} (2004) 365
  [arXiv:hep-ph/0308216].


\bibitem{Anastasiou:2003ds}
  C.~Anastasiou, L.~J.~Dixon, K.~Melnikov and F.~Petriello,
  Phys.\ Rev.\ D {\bf 69} (2004) 094008
  [arXiv:hep-ph/0312266].


\bibitem{Chetyrkin:2004fq}
  K.~G.~Chetyrkin, J.~H.~Kuhn, P.~Mastrolia and C.~Sturm,
  %
  Eur.\ Phys.\ J.\ C {\bf 40} (2005) 361
  [arXiv:hep-ph/0412055].

\bibitem{Schroder:2005va}
  Y.~Schroder and A.~Vuorinen,
  JHEP {\bf 0506}, 051 (2005)
  [arXiv:hep-ph/0503209].

\bibitem{Faisst:2006sr}
  M.~Faisst, P.~Maierhoefer and C.~Sturm,
  arXiv:hep-ph/0611244.






\bibitem{Neubert:1993mb}
  M.~Neubert,
  Phys.\ Rept.\  {\bf 245} (1994) 259
  [arXiv:hep-ph/9306320].


\bibitem{Juste:2006sv}
  A.~Juste {\it et al.},
  arXiv:hep-ph/0601112.


\bibitem{Bernreuther:2006yt}
  W.~Bernreuther, R.~Bonciani, T.~Gehrmann, R.~Heinesch, T.~Leineweber, P.~Mastrolia and E.~Remiddi,
  PoS {\bf HEP2005}, 229 (2006)
  [arXiv:hep-ph/0601207].

\bibitem{Bernreuther:2006vp}
  W.~Bernreuther, R.~Bonciani, T.~Gehrmann, R.~Heinesch, T.~Leineweber, P.~Mastrolia and E.~Remiddi,
  %
  arXiv:hep-ph/0604031.


\bibitem{Bonciani:2006eu}
  R.~Bonciani,
  arXiv:hep-ph/0607037.





\bibitem{Aglietti:2003yc}
  U.~Aglietti and R.~Bonciani,
  Nucl.\ Phys.\ B {\bf 668}, 3 (2003)
  [arXiv:hep-ph/0304028].

\bibitem{Aglietti:2004tq}
  U.~Aglietti and R.~Bonciani,
  Nucl.\ Phys.\ B {\bf 698}, 277 (2004)
  [arXiv:hep-ph/0401193].

\bibitem{Aglietti:2007as}
  U.~Aglietti, R.~Bonciani, L.~Grassi and E.~Remiddi,
  arXiv:0705.2616 [hep-ph].



\bibitem{Maher:1993vj}
  P.~N.~Maher, L.~Durand and K.~Riesselmann,
  Phys.\ Rev.\ D {\bf 48} (1993) 1061
  [Erratum-ibid.\ D {\bf 52} (1995) 553]
  [arXiv:hep-ph/9303233].



\bibitem{Aglietti:2004nj}
  U.~Aglietti, R.~Bonciani, G.~Degrassi and A.~Vicini,
  Phys.\ Lett.\ B {\bf 595}, 432 (2004)
  [arXiv:hep-ph/0404071].

\bibitem{Aglietti:2004ki}
  U.~Aglietti, R.~Bonciani, G.~Degrassi and A.~Vicini,
  Phys.\ Lett.\ B {\bf 600}, 57 (2004)
  [arXiv:hep-ph/0407162].

\bibitem{Bernreuther:2005gw}
  W.~Bernreuther, R.~Bonciani, T.~Gehrmann, R.~Heinesch, P.~Mastrolia and E.~Remiddi,
  Phys.\ Rev.\ D {\bf 72}, 096002 (2005)
  [arXiv:hep-ph/0508254].

\bibitem{Anastasiou:2006hc}
  C.~Anastasiou, S.~Beerli, S.~Bucherer, A.~Daleo and Z.~Kunszt,
  arXiv:hep-ph/0611236.

\bibitem{Aglietti:2006tp}
  U.~Aglietti, R.~Bonciani, G.~Degrassi and A.~Vicini,
  arXiv:hep-ph/0611266.





\bibitem{Smirnov:2001cm}
  V.~A.~Smirnov,
  Phys.\ Lett.\ B {\bf 524}, 129 (2002)
  [arXiv:hep-ph/0111160].

\bibitem{Bonciani:2003cj}
  R.~Bonciani, A.~Ferroglia, P.~Mastrolia, E.~Remiddi and J.~J.~van der Bij,
  Nucl.\ Phys.\ B {\bf 681}, 261 (2004)
  [Erratum-ibid.\ B {\bf 702}, 364 (2004)]
  [arXiv:hep-ph/0310333].

\bibitem{Bonciani:2004gi}
  R.~Bonciani, A.~Ferroglia, P.~Mastrolia, E.~Remiddi and J.~J.~van der Bij,
  Nucl.\ Phys.\ B {\bf 701}, 121 (2004)
  [arXiv:hep-ph/0405275].

\bibitem{Heinrich:2004iq}
  G.~Heinrich and V.~A.~Smirnov,
  Phys.\ Lett.\ B {\bf 598}, 55 (2004)
  [arXiv:hep-ph/0406053].


\bibitem{Bonciani:2004qt}
  R.~Bonciani, A.~Ferroglia, P.~Mastrolia, E.~Remiddi and J.~J.~van der Bij,
  Nucl.\ Phys.\ B {\bf 716}, 280 (2005)
  [arXiv:hep-ph/0411321].

\bibitem{Czakon:2004wm}
  M.~Czakon, J.~Gluza and T.~Riemann,
  Phys.\ Rev.\ D {\bf 71}, 073009 (2005)
  [arXiv:hep-ph/0412164].

\bibitem{Bonciani:2005im}
  R.~Bonciani and A.~Ferroglia,
  Phys.\ Rev.\ D {\bf 72}, 056004 (2005)
  [arXiv:hep-ph/0507047].

\bibitem{Bonciani:2006qu}
  R.~Bonciani and A.~Ferroglia,
  Nucl.\ Phys.\ Proc.\ Suppl.\  {\bf 157}, 11 (2006)
  [arXiv:hep-ph/0601246].

\bibitem{Czakon:2006hb}
  M.~Czakon, J.~Gluza, K.~Kajda and T.~Riemann,
  Nucl.\ Phys.\ Proc.\ Suppl.\  {\bf 157}, 16 (2006)
  [arXiv:hep-ph/0602102].






\bibitem{Seidel:2004jh}
  D.~Seidel,
  Phys.\ Rev.\ D {\bf 70}, 094038 (2004)
  [arXiv:hep-ph/0403185].

\bibitem{Melnikov:2005bx}
 K.~Melnikov and A.~Mitov,
 Phys.\ Lett.\  B {\bf 620}, 69 (2005)
 [arXiv:hep-ph/0505097].

\bibitem{Asatrian:2006sm}
  H.~M.~Asatrian, T.~Ewerth, A.~Ferroglia, P.~Gambino and C.~Greub,
  Nucl.\ Phys.\  B {\bf 762} (2007) 212
  [arXiv:hep-ph/0607316].





\bibitem{DeFazio:2000up}
  F.~De Fazio,
  arXiv:hep-ph/0010007.


\bibitem{Jegerlehner:2002em}
  F.~Jegerlehner, M.~Y.~Kalmykov and O.~Veretin,
  Nucl.\ Phys.\ B {\bf 658} (2003) 49
  [arXiv:hep-ph/0212319].



\bibitem{Awramik:2004ge}
  M.~Awramik, M.~Czakon, A.~Freitas and G.~Weiglein,
  Phys.\ Rev.\ Lett.\  {\bf 93}, 201805 (2004)
  [arXiv:hep-ph/0407317].

\bibitem{Awramik:2004qv}
  M.~Awramik, M.~Czakon, A.~Freitas and G.~Weiglein,
  Nucl.\ Phys.\ Proc.\ Suppl.\  {\bf 135}, 119 (2004)
  [arXiv:hep-ph/0408207].

\bibitem{Freitas:2005vb}
  A.~Freitas, M.~Awramik and M.~Czakon,
  arXiv:hep-ph/0507159.

\bibitem{Awramik:2006ar}
  M.~Awramik, M.~Czakon and A.~Freitas,
  arXiv:hep-ph/0605339.






\bibitem{Broadhurst:1998ke}
  D.~J.~Broadhurst,
  arXiv:hep-th/9806174.


\bibitem{Davydychev:1998fk}
  A.~I.~Davydychev and R.~Delbourgo,
  Acta Phys.\ Polon.\ B {\bf 29}, 2891 (1998)
  [arXiv:hep-th/9806248].



\bibitem{HPL}{
Remiddi, E. and Vermaseren, J. A. M.,
Int. J. Mod. Phys. A {\bf 15} (2000) 725-754 
[hep-ph/9905237].

Gehrmann, T. and Remiddi, E.,
Comput. Phys. Commun. {\bf 141} (2001) 296 
[hep-ph/0107173] ;
%
Comput. Phys. Commun. {\bf 144} (2002) 200 
[hep-ph/0111255];
%
Nucl. Phys. B {\bf 640} (2002) 379 
[hep-ph/0207020].
}

\bibitem{Maitre:2005uu}
  D.~Maitre,
  Comput.\ Phys.\ Commun.\  {\bf 174} (2006) 222
  [arXiv:hep-ph/0507152].

\bibitem{Maitre:2007kp}
  D.~Maitre,
  arXiv:hep-ph/0703052.



\bibitem{Moch:2001zr}
Moch, S. and Uwer, P. and Weinzierl, S.,
J. Math. Phys. {\bf 43} (2002) 3363 
[hep-ph/0110083].

\bibitem{Weinzierl:2002hv}
Weinzierl, S.,
Comput. Phys. Commun. {\bf 145} (2002) 357
[math-ph/0201011]

\bibitem{Blumlein:2003gb}
Blumlein, J.,
(2003) hep-ph/0311046.

\bibitem{Kalmykov:2004kg}
  M.~Y.~Kalmykov,
  Nucl.\ Phys.\ Proc.\ Suppl.\  {\bf 135}, 280 (2004)
  [arXiv:hep-th/0406269].

\bibitem{Kalmykov:2006hu}
  M.~Y.~Kalmykov, B.~F.~L.~Ward and S.~Yost,
  arXiv:hep-th/0612240.



\bibitem{Dunne:2003tr}
  G.~V.~Dunne,
  JHEP {\bf 0402}, 013 (2004)
  [arXiv:hep-th/0311167].

\bibitem{Martin:2003it}
  S.~P.~Martin,
  Phys.\ Rev.\ D {\bf 70}, 016005 (2004)
  [arXiv:hep-ph/0312092].

\bibitem{Martin:2005qm}
  S.~P.~Martin and D.~G.~Robertson,
  Comput.\ Phys.\ Commun.\  {\bf 174}, 133 (2006)
  [arXiv:hep-ph/0501132].

\bibitem{Dunne:2006sx}
  G.~V.~Dunne and M.~Krasnansky,
  JHEP {\bf 0604} (2006) 020
  [arXiv:hep-th/0602216].


\bibitem{Gracey:2006jc}
  J.~A.~Gracey,
  arXiv:hep-th/0605037.




\bibitem{KS} 
G. K\"allen, A. Sabry, \emph{Da. Mat. Fys. Medd.} 29, No.17 (1955) 1.

\bibitem{Djouadi} 
A. Djouadi, P.Gambino \emph{Phys. Rev.} {\bf{D49}} (1994) 3499

\bibitem{BroFleTar} 
D.J. Broadhurst, J. Fleischer, O.V. Tarasov, \emph{Z. f\"ur Physik} {\bf{C60}} (1993) 287 .












\bibitem{Argeri:2002wz}
  M.~Argeri, P.~Mastrolia and E.~Remiddi,
  %
  Nucl.\ Phys.\ B {\bf 631} (2002) 388
  [arXiv:hep-ph/0202123].

\bibitem{Mastrolia:2002tv}
  P.~Mastrolia and E.~Remiddi,
  Nucl.\ Phys.\ B {\bf 657} (2003) 397
  [arXiv:hep-ph/0211451].

\bibitem{Laporta:2003xa}
  S.~Laporta, P.~Mastrolia and E.~Remiddi,
  Nucl.\ Phys.\ B {\bf 688} (2004) 165
  [arXiv:hep-ph/0311255].

\bibitem{Laporta:2004rb}
  S.~Laporta and E.~Remiddi,
  Nucl.\ Phys.\  B {\bf 704} (2005) 349
  [arXiv:hep-ph/0406160].



\bibitem{ibp}{
   Chetyrkin, K. G. and Tkachov, F. V.,
   Nucl. \ Phys. B {\bf 192} (1981) 159.

   Tkachov, F. V.,
   Phys. \ Lett. B {\bf 100} (1981) 65.
}

\bibitem{Laporta:1996mq}
  S.~Laporta and E.~Remiddi,
  Phys.\ Lett.\  B {\bf 379} (1996) 283
  [arXiv:hep-ph/9602417].

\bibitem{Mastrolia:2000va}
  P.~Mastrolia and E.~Remiddi,
  Nucl.\ Phys.\ Proc.\ Suppl.\  {\bf 89} (2000) 76.



\bibitem{ibpsolver}{

{\tt SOLVE}, by E.~Remiddi, available from the author.

{\tt SYS}, by S.~Laporta, \cite{Laporta:2001dd} .

{\tt IdSolver}, by M.~Czakon.

{\tt AIR}, by C.~Anastasiou and A.~Lazopoulos, 
  JHEP {\bf 0407} (2004) 046
  [arXiv:hep-ph/0404258].
}


\bibitem{Baikov:2005nv}
  P.~A.~Baikov,
  Phys.\ Lett.\  B {\bf 634} (2006) 325
  [arXiv:hep-ph/0507053].

\bibitem{Smirnov:2005ky}
  A.~V.~Smirnov and V.~A.~Smirnov,
  JHEP {\bf 0601} (2006) 001
  [arXiv:hep-lat/0509187].

\bibitem{SMatrix}
 R. J. Eden, P. V. Landshoff, D. I. Olive, and J. C. Polkinghorne 
  ``The Analytic S-Matrix,''
{\it  Cambridge University Press (1966) 287 p}



\bibitem{DimReg}{
t Hooft, G. and Veltman, M. J. G.,
Nucl. Phys.",
B {\bf 44} (1972) 189-213.

Bollini, C. G. and Giambiagi, J. J.,
Phys. Lett. B {\bf 40} (1972) 566-568;
Nuovo Cim. B {\bf 12} (1972) 20-25.

Ashmore, J. F.,
Lett. Nuovo Cim. 4 (1972) 289-290.

Cicuta, G. M. and Montaldi, E.,
Nuovo Cim. Lett. 4 (1972) 329-332.

Gastmans, R. and Meuldermans, R.,
Nucl. Phys. {\bf B} 63 (1973) 277-284.

Collins, J. C.,
{\it Renormalization}, "Cambridge University Press", (1987).

}



\end{thebibliography}
\end{document}